\documentclass[11 pt]{report}

\usepackage{gmudissertation}

\usepackage{graphicx}                    
\usepackage{amsmath}                     
\usepackage{amsfonts}                    
\usepackage{amssymb}                     
\usepackage{amsthm}                      
\usepackage[normalem]{ulem}              
\usepackage[noadjust,verbose,sort]{cite} 
\usepackage{setspace}                    
\usepackage{natbib}			 
\usepackage{aas_macros}
\usepackage{subfig}
\usepackage{pdflscape}
\usepackage{threeparttable}
\usepackage{color}

\beforedoc

\begin{document}

\title{Investigation of the forces that govern the three-dimensional \\
propagation and expansion of coronal mass ejections\\
from Sun to Earth}
\onelinetitle{Investigation of the Forces that Govern the Three-Dimensional Propagation and Expansion of Coronal Mass Ejections from Sun to Earth}
\author{Robin C. Colaninno}
\degree{Doctor of Philosophy}
\doctype{Dissertation}
\dept{Physics}
\discipline{Physics}

\seconddeg{Master of Science}
\seconddegschool{My Former School}
\seconddegyear{Year of second degree}

\firstdeg{Bachelor of Science}
\firstdegschool{Guilford College}
\firstdegyear{2002}

\degreeyear{2012}

\degreesemester{Spring Semester}

\advisor{Dr. Angelos Vourlidas}

\firstmember{Dr. Robert Weigel}

\secondmember{Dr. Arthur Poland}

\thirdmember{Dr. Jie Zhang}

\depthead{Dr. Michael Summers}

\assdean{Dr. Timothy L. Born}
\deanITE{Dr. Vikas Chandhoke}


\signaturepage

\titlepage

\copyrightpage


\dedicationpage

\noindent I dedicate this dissertation to my husband for all his love and support.


\acknowledgementspage

\noindent I would like to thank all my colleagues that have provided assistance and collaboration on this work. I would particularly like to acknowledge Dr. Teresa Nieves-Chinchilla for all her help with the in situ data and analysis. I would also like to acknowledge Dr. Chin-Chun Wu, Dr. Brian Wood, Dr. Oscar Olmedo and Dr. Watanachak Poomvises for their assistance.

\tableofcontents

\listoftables

\listoffigures

\abstractpage

In the last few decades, we have discovered that the environment of our solar system is as dynamic as terrestrial weather. The source of space weather is the Sun which produces winds and storms that effect modern human systems. The most geo-effective aspect of space weather is Coronal Mass Ejections (CMEs) which are analogous to terrestrial hurricanes.  These powerful storms, comprised of plasma and magnetic fields, can significantly disrupt Earth's magnetic field and cause a range of terrestrial effects from the aurora to the destruction of technological infrastructure. It is only recently that we have been able to continuously monitor CMEs as they progress from the Sun to Earth. Even with continuous monitoring with remote sensing observations, we are still unable to accurately predict the arrival or terrestrial impact of a CME. There is much about the evolution of CMEs we do not understand. 

In this study, we analyze nine CMEs from the Sun to Earth as observed in both the remote sensing and in situ data sets. To date, this is the largest study of Earth impacting CMEs using the multi-view point remote sensing and in situ data. However, the remote sensing and in situ data of the same CME cannot be directly compared. Thus, we use 

\abstractmultiplepage
\noindent several models to parameterize the two data sets. We are able to compare the arrival time, Earth impact speed, internal magnetic field, size and orientation as derived from the remote sensing and in situ methods. By comparing these results, we hope to gain a more comprehensive understanding of the inner heliospheric evolution of CMEs. 

In this study, we track CMEs from the Sun to 70\% - 98\% of the distance to Earth with the remote sensing data.  We analyze the propagation and expansion of the CMEs by applying two geometric models of their structure. From the derived kinematics, we compare the predicted arrival times and impact velocities with the in situ data. We find that even with nearly continuous observations and the best available model of the CME structure, there is still a significant error in the predicted values. We discuss the possible causes of these errors.

To investigate the drivers of the CME, we use the derived kinematics and mass of the CMEs to calculate the net force. We estimate the various forces acting on the CME as predicted by three theoretical models of CME propagation and expansion and compare these results with the observational results. We find that the flux rope model of \citet{1989ApJ...338..453C} provides the best agreement with the observations. With the flux rope model, we are able to predict the internal magnetic field of the CME near Earth from the remote sensing data to an order of magnitude. This result is of great importance to space weather predictions since the strength of the internal magnetic field is a major factor in the geo-effectiveness of a CME. We discuss a possible method of better fitting the flux rope model to remote sensing observations with the goal of improving the accuracy of the predicted magnetic field values. 

Finally, we compare the size and orientation of the CMEs as predicted from the remote sensing and in situ data. We find very little agreement between the values derived from the two data sets. This aspect of the study also has large implications to space weather since the orientation of the CME's magnetic field is the other major factor in geo-effectiveness. Since the in situ data is our only source of direct information of the magnetic field and remote sensing data the main means of prediction, it is essential that the analysis of these two data sets is reconciled.

\startofchapters

\chapter[Introduction]{Introduction}

\begin{figure}
\begin{center}
\includegraphics[width=3in]{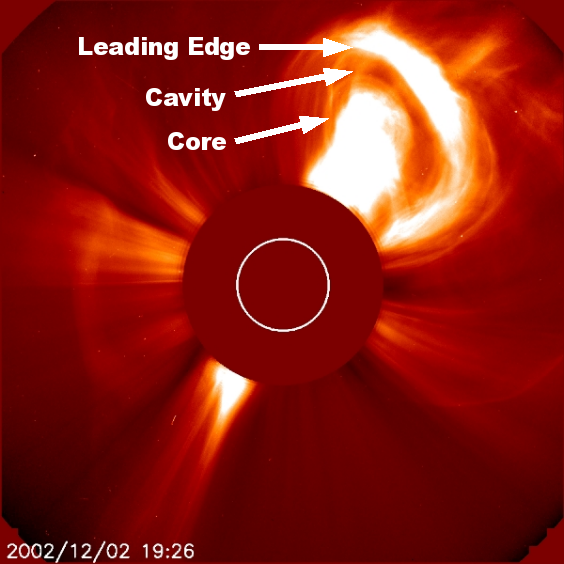}
\caption{A three-part structure CME showing the loop like leading edge, dark cavity and bright core as seen by SOHO-LASCO C2 on December 2, 2002.}
\label{three_part_structure}
\end{center} 
\end{figure}

Coronal mass ejections (CMEs) are complex, dynamic, three-dimensional structures comprised of energetic particles, plasma and magnetic field that are ejected away from the Sun. In coronagraphic and heliospheric image data, CMEs appear as a structured brightening in the corona moving away from the Sun. Many large CMEs have a three-part structure which includes a bright leading edge, a dark void, and a bright complex core, see figure \ref{three_part_structure} \citep{1985JGR....90..275I}. The bright relatively smooth leading edge is created by coronal material displaced by the motion of the CME. The leading edge demarcates the boundary of the magnetically closed region of the CME. A shock is sometimes associated with the front of the CME. The shock forms ahead of the CME leading edge and is much dimmer than the leading edge. Near the center of the CME envelope is a bright core often with a complex, twisted structure. The CME core is associated with material from an eruptive prominence. Between the leading edge and the core is a dark region comprised of rarefied material suggesting a magnetically closed cavity. 

\begin{figure}
\begin{center}
\includegraphics[width=5in]{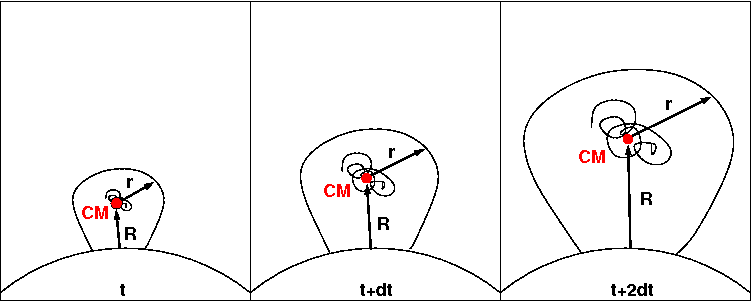}
\caption{As the CME moves away from the Sun, the center of motion ({\bf CM}) increases in height ({\bf R}). The change in {\bf R} with time is the propagation speed. At the same time the distance between the {\bf CM} and leading edge increases ({\bf r}). The change in the CME radius, {\bf r}, is the expansion speed.}
\label{expan_prop}
\end{center}
\end{figure}

The kinematics of CMEs is driven by the forces exerted by the magnetic field and gas pressure both internal and external to the CME. Internal to the CME, its magnetic structure and gas pressure drive the motion of the CME. These internal forces interact with each other and with the external magnetic field and gas of the solar wind. As a CME leaves the Sun, its evolution can be separated into two components; propagation and expansion.  The propagation speed is the change in height of the CME's center of motion above the solar limb. The expansion motion is the movement of the leading edge way from the center of motion \citep{2006ApJ...652.1747C}. Figure \ref{expan_prop} shows a diagram of the change in height and radius of the CME as it moves in the coronagraph's field of view.  A parameter often used to quantify CME motion is the speed of the leading edge. Ideally, the leading edge speed is the maximum of the superposition of the propagation and expansion speeds. Analytical models can be used to describe how the magnetic field and gas pressure drive the propagation and expansion velocity of CMEs \citep{1984SoPh...94..387P, 2006SoPh..239..293S, 1982ApJ...254..796L, 1989ApJ...338..453C}. 

Near Earth, magnetic and gas parameters can be detected by in situ instruments. Several spacecraft have instruments that detect the electromagnetic field, particle composition, speed, pressure, temperature and density of the near Earth space environment. Signatures in the interplanetary medium have been linked to the passage of the spacecraft through a CME. These signatures seen in the in situ data are referred to as interplanetary CMEs (ICMEs). Some ICMEs have a particular well-organized structure called a magnetic cloud (MC). The signature of a MC is defined by enhanced magnetic fields, a smooth rotation of the magnetic field vector, and a low proton density and temperature \citep{1981JGR....86.6673B}.

The magnitude and direction of the magnetic field of the ICME as it impinges on Earth's magnetosphere is the main driver of space weather and geomagnetic storms. Several studies found that large geomagnetic storms are caused by ICMEs \citep {1991JGR....96.7831G, 2002JGRA..107.1187R}. However, due to the lack of observational data and the complex nature of interplanetary interactions, the development of CMEs into ICMEs is unclear. Based on observations between 1996 and 2000, unique CME-ICME associations have been made for approximately half off all ICME events \citep{2003ApJ...582..520Z}.

\section{Observations of CMEs}

In this study, we use two types of data to study CMEs in the corona and heliosphere; remote sensing and in situ data.  Remote sensing observations are made by telescopes that image solar light scattered by electrons in the corona. In 1972, the first CME was observed in remote sensing data \citep{1972BAAS....4R.394T} with the coronagraph aboard the Orbiting Solar Observatory-7 \citep{1974SoPh...34..447K}. Since then, CMEs have been observed by a number of space-borne coronagraphs. Here, we use remote sensing data from two spacecraft; Solar TErrestrial RElations Observatory  (STEREO; \citealt{2008SSRv..136....5K}) and SOlar and Heliospheric Observatory (SOHO; \citealt{1995SoPh..162....1D}). In situ observations are made by instruments in the heliosphere measuring properties of the corona plasma as it passes over them. \citet{1982GeoRL...9.1317B} was the first to associate a magnetic cloud measured in situ by Helios 1 spacecraft with a CME seen in remote sensing data with the Solwind coronagraph on the spacecraft P78-1 \citet{1980ApJ...237L..99S}. Here, we use in situ data from the WIND mission to find the corona plasma signatures of CMEs near Earth.

\subsection{Remote Sensing}
\begin{figure}
 \begin{center}
\includegraphics[width=5in]{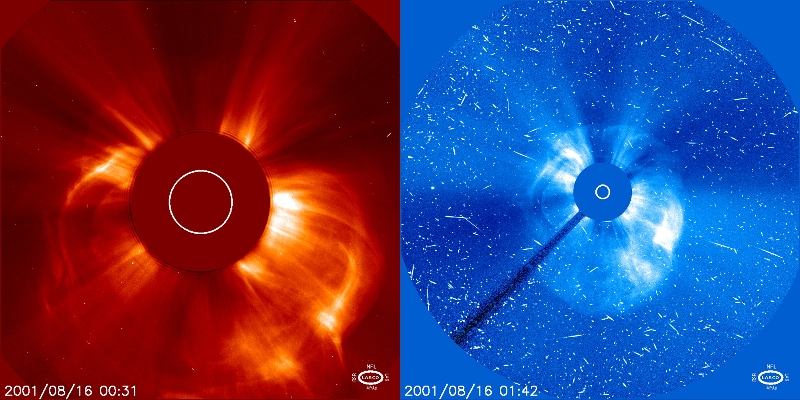}
\caption{Shows LASCO C2 (left) and LASCO C3 (right) observations of a large CME on 16 August 2001. The LASCO C3 image also shows the impact of fast, high energy protons that struck the SOHO spacecraft's detectors.}
\label{lascoC2C3}
\end{center}
\end{figure}

 Since its launch in 1996, the Large Angle Spectroscopic Coronagraph (LASCO) aboard the SOHO mission has been observing the corona, almost continually.  LASCO observes the corona from the L-1 Lagrange point along the Sun-Earth line. LASCO is comprised of two operating coronagraphs; C2 and C3. The LASCO C2 has a field of view from 2.2 to 6 $R_\odot$. LASCO C3 has an overlapping field of view from 3.8 to 32 $R_\odot$. Figure \ref{lascoC2C3} shows an example of the LASCO C2 and C3 observations during a large CME.

\begin{figure}
\begin{center}
\includegraphics[width=5in]{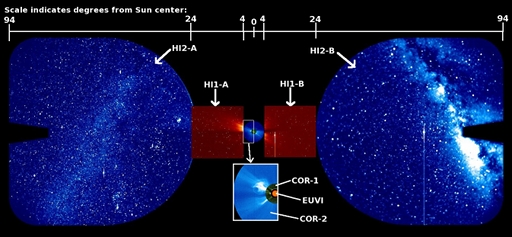}
\caption{Shows SECCHI instrument suite's field of view along the ecliptic plane from the Sun to Earth for a single point of view. The insert shows the EUVI, COR1 and COR2 fields of view which cover the corona from the solar disk to 15 $R_\odot$ approximately equal to 4$^o$ elongation. The HI-1 field of view begins at 4$^o$ and is 20$^o$ by 20$^o$.  The HI-2 field of view is 90$^o$ square and a slightly overlaps the HI-1 field of view. The field of view of both HI instruments are centered on the ecliptic plane.  The trapezoidal occulter at the end of the HI-2 field of view is to block Earth shine.}
\label{fov}
\end{center}
\end{figure}

The Sun-Earth Connection Coronal and Heliospheric Investigation (SECCHI) instrument suite abroad the STEREO mission offers a novel view of the Sun-Earth system. STEREO, launched on October 25, 2006, is comprised of two spacecraft with nearly identical instrumentation. The two STEREO spacecraft orbit the Sun; the Ahead (A) spacecraft slightly faster than Earth and the Behind (B) spacecraft slightly slower. The two spacecraft are separating from Earth at a rate of 22.5$^o$ per year. SECCHI is the imaging instrument suite of the mission. The SECCHI suite includes an extreme-ultraviolet imager, two coronagraphs and two heliospheric imagers. Together these instruments image the inner heliosphere from the Sun to Earth from two viewpoints. The two coronagraphs in the SECCHI suite are COR1 with a field of view from 1.5 to 4 $R_\odot$ and COR2 from 2.5 to 15 $R_\odot$ \citep{2008SSRv..136...67H}. SECCHI also includes two heliospheric imagers (HI-1, HI-2) which are similar to coronagraphs but have no occulter and a field of view off pointed from the center of the Sun. The heliospheric imagers view one side of the corona along the Sun-Earth line. HI-1 and HI-2 have square fields of view centered on the elliptic plane from 15 to 84 $R_\odot$ (20$^o$)  and 66 $R_\odot$ to 1 AU (70$^o$), respectively \citep{2008SSRv..136...67H}. Figure \ref{fov} shows the combined fields of view of the SECCHI instrument suite.

When studying CMEs with remote sensing data it is very advantageous to combine the data from SOHO and STEREO. 
The LASCO data set has offered abundant insight to the nature of CMEs over the solar cycle. From the LASCO data, physical parameters such as position, angular width, speed, mass and acceleration have been cataloged in the CDAW database for all observed CMEs \citep{2004JGRA..10907105Y}. However, the LASCO CME measurements were made using data from a single view point and are two-dimensional projections of three-dimensional values. With the two views of SECCHI, we are able to derive the true three-dimensional position, velocity and mass of CMEs \citep{2010AnGeo..28..203M, 2009ApJ...698..852C, 2009SoPh..256..111T}. With the LASCO data, CMEs have not been measured continually from the Sun to 1 AU. Thus the propagation of CMEs in the inner heliosphere and three-dimensional structure and dynamics of CMEs has only become available with the SECCHI data. For the first time, we can observed CMEs from the Sun to 1 AU from two viewpoints away from the Sun-Earth line. 

\subsection{In Situ}
The WIND spacecraft was launched on 1 November 1994 with the primary science objectives to provide data for magnetospheric and ionospheric studies and investigate basic plasma processes occurring in the near-Earth solar wind. The WIND spacecraft has a suite of several instruments to collect data in situ. In this study, we will use data from two of these instruments; Magnetic Field Investigation  (MFI; \citealt{1995SSRv...71..207L}) and Solar Wind Experiment (SWE; \citealt{1995SSRv...71...55O}). The MFI instrument is a triaxial magnetometer which provides the magnitude and direction of the solar wind's magnetic field. The MFI instrument has a dynamic range of $\pm$4 nT to $\pm$ 65,536 nT with a sensitivity of 0.008 nT/step quantization. The SWE instrument includes two Faraday cup sensors and an electrostatic analyzer. The Faraday cup sensors provide the density, velocity and temperature of the ions of the solar wind. The electrostatic analyzer measures the solar wind electron pitch-angle distribution function. With the data from these two instruments, we can detect CMEs and MCs as they pass over the WIND spacecraft. 

\section{Theoretical Models of CME Structures}\label{Structures Models}

In this study, we use three theoretical models to analyze the three-dimensional structure of CMEs from the observational data. We use two models to fit the remote sensing data; the Elliptical Model and the Graduated Cylindrical Shell Model. To model the in situ data, we use the Force-Free Elliptical Flux Rope Model. We have chosen to use the Elliptical Model because the remote sensing observations show the cross-section of the CME in the inner heliosphere to be elliptical.  For the same reason, we have chosen an in situ model which also has a elliptic cross section. We expect that as the CME expands latitudinally the trajectory of an in situ spacecraft through the CME could be approximated by an elliptical cylinder. However, the Elliptical Model is very limited especially in dealing with the projection effects in the remote sensing data. Thus, we also use the more sophisticated Graduated Cylindrical Shell Model to determine the trajectory of the CME. The drawback of the Graduated Cylindrical Shell Model is it has a circular cross-section. Thus, we can not use this model to analyze the expansion of the CME in different directions. 

\subsection{Elliptical Model}\label{elliptical_model}

The most common method of deriving physical parameters from remote sensing images is direct measurements made of features in the image. These measurements are made by an observer choosing a feature in the image and tracking that feature over many images. The data from the SOHO LASCO CME Catalog \citep{2004JGRA..10907105Y} is collected in this way where the speed, acceleration, position angle, and width are determined by an observer selecting the pixel locations of the leading edge height and angular extent of the CME over three or more images. 

\begin{figure}
\begin{center}
\includegraphics[width=0.7\textwidth]{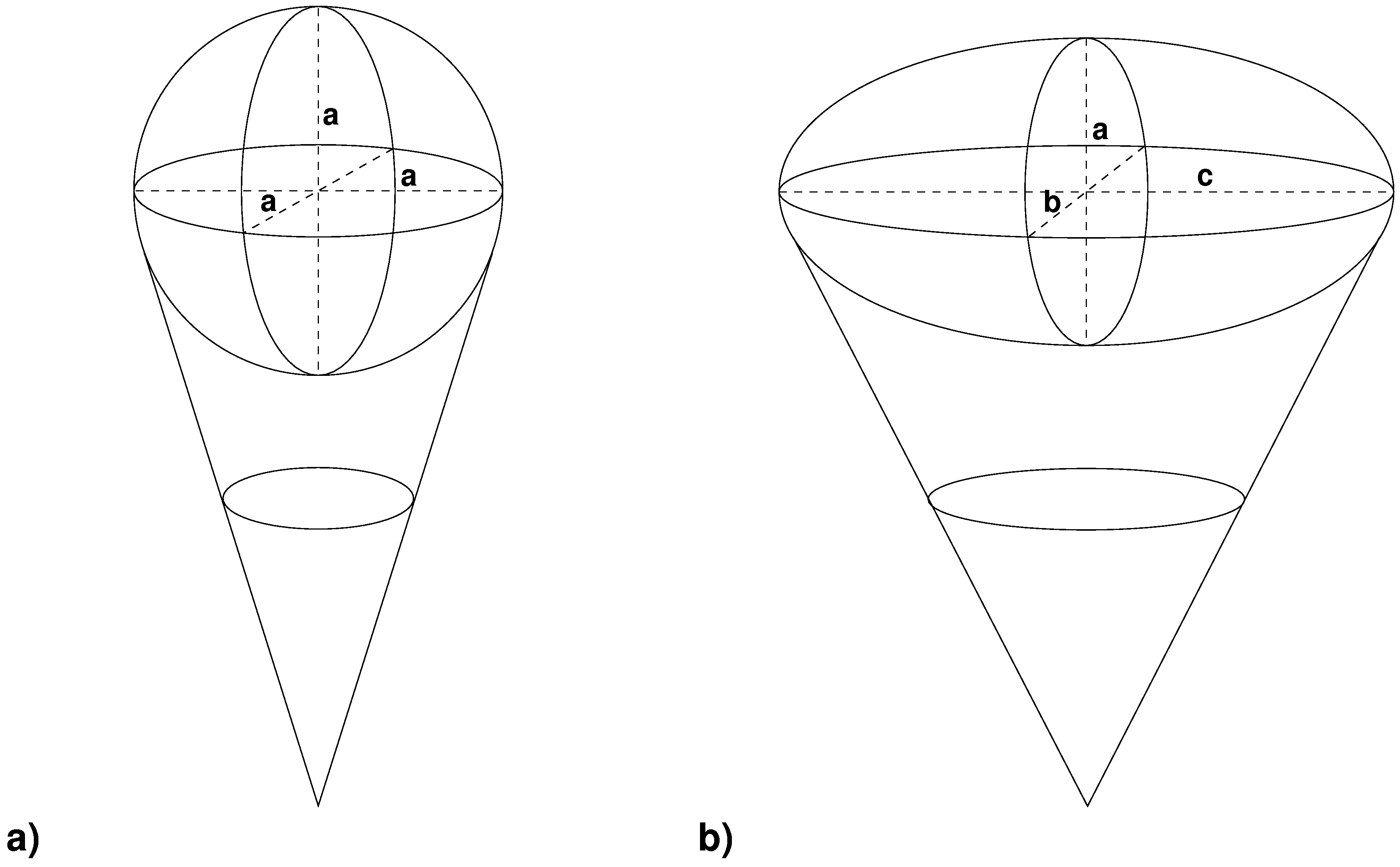}              
\caption{Geometric representation of the (a) Cone Model and the (b) Ellipse Model. For the Cone Model the radii of all the axes are equal. For the Ellipse Model the three axes can be of different lengths. }
\label{Fig:ellipse_cone_model}
\end{center}
\end{figure}

The structure of the CME can also be analyzed using direct measurements by selecting the points in the image which define the envelope of the CME  and fitting a geometric structure to these points. The geometric structure most widely used in the literature is the Cone Model.  \citet{1982ApJ...263L.101H} introduced the Cone Model and it has been widely used to analyze the structure of CMEs \citep[e.g.][]{2002JGRA..107.1223Z, 2006SpWea...410002X}. The Cone Model subscribes an ice cream cone like structure to CMEs where the cavity of the CME is described by a sphere and is attached back to the surface of the Sun by a spherical cone. The geometry of the Cone Model is represented in figure \ref{Fig:ellipse_cone_model}(a). The Ellipse Model used here is a two dimensional variation of the Ellipsoid Model were the cavity of CME is described with an elliptical cross-section. Figure \ref{Fig:ellipse_cone_model}(b) shows the geometry of the Ellipse Model. Of course this model is easily reduced to the Cone Model where the length of all the axes are equal.

The Elliptical Model is implemented by making direct measurements of the outer envelope of the CME structure in the image. The trailing (sunward) edge of the CME is the most difficult to identify. We define the trailing edge as the apex of the V feature of the CME \citep{2007JGRA..11209103K}.  If there is no clear V feature, we identify any trailing feature that is easily track in the images. After six points have been selected, an ellipse of the form equation \ref{ellipse} is fitted to the direct measurements and over plotted on the image. The observer can then acept the fit or add more points until a satisfactory fit is obtained. Thus the number of points used for each ellipse fit is determined by the observer. 
\begin{align}
x(\theta) &= x_c + a \cos \theta \cos \phi - b \sin \theta \sin \phi \nonumber \\
y(\theta) &= y_c + a \cos\theta \sin \phi + b \sin \theta \cos \phi
\label{ellipse}
\end{align}
For each fit, we get the center of the structure, $(x_c, y_c)$, the radii of the major and minor axis, ($a, b$) and the tilt angle ($\phi$).  We define the minor axis as the parallel radius and the major axis and the perpendicular radius where parallel and perpendicular are defined with respect to direction of the CME propagation.
  
The Elliptical model is a very basic first order approximation of the three-dimensional geometry of a CME. While the simplicity of the model is compelling, it is merely a geometric construct of the density and has no theoretical explanation. Also the model does not address the effects of projection in the images since it relies  on direct measures.  Direct measurements of CMEs are useful because of their simplicity of collection. However, direct measurements must be interpreted with care because of the projection effects that are dependent on the orientation of the CME.  All measurements made directly from remote sensing data are projected on the plane of the observation. For example, the perpendicular diameter from the direct measurements is an upper limit to the cross-section of the CME. If the CME is oriented face-on then the perpendicular diameter is the width of the CME. Only if the CME is oriented edge-on is the perpendicular diameter the cross-section of the CME. For any orientation of the  CME between these two extremes, the perpendicular diameter will be larger than the cross-section.  It is difficult to correct for this enlargement since, we do not know which part of the CME will be projected onto the plane of the sky as the furthest extent of the flank. A similar argument can be made about the parallel diameter from the direct measurements. However, in contrast to the perpendicular diameter, the parallel diameter is the lower bound of the CME  cross-section. If the CME is oriented in the plane of the image then the parallel diameter will be the cross-section. If the CME is oriented out of the plane of the image then the parallel diameter will be shortened by the projection onto the plane of the image. Thus the parallel diameter will always be less then or equal to the true cross-section.

\subsection{Graduated Cylindrical Shell Model}\label{gcs_model}
\begin{figure}
  \centering
  \subfloat{\includegraphics[width=0.55\textwidth]{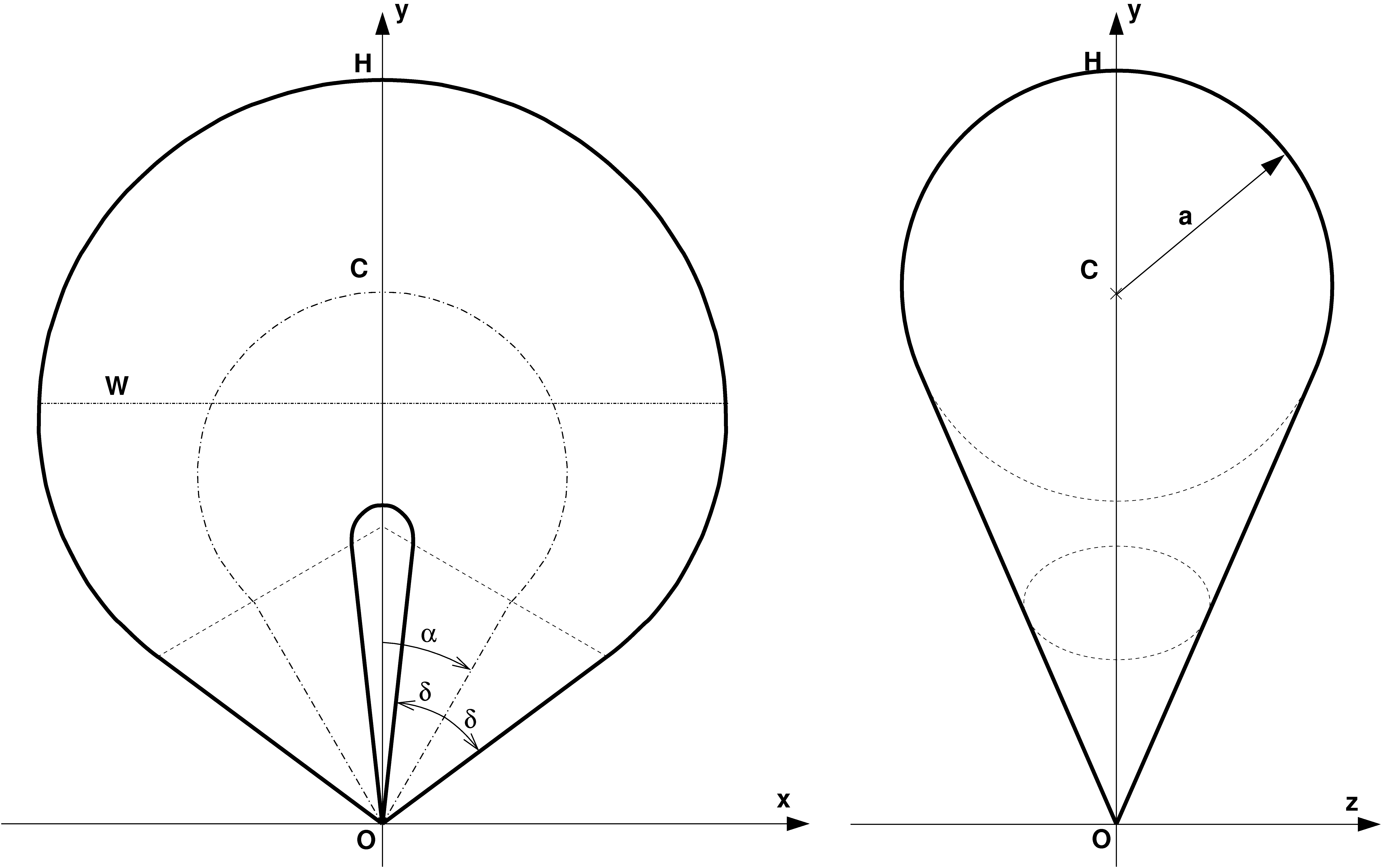}}                
  \subfloat{\includegraphics[width=0.45\textwidth]{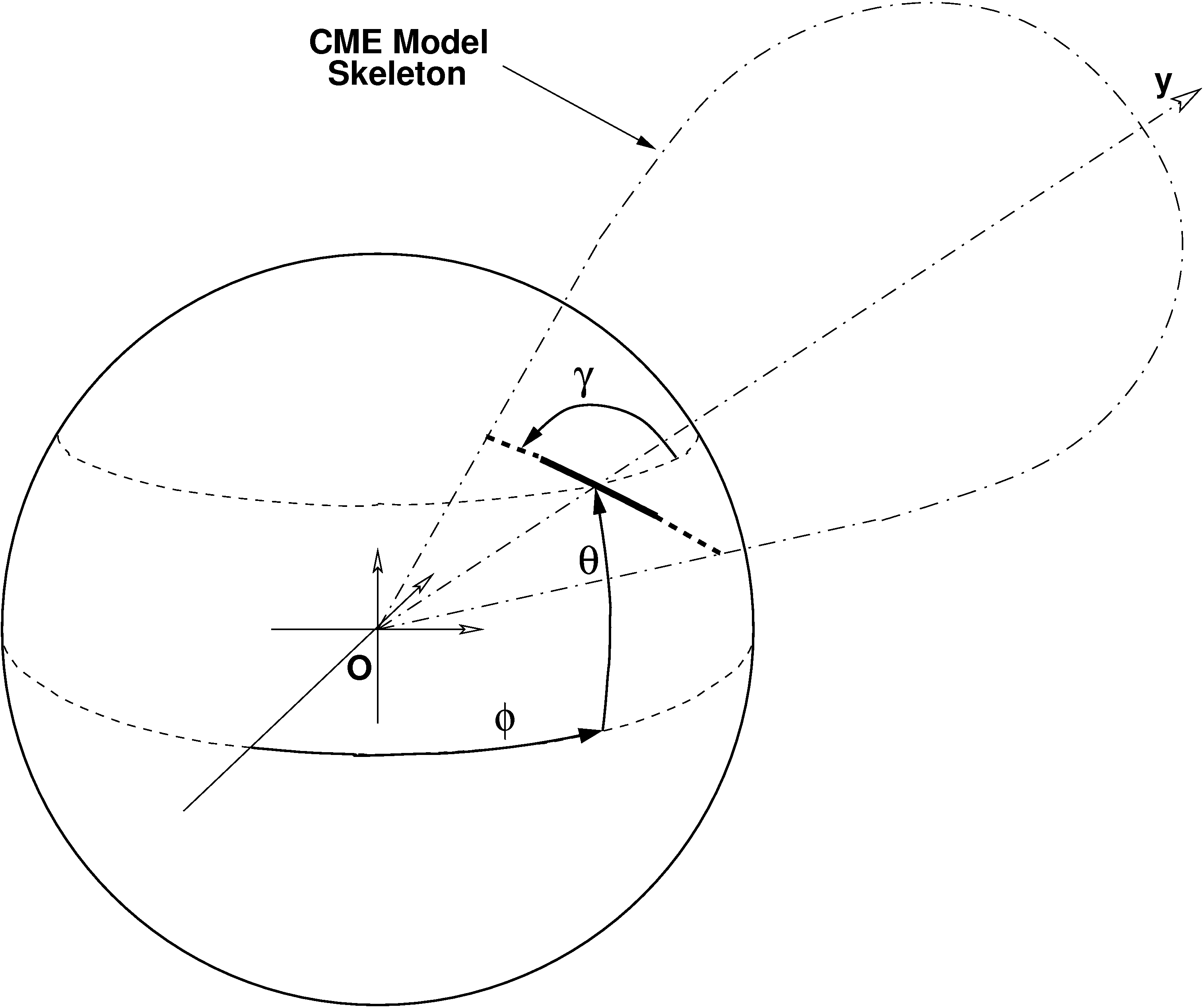}}     
  \caption{Representation of the graduated cylindrical shell (GCS) model (a) face-on and (b) edge on. The dash-dotted line is the axis through the center of the shell, C. The solid line represents a planar cut through the cylindrical shell at the origin. The width, W, of the model is defined as the largest vertical extent, dotted line. The radius, a, defines the circular cross-section. These values are controlled by the model fitting parameters height (H), ratio ($\delta$) and half angle ($\alpha$) \citep{2009SoPh..256..111T}. }
  \label{Fig:GCS}
\end{figure}

The graduated cylindrical shell model (GCS) was developed by \citet{2006ApJ...652..763T} and \citet{2009SoPh..256..111T}. This model provides a method for analysis of the three-dimensional morphology, position and kinematics of CMEs in white-light remote sensing observations. The GCS model uses forward-modeling techniques which allows the user to fit a geometric repesentation of a flux rope to CME observations. The geometry of the empirically flux rope model is represented in figure \ref{Fig:GCS}. The model is positioned using the longitude, latitude  and the rotation parameters. The origin of the model is fixed at the center of the Sun. The size of the flux rope model is controlled using three parameters which define the apex height, foot point separation and the radius of the shell. The details of the model as well as the derivation of many secondary parameters used in this study are discussed by \citet{2011ApJS..194...33T}.

The GCS model has been used in several studies of various aspects of the three-dimensional structure and kinematics of CME in the heliosphere. \citet{2010JGRA..11507106L} successfully used the GCS model to derive the orientation of the flux rope in remote sensing data for comparison with in situ observation.   \citet{2011ApJ...733L..23V} used this model to show the rapid rotation of a CME in the middle corona. Also, \citet{2010PhDT.......182P} used the GCS model to analyze the propagation and expansion of several CMEs with the goal of deriving a solar wind drag force.

This model is implemented by over plotting the projection of the cylindrical shell structure onto each image.  The observer then adjusts the six parameters of the model until a best fit with the data is achieved. We fit the GCS model to all available images from the CME's emergence in the SECCHI COR2 and LASCO C2 fields of view until the SECCHI HI-2 images. We analyze the variation of the model fit parameters over time to derive the kinematics of the CME.  

The GCS model is a more sophisticated method of modeling the three-dimensional structure of the CME when compared with the Ellipse model. The GCS model correctly handles the effects of projection for each image's point of view. Also the geometry of the GCS model is a good proxy for the flux rope like magnetic structure of CMEs. However, this technique is limited in that the empirically defined model has a circular cross-section. Thus we are only able to fit the overall structure of the CME and cannot model any distortion in the interplanetary space.

\subsection{Non Force-Free Elliptical Flux Rope Model}\label{EFR}
Since \citet{1981JGR....86.6673B} identified the signatures of magnetic clouds, several models have been presented in the literature for  reconstructing their magnetic structure. The most common is the model of \citet{1990JGR....9511957L}  which solves a force free-form of Maxwell's equations in cylindrical coordinates. The force-free cylindrical torus is the basic framework for understanding the interplanetary magnetic cloud. It has provided an approach for the interpretation for the magnetic topology and orientation of the magnetic cloud from in situ data. 

The analytical model used in this study with the in situ data was published by \citet{2002GeoRL..29m..15H} and developed by \citet{2009EM&P..104..109N} with successive improvements in order to facilitate implementation. It is similar to the \citet{1990JGR....9511957L} with two changes that make it more general: 1) a non force-free condition is assumed; and, 2) the coordinate system chosen to resolve the Maxwell equations is elliptical cylindrical. Both of these conditions have consequences in the physical understanding of the magnetic cloud. The first condition has consequences in the overall picture of the CME evolution in the interplanetary medium. Locally, at 1 AU, the assumption that the system is under a force-free condition is not correct. The second condition has a direct impact on the geometry of the magnetic cloud. The elliptic cylindrical coordinate system allows the magnetic cloud to have an elliptic cross-section as indicated by the remote sensing data. The relation of elliptical cylindrical coordinates to cartesian is given by
\begin{align}
x& = r \cosh \eta \cos \varphi \nonumber \\
y& = y \nonumber \\
z& = r \sinh \eta \sin \varphi
\label{elliptic}
\end{align}
where $r$ is the radial coordinate, $\varphi$ is the angular coordinate where $\varphi \in [0,2\pi)$ and $\eta$ is the coordinate that controls the eccentricity of the cross-section. 

The non force-free Elliptical Flux Rope model can generate many parameters to describe the CME. In this study, we are focused on comparing the orientation and size of the magnetic cloud with the remote sensing data. From the fitting of the model to the experimental data, we get several parameters related to the orientation of the flux rope. The three angles, ($\theta$, $\phi$, $\xi$) are the Euler angles relating the Y, X and Z axis, respectively, of the magnetic cloud to Geocentric Solar Ecliptic (GSE) coordinates. Also the vector, $Y_o$ gives the impact parameter, distance of the spacecraft crossing from the center of the flux rope. The shape of the flux rope is given by two parameters, radius of the major axis and the ratio of the major and minor axis given as a percentage. 
\begin{figure}
  \centering
  \subfloat[]{\includegraphics[width=0.325\textwidth]{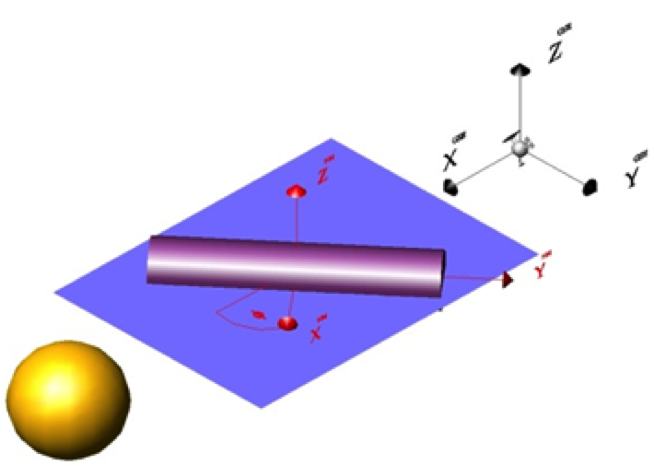}}                
  \subfloat[]{\includegraphics[width=0.25\textwidth]{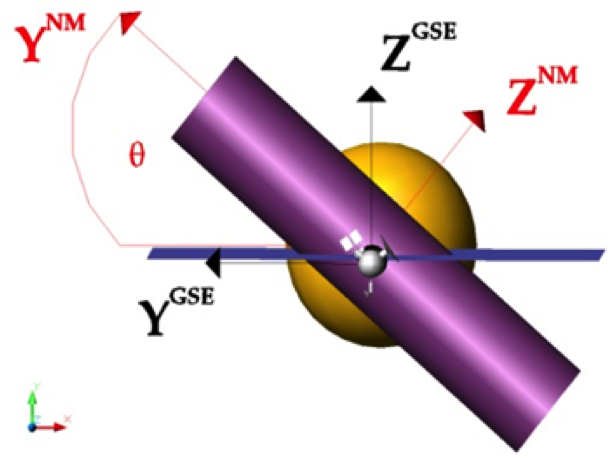}}
   \subfloat[]{\includegraphics[width=0.325\textwidth]{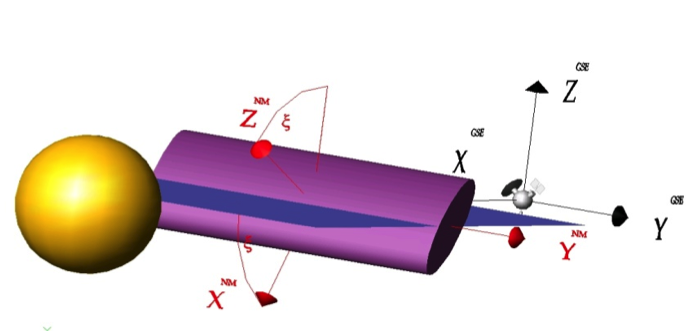}}
  \caption{The relation between the coordinates system of the magnetic cloud (X$^{NM}$,Y$^{NM}$,Z$^{NM}$) and the GSE coordinates system (X$^{GSE}$,Y$^{GSE}$,Z$^{GSE}$). The angles given by the model are (a) the longitude ($\theta$), (b) the latitude ($\phi$) and (c) the tilt angle ($\xi$) \citep{2005SoPh..232..105N}. }
  \label{Fig:MCangles}
\end{figure}

\section{Theoretical Models of CME Propagation and Expansion}\label{intro:theory}

Even with the availability of data from remote sensing and in situ instruments, without the application of analytical models, we still do not have access to the underlying physics of CME dynamics. Many theoretical models of CMEs allow us to relate their propagation and expansion back to unobservable quantities of the CME such as the mass, dimensions and kinematics. The three propagation and expansion models we will use here are; kinematic, self-similar MHD and flux rope. Each of these models takes a very different approach to solving for the propagation and expansion of CMEs. The models also have varying levels of complexity and physical insight.

\subsection{Basic Equations of Magnetohydrodynamics}
To begin reviewing the theoretical models of propagation and expansion; we must first review the basic equations of magnetohydrodynamics. The movement of a CME is controlled by the balance of forces interior and exterior to the CME. The two main forces that are interacting in the solar corona are the gas pressure and magnetic field forces. Due to the high coronal temperatures, the corona is completely ionized. The corona is a magnetized plasma and can be described by magnetohydrodynamics (MHD) to the first order. MHD is the combination of Maxwell's equations, Ohm's law, and the conservation equations of fluid dynamics to describe the bulk properties of a one-fluid plasma. Ideal MHD assumes that the plasma has infinite conductivity (or no resistivity). This assumption allows us to eliminate the electric field and current density terms from Maxwell's equations. Thus the plasma is parametrized by its density, pressure, magnetic field and velocity.

All of the analytical models reviewed here, to some extent draw on the basic equations of ideal MHD. Here we will briefly review the basic equations of MHD for later use in the CME propagation and expansion models. The self-similar MHD model solves the ideal MHD equations directly while the kinematic and flux rope models use the same simplification of Maxwell's equations used to derive the ideal MHD equations.

Maxwell's equations describe the evolution of the electric field, ${\bf E}({\bf r},t)$, and magnetic field, ${\bf B}({\bf r},t)$ in response to the current density ${\bf J}({\bf r},t)$ and the electric change density $q({\bf r},t)$:
\begin{align}
&\text{Faraday's induction equation :} &&\nabla \times {\bf E}  = -\frac{1}{c} \frac{\partial {\bf B}}{\partial t} \label{faraday1} \\
&\text{Amp\'ere's law :} &&\nabla \times {\bf B}  = 4 \pi {\bf J} + \frac{1}{c} \frac{\partial {\bf E}}{\partial t}\label{ampere1}\\
&\text{Gauss's law:} &&\nabla \cdot {\bf E} = 4 \pi q\\
&\text{Gauss's law for magnetism :} &&\nabla \cdot {\bf B} = 0.
\end{align}
To relate the motion of the plasma with Maxwell's equations, we also need to include Ohm's law
\begin{equation}\label{ohm1}
{\bf E} + \frac{1}{c} ({\bf v} \times {\bf B}) = \eta {\bf J}.
\end{equation}
where $\eta$ is the electrical resistivity and $\bf{v}$ is the mean velocity of the plasma. 
The equations of gas dynamics for the the evolution of the density, $\rho({\bf r},t)$, and the pressure, $p({\bf r},t)$ are:
\begin{align}
&\text{continuity equation :} && \frac{\partial \rho}{\partial t} + \nabla \cdot (\rho {\bf v}) = 0 \label{mass1} \\
&\text{equation of motion :} &&\rho \left( \frac{\partial {\bf v}}{\partial t} + {\bf v} \cdot \nabla {\bf v}\right) = -\nabla p + {\bf J} \times {\bf B} + \rho {\bf g} \label{motion1} \\
&\text{equation of energy :} && \frac{d}{dt} \left( \frac{p}{\rho^\gamma} \right) = 0 \label{state1} 
\end{align}
where ${\bf g}$ is the acceleration due to gravity and $\gamma$ is the adiabatic index.

However, these equations are not yet in the most convenient and compact form for the self-consistent study of plasma. We begin reducing the equations by assuming the ideal case where the resistivity in equation \ref{ohm1} is $\eta=0$. Thus, we can substitute equation \ref{ohm1} into equation \ref{faraday1} to eliminate the electric field term,
\begin{equation}\label{diffb1}
\frac{\partial {\bf B}}{\partial t} = \nabla \times ({\bf v} \times {\bf B}). 
\end{equation}
We can easily assume that the plasma is non-relativistic, $v \ll c$. We can then estimate the order of magnitude the electric and magnetic field terms in equation \ref{ampere1} using equation \ref{ohm1}.
\begin{equation}
{\frac{1}{c} \left | \frac{\partial {\bf E}}{\partial t} \right |} 
\sim {\frac{v}{c} \frac{E}{l_o}} 
\sim {\frac{v^2}{c^2} \frac{B}{l_o}} 
\ll {\left | \nabla \times {\bf B} \right | \sim \frac{B}{l_o}}
\end{equation}
where $t_o$ and $l_o$ are the time and length scales of the plasma. Thus equation \ref{ampere1} can be reduced to :
\begin{equation}\label{ampere2}
\nabla \times {\bf B} = 4 \pi {\bf J}. 
\end{equation}
Solving equation \ref{ampere2} for the current and substituting the result in equation \ref{motion1}, we have
\begin{equation}\label{motion2}
{\rho \left({\frac{\partial {\bf v}}{\partial t} + {\bf v} \cdot \nabla {\bf v}}\right) = -\nabla p 
+ {\frac{1}{4 \pi}(\nabla \times {\bf B}) \times {\bf B}} + \rho {\bf g}}.
\end{equation}
Thus, the basic equations of ideal MHD are :
\begin{align}
&\frac{\partial \rho}{\partial t} + \nabla \cdot (\rho {\bf v}) = 0 \label{mass} \\
&{\rho \left({\frac{\partial {\bf v}}{\partial t} + {\bf v} \cdot \nabla {\bf v}}\right) = -\nabla p 
+ {\frac{1}{4 \pi}(\nabla \times {\bf B}) \times {\bf B}} + \rho {\bf g}} \label{motion} \\
&\frac{\partial {\bf B}}{\partial t} = \nabla \times ({\bf v} \times {\bf B}) \label{induction} \\
&{\nabla \cdot {\bf B} = 0} \label{divb} \\
&{\frac{\partial}{\partial t} \left(\frac{p}{\rho^\gamma}\right) + {\bf v} \cdot \nabla \left(\frac{p}{\rho^\gamma}\right) = 0} \label{state}
\end{align}

Another equation often used by the models is the Lorentz force. The Lorentz force for a infinitely conducting plasma can be divided into two terms, magnetic pressure and magnetic tension.
\begin{equation}\label{pressure_tension}
{\bf F_L} = \frac{1}{4 \pi} (\nabla \times {\bf B}) \times {\bf B}  = \frac{1}{4 \pi} {\bf B} \cdot \nabla  {\bf B} - \frac{1}{8 \pi} \nabla B^2
\end{equation}
where ${\bf B} \cdot \nabla  {\bf B} $ is the magnetic tension and $ \nabla B^2$ is the magnetic pressure. It is often useful to express the Lorentz force in this way because the magnetic pressure is analogous to the gas pressure. Another useful plasma parameter is the Alfv\'en speed, 
\begin{equation}
v_A = \frac{B}{\sqrt{4 \pi \rho}}
\end{equation}
It is the electrodynamic equivalent of the sound speed in a gas.

\subsection{Kinematic Model}\label{intro:kinematic}

The kinematic model is a simple view of CME propagation and expansion. In the kinematic model the CME is considered as defined volume enclosing a fixed mass. Thus, the forces on the CME volume can be solved for by integrating over the surface of the volume. Here, we will review two example of kinematic models; melon-seed and melon-seed overpressure. The kinematic model offers little insight to the physical properties of the CME since all motion is being driven by forces outside of the CME. However, the simplicity of the approach is attractive. If this simple approach can describe the expansion and propagation of the CME then perhaps the output of the kinematic model could be used in more sophisticated models to understand physical parameters of the CME.


\subsubsection{Melon-Seed Model}

In the melon-seed model, \citet{1984SoPh...94..387P} describes the CME as a magnetically isolated structure in the ambient solar coronal magnetic field. The magnetic structure of the CME defines a volume of constant mass. The forces acting on the volume of the CME are the ambient magnetic field, the ambient gas pressure and the solar gravity. The Lorentz force of the external magnetic field exerts both a magnetic tension and pressure force on the CME. The magnetic tension on the CME tethers it to the solar surface.  The gradient of the ambient gas pressure and the magnetic pressure of the coronal field are pushing the CME away.  The gradient of the gas pressure is assumed to be insignificant compared to the magnetic pressure.  Thus the visually descriptive name of the model comes from the coronal magnetic field squeezing the CME away from the Sun.  

From the ideal MHD equation, the contributions of the gas pressure and Lorentz force can be written as
\begin{equation}
{\bf F} =  -\nabla p + {\frac{1}{4 \pi}(\nabla \times {\bf B}) \times {\bf B}}.
\end{equation}
We can integrate the force expression over the entire volume of the CME structure, thus the net force on the center of mass can be expressed in the form of a surface integral 
\begin{equation} 
{\bf F} =  \iint \limits_S \left [  \frac{1}{4 \pi} ({\bf B_e} \cdot d{\bf s}){\bf B_e} - \left( P_e + \frac{B_e^2}{8 \pi} \right)d{\bf s} \right],
\end{equation}
where the subscript $e$ refers to the conditions external to the CME. The external Lorentz force has been expanded using equation \ref{pressure_tension}.  The magnetic pressure force can be converted into a volume integral of the form
\begin{equation}
{\bf F}_m = \iiint \limits_V  - \nabla \left( \frac{B_e^2}{8 \pi} \right) dV.
\end{equation} 
If we assume that the ambient field outside the CME can be approximated as constant around the CME, then the magnetic pressure force is
 \begin{equation}\label{pneuman_equ3}
{\bf F}_m \approx -V  \cdot \nabla \left( \frac{B_e^2}{8 \pi} \right).
\end{equation} 
Thus it is the total volume of the CME, not the specific shape, which is most important in determining the net force. 

\begin{figure}
\begin{center}
\includegraphics[width=60mm]{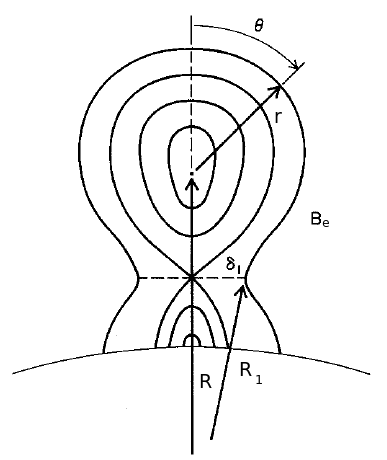}
\caption{Geometry of magnetic structure \citet{1984SoPh...94..387P}. The magnetic field lines connected to the photosphere tether the CME to the Sun while the magnetic pressure lifts the CME upward.}
\label{pneuman_fig1}
\end{center}
\end{figure} 
 
The derivation of the magnetic tension force is much more involved and does not effect the CME expansion. For a magnetic structure like that shown in figure \ref{pneuman_fig1}, the top section ($R<R_1$) of the CME does not contribute to the magnetic tension.  The only contribution to the magnetic tension is from the field lines less then $R_1$ which are tethering the CME to the solar surface.  

\citet{1984SoPh...94..387P} goes on to estimate the expansion of the CME with respect to the height for a pressure dominated and magnetic dominated interior assuming a force balance at the surface of the CME.  For the pressure dominated case, the internal gas pressure must balance the external magnetic pressure. Thus, we can write the volume of the CME as 
\begin{equation}
V = \pi r^2 R \phi = \frac{M}{\rho} = \frac{M v_s^2 }{p} \cong \frac{8 \pi M v_s^2}{B_e^2}
\end{equation}
where $M$ is the mass of the CME, $v_s$ is the internal sound speed, and $\phi$ is the angular extent of the arcade (perpendicular to figure \ref{pneuman_fig1}). If we solve for $r$ and express the magnetic field as radial $B_e = B_o R^2_\odot / R^2$ where $B_o$ is the field strength at the solar surface, then the radius of the CME goes as 
\begin{equation}
r = \left( \frac{8 M v_s^2}{\phi} \right )^{1/2} \frac{R^{3/2}}{B_o R^2_\odot}.   
\label{eq:pressure_case}
\end{equation}
Thus for the internally pressure dominated case, the CME expands faster than it propagates away from the Sun. For the other limiting case, the magnetically dominated interior, the radius goes as 
\begin{equation}
r = r_o \left ( \frac{R}{R_\odot} \right ),
\end{equation}
where $r_o$ is an arbitrary function. From observations of ICMEs, we see that the pressure dominated case is an incorrect description of the kinematics. In \citet{2008AnGeo..26.1919L}, the average speed of the MC is 458 km s$^{-1}$ which is an order of magnitude larger then the expansion speed. Thus the approximation that the gas pressure dominates the interior of the CME seems unlikely at 1 AU.  However, \citet{2006ApJ...652.1747C} shows that low in the corona some CMEs have bulk speed less then the expansion speed.


\subsubsection{Melon-Seed Overpressure Expansion Model }

The melon-seed overpressure expansion (MSOE) model \citep{2006SoPh..239..293S}, uses a very similar derivation to find the propagation and expansion of the CME. In the MSOE model, as in \citet{1984SoPh...94..387P}, the gradient of the ambient magnetic field pushes the CME away from the Sun as well as the external gas pressure.  The MSOE model describes fast CMEs that have been magnetically released from the photosphere and ignores any remaining magnetic tethering forces. The model is parametrized by the magnetic force pushing the CME up opposed by gravity and drag pulling it down. The radial expansion is parametrized by the overpressure interaction of the internal and external media using standard fluid dynamics.  This model strives to construct a minimal set of equations for which the propagation and expansion speeds are the outputs.

The propagation equation is a force equation with four terms that characterize the forces due to magnetic repulsion, inertia, gravity, and drag
\begin{equation}\label{siscoe_equ5}
F_{net} =  F_m - F_g - F_d. 
\end{equation} 
The force due to gravity is 
\begin{equation}\label{eq:Fg}
F_g =  M g_o \frac{R_o^2}{R^2},
\end{equation}
where M is the mass of the CME, R is the height of the CME center, $g_o$ is the gravitational acceleration at the solar surface and $R_o$ is the initial height of the CME.
The magnetic repulsion force is,
\begin{equation}\label{eq:Fm}
F_m =  Vol B_o^2 \frac{R_o^4}{R^5}
\end{equation}
where Vol is the volume of the CME and B$_o$ is the magnetic field at the surface.
The final term of the propagation equation is the drag force due to the solar wind,
\begin{equation}\label{eq:Fd}
F_d =  \frac{1}{2}Cd(R) Area(r) \rho_a (v-v_{sw})^2,
\end{equation}
where $v_{sw}$ is the solar wind velocity and $Cd(R)$ is the drag coefficient of solar wind. Though not denoted in \citet{2006SoPh..239..293S}, this force also requires a surface area which would be the surface area of the CME.

To calculate the expansion rate of the CME, \citet{2006SoPh..239..293S} borrows directly from classical hydrodynamics.  The magnetic field inside the CME is assumed to be stronger then the external field of the ambient corona and causes the CME to expand radially in all directions. The CME is magnetically isolated from the surrounding field, thus to conserve magnetic flux the magnetic pressure must decrease as
\begin{equation}
p_B = p_{B_o} \left ( \frac{r_o}{r} \right )^4.
\end{equation}  
The expansion is taken directly from classical hydrodynamics, only replacing the adiabatic gas pressure with the magnetic overpressure.  Thus the CME expansion is, 
\begin{equation}\label{siscoe_equ6}
r(t) \frac{d^2 r(t)}{dt^2} = - \frac{3}{2} \left( \frac{d r(t)}{dt}\right)^2 + \left ( \frac{r_o}{r(t)}\right)^4 \frac{
(A_{ho}^2 - A_{ao}^2)}{2} \frac{\rho_o}{\rho_a}
\end{equation}  
 where $A_{ho}$ is a hybrid Alfv\'en speed based on the initial overpressure field strength within the CME and the initial ambient density, $\rho_{ao}$ outside the CME. 
 
This formulation assumes that the radius of the CME is smaller than the scale height of the density of the ambient corona. In other words, the ambient density can be approximated as a constant. Also this formulation assumes that the ambient medium is incompressible. While this is not a good approximation, it mostly effects the kinematics of the surrounding medium which were are not addressing here.  

\begin{figure}
\begin{center}
\includegraphics[width=60mm]{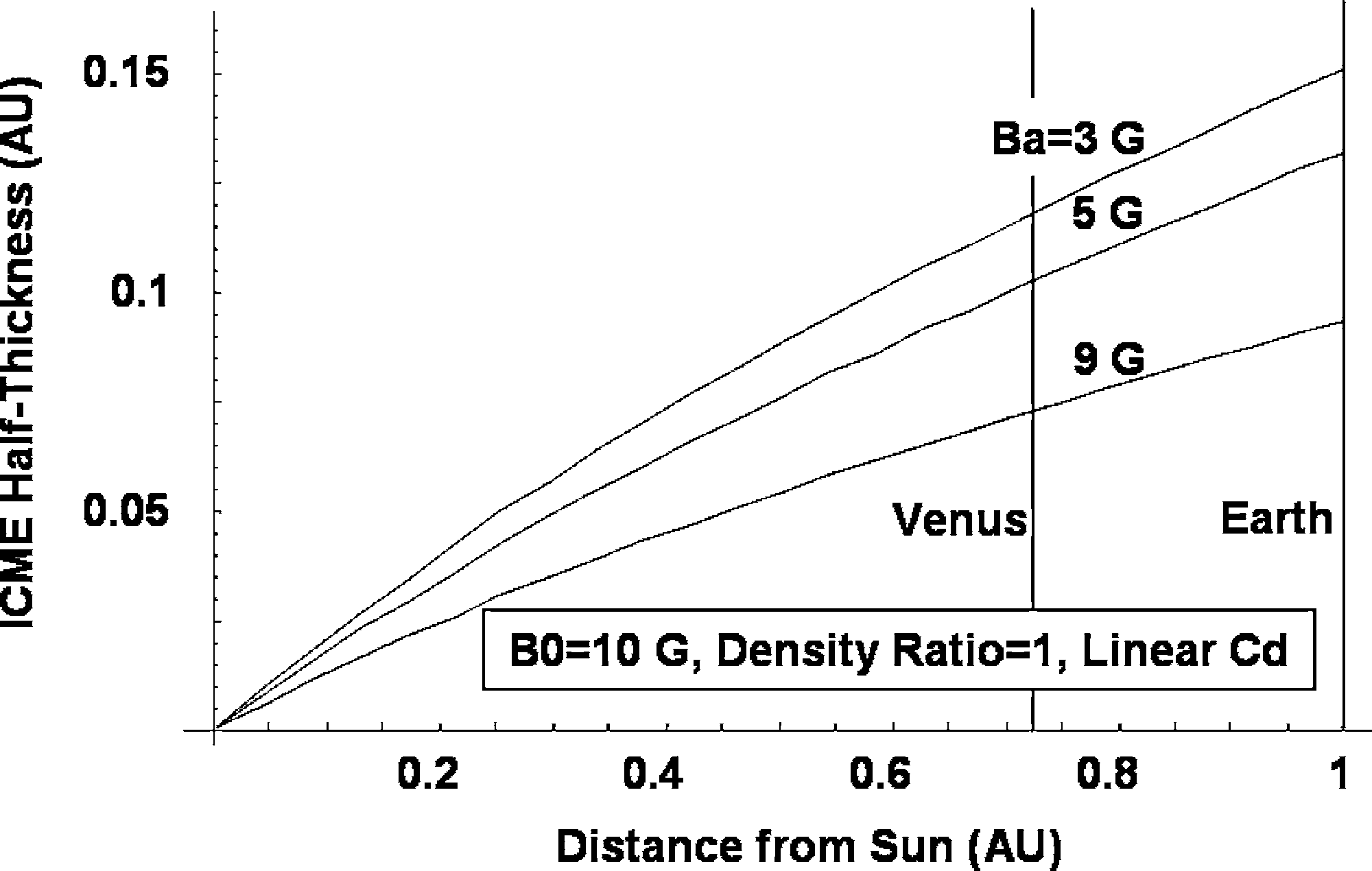}
\caption{ICME half-thickness as a function of distance from the Sun calculated for a set of initial field strengths \citep{2006SoPh..239..293S}.}
\label{siscoe_fig1}
\end{center}
\end{figure} 

With the addition of a solar wind model and a coronal density model, equations \ref{siscoe_equ5} and \ref{siscoe_equ6} form a set of coupled ordinary differential equations. The only input parameters that are needed are the external field strength, $B_a$, internal field strength, $B_o$, and the CME mass.  For nominal  input parameters, \citep{2006SoPh..239..293S} modeled the radius of the CME as it expands out into the interplanetary medium. For three values of the internal magnetic field, the radii obtained (figure \ref{siscoe_fig1}) compare well with measurements of ICME half-thicknesses at Earth and Venus \citep{2005JGRA..11001105O, 2002AdSpR..29..301R}.

The MSOE model uses the strength of both the internal and external magnetic fields, unlike  \citet{1984SoPh...94..387P} where only the external field is considered. Thus this formulation of the kinematic model offers more insight to the physical properties of the CME. However, many more inputs are necessary  since the coronal density and solar wind drag are included.


\subsection{Self-Similar Magnetohydrodynamics Model}\label{intro:ssmhd}

In contrast to the Kinematic models, the self-similar models focus on the magnetic field internal to the CME. The solutions derived from self-similar MHD can be applied to a variety of magnetic structures. Most offend the force-free solution is applied to the cylindrical of the magnetic cloud \citep{1981JGR....86.6673B}. Self-similar MHD models rely on directly solving the MHD equation for an azimuthally symmetric magnetic field that expands in a self-similar way. Self-similar solutions are often the leading term in an asymptotic expansion of a non-self-similar evolution, in a regime far from the initial conditions and far from the influence of boundary conditions \citep{1982ApJ...254..796L}.  For this model, we will solve it for two different sets of initial conditions. First we will assume a force-free configuration where there are no forces acting on the plasma of the CME.  This scenario is most applicable to magnetic clouds or ICMEs that have propagated away from the Sun into the interplanetary medium. Closer to the Sun, these assumptions are less applicable. The more general solution of the model is to assume that there is pressure gradient and gravity acting on the CME.  We will derive the solution for the self-similar MHD model for both situations. 

The self-similar MHD model assumes a force-free magnetic field, define as  
\begin{equation}
(\nabla \times {\bf B}) \times {\bf B} = 0.
\end{equation}
The force-free condition requires that
\begin{equation}\label{force-free}
(\nabla \times {\bf B}) = \alpha {\bf B} .
\end{equation}
where $\alpha$ is some function of position. The self-similar MHD model assumes the case where ${\bf B}$ is symmetric about the azimuthal axis and $\alpha$ is constant. In cylindrical coordinates, a force-free magnetic field with no azimuthal dependence can be written as,
\begin{equation}
{\bf B} = \hat{z} \times r B_t + \nabla \times (\hat{z} \times r B_p)
\end{equation}
the sum of a toroidal and poloidal magnetic field which do not depend on the azimuthal angle, $B_t(r, z)$ and $B_p(r, z)$ \citep{1956PNAS...42....1C}.  Thus the magnetic field is
\begin{equation}\label{cyclical_field}
{\bf B} = - r \frac{\partial }{\partial z} B_p \hat{r} + r B_t \hat{\theta} + \frac{1}{r} \frac{\partial }{\partial r} [r^2 B_p]\hat{z}
\end{equation}
and the curl of the magnetic field is 
\begin{equation}
\nabla \times {\bf B} = - r \frac{\partial }{\partial z} B_t \hat{r} + r \Delta_5 B_p \hat{\theta} + \frac{1}{r} \frac{\partial }{\partial r} [r^2 B_t]\hat{z}
\end{equation}
where
\begin{equation}
\Delta_5 = \left[ \frac{3}{r}\frac{\partial }{\partial r} +\frac{\partial^2 }{\partial r^2} + \frac{\partial^2 }{\partial z^2}\right ].
\end{equation}
For equation \ref{force-free} to be satisfied 
\begin{equation}
\nabla r^2 B_p = \nabla r^2 B_t.
\end{equation}
Thus the magnetic field is force-free if both the toroidal and poloidal fields are scalar functions.


\subsubsection{Free Expansion : Magnetic Clouds and ICMEs}

To formulate the self-similar MHD solution for a magnetic cloud or ICME, we begin by assuming that the influence of solar gravity is negligible.  If, as we showed in the previous section, the magnetic field configuration is force-free and the gas pressure gradient is also negligible then we can set all the force terms in the MHD equation of motion (equation \ref{motion}) to zero,
\begin{equation}\label{farrugia_equ2}
{\rho \left({\frac{\partial {\bf v}}{\partial t} + {\bf v} \cdot \nabla {\bf v}}\right) = -\nabla p 
+ {\frac{1}{4 \pi}(\nabla \times {\bf B}) \times {\bf B}} + \rho {\bf g}} = 0.
\end{equation}
If we assume that the velocity of the CME is radial, ${\bf v} = v(r, z, t) \hat{r}$,  then equation \ref{farrugia_equ2} can be reduced to 
\begin{equation}\label{free_expansion}
\frac{\partial v}{\partial t} + v\frac{\partial v}{\partial r} = 0.
\end{equation}
Equation \ref{free_expansion} is what \citet{1993JGR....98.7621F} calls free expansion.

If we apply a cylindrically symmetric, force-free magnetic field as defined by equation \ref{cyclical_field} to the MHD induction equation,  
\begin{equation}
\frac{\partial {\bf B}}{\partial t} = \nabla \times ({\bf v} \times {\bf B}), 
\end{equation}
for a radial velocity, the z-component of the induction equation is
\begin{equation}\label{farrugia_equ16}
\frac{\partial B_t}{\partial t} = - v \frac{\partial B_t}{\partial r.}
\end{equation}
We now introduce the assumption of self-similarity, which means that the magnetic field is a function of only one variable, $\zeta$, rather than radius and time, where 
\begin{equation}
\zeta = \frac{r}{\Phi},
\end{equation}   
and $\Phi(t)$ is a function of time only. We will show that the magnetic field is a function of $\zeta$ only more rigorously  in the next section.

Solving equation \ref{farrugia_equ16} for $v$, using $B_t (\zeta)$ gives
\begin{equation}\label{farrugia_equ18}
v =  \frac{r}{\Phi}\frac{d \Phi}{dt}
\end{equation}
The function of $\Phi(t)$ can now be solved for by substituting equation \ref{farrugia_equ18} into the free expansion equation \ref{free_expansion}, obtaining
\begin{equation}
\frac{\partial}{\partial t} \left( \frac{1}{\Phi} \frac{\partial \Phi}{\partial t} \right) - \left( \frac{\partial \Phi}{\partial t} \right)^2 = 0.
\end{equation}
The general solution to this equation is
\begin{equation}
\frac{1}{\Phi} \frac{\partial \Phi}{\partial t} = (t - t_o)^{-1}
\end{equation}
where $t_o$ is a constant of integration. Thus the velocity during free expansion is
\begin{equation}
v = \frac{r}{(t + t_o)}
\end{equation}

If an ICME is moving with a bulk flow speed of $U$ reaches the position of a spacecraft at 1 AU at time $t_o$, then let $t$ be the time that the spacecraft is inside the ICME.  Also let $r_o$ be the radius of the ICME at $t = 0$. Then as $t$ increases, the distance of the spacecraft from the axis of the ICME is $r_o - Ut$. The relative speed in the inertial frame is
\begin{equation}
v = U +\frac{(r_o + Ut)}{(t-t_o)}
\end{equation}

By fitting this equation to the in situ velocity measurements of the magnetic cloud, \citet{1993JGR....98.7621F} was able to find the propagation time, $t_o$, and initial radius, $r_o$. Using these parameters \citet{1993JGR....98.7621F} further found that some of the decrease in the magnetic field strength seen in the data can be attributed to the expansion of the magnetic cloud. \citet{2008AnGeo..26.1919L} uses this model to calculate the linear expansion speeds of magnetic clouds.


\subsubsection{General Solution}

The general solution of the self-similar MHD model does not require that the gas pressure gradient and gravity be negligible. Low published a series of articles outlining a general self-similar magnetohydrodynamics (MHD) model \citep{1982ApJ...254..796L, 1982ApJ...261..351L,1984ApJ...281..381L, 1984ApJ...281..392L}. In the first paper of the series, \citet{1982ApJ...254..796L} derives an expression for the Lagrangian velocity of the self-similar expansion from the MHD equations. We will see that the general solution is similar to the free-expansion model.  However, more information relating the expansion speed back to other physical parameters are found.

Using spherical coordinates $(r, \theta, \phi)$, we can express the azimuthally symmetric, force-free magnetic field in terms of a toroidal, $B_t(r, \theta, t)$, and poloidal $B_p(r, \theta, t)$ fields of the form:
\begin{equation}\label{Bfield}
{{\bf B} = \frac{1}{r sin \theta} \left({\frac{1}{r} \frac{\partial B_p}{\partial r}}\hat{r} 
- {\frac{\partial B_p}{\partial r}}\hat{\theta}  
+ B_t \hat{\phi} \right)   }
\end{equation}
thus the Lorentz force becomes,
\begin{equation}\label{lorentz_force}\begin{split}
{\frac{1}{4 \pi} ({\nabla \times {\bf B}}) \times {\bf B} } &={\frac{-1}{4\pi r^2 sin^2 \theta}}
\left[ \left({ \mathcal{L} B_p \frac{\partial B_p}{\partial r} + B_t \frac{\partial B_t}{\partial r} }\right) \right. \hat{r}\\
& +  \left({ \mathcal{L} B_p \frac{1}{r} \frac{\partial B_p}{\partial \theta} + B_t \frac{1}{r} \frac{\partial B_t}{\partial \theta} }\right) \hat{\theta}
\left. + \left({ \frac{1}{r} \frac{\partial (B_t,B_p)}{\partial (r,\theta)} }\right) \hat{\phi} \right] \end{split}
\end{equation}
where we have introduced the operator
\begin{equation}\label{operator}
\mathcal{L} = \frac{\partial^2}{\partial  r^2} + \frac{sin \theta}{r^2} \frac{\partial}{\partial \theta} 
\left( \frac{1}{sin \theta} \frac{\partial}{\partial \theta}\right).
\end{equation}
As before, we assume that the velocity is only in the radial direction ${\bf v} = v(r, \theta, t)\hat{r} $. The $(r, \theta, \phi)$-components of the MHD equation of motion, \ref{motion}, are :
\begin{equation}\label{motion_r}
{\rho \left(\frac{\partial v}{\partial t} + v \frac{\partial v}{\partial r} \right) = {\frac{-1}{4\pi r^2 sin^2 \theta}} \left({ \mathcal{L} B_p \frac{\partial B_p}{\partial r} + B_t \frac{\partial B_t}{\partial r} }\right) - \frac{\partial p}{\partial r} -{\rho \frac{G M_\odot}{r^2}}} 
\end{equation}
\begin{equation}\label{motion_theta}
0 = {\frac{1}{4\pi r^2 sin^2 \theta}} \left({ \mathcal{L} B_p \frac{1}{r} \frac{\partial B_p}{\partial \theta} + B_t \frac{1}{r} \frac{\partial B_t}{\partial \theta} }\right) + \frac{1}{r} \frac{\partial p}{\partial \theta}
\end{equation}
\begin{equation}\label{motion_phi}
0 = { \frac{1}{r} \frac{\partial (B_t,B_p)}{\partial (r,\theta)} }.
\end{equation}
If we apply the magnetic field in equation \ref{Bfield} to the MHD induction equation \ref{induction}, the components are :
\begin{equation}\label{induction_p}
\frac{\partial B_p}{\partial t} + v \frac{\partial B_p}{\partial r} = 0
\end{equation}
\begin{equation}\label{induction_t}
\frac{\partial B_t}{\partial t} + \frac{\partial}{\partial r}(v B_t) = 0
\end{equation}
where the $r$ and $\theta$-components have reduced to the same equation.
The MHD conservation of mass equation \ref{mass} becomes :
\begin{equation}\label{mass_r}
{{\frac{\partial \rho}{\partial t}} + \frac{1}{r^2} \frac{\partial}{\partial t} (r^2 \rho v) = 0}
\end{equation}
And finally, assuming that the adiabatic index, $\gamma = 4/3$, the MHD equation of state \ref{state} becomes :
\begin{equation}\label{state_r}
{\frac{\partial}{\partial t} \left(\frac{p}{\rho^{4/3}}\right) + v \frac{\partial}{\partial r} \left(\frac{p}{\rho^{4/3}}\right) = 0}
\end{equation}

Again we introduce the assumption of self-similarity. We eliminate the $r$ and $t$ dependence of the equations by defining the control variable
\begin{equation}\label{control}
{\zeta = \frac{r}{\Phi}}
\end{equation}
where  $\Phi(t)$ is a function of time only. If we assume that poloidal magnetic field varies with time and radial distance through the self-similar control variable $\zeta$ then :
\begin{equation}
B_p(r, \theta, t) = B_p(\zeta, \theta)
\end{equation}
Transforming the variables in equation \ref{induction_p}, we have :
\begin{equation}\label{induction_zeta}
-\frac{r}{\Phi^2}\frac{d \Phi}{dt} \frac{\partial}{\partial \zeta}B_p(\zeta, \theta) + v \frac{1}{\Phi}\frac{\partial}{\partial \zeta}B_p(\zeta, \theta) = 0.
\end{equation}
Solving equation \ref{induction_zeta} for the velocity $v$, we find :
\begin{equation}
v = \frac{r}{\Phi}\frac{d \Phi}{dt}
\end{equation}
We have found the same solution as we did for the free expansion, self-similar MHD model equation \ref{farrugia_equ16}. However, for this solution we did not assume that the gas pressure gradient and gravity were negligible.  This form of the velocity is general to the self-similar model assuming a radial velocity and azimuthal symmetry. 

If we make the change of variables in $\phi$-component of the induction equation \ref{induction_t} using this result, we have :
\begin{equation}
-\frac{r}{\Phi^2}\frac{d \Phi}{dt} \frac{\partial}{\partial \zeta}B_t(\zeta, \theta) + \frac{1}{\Phi}B_t(\zeta, \theta) + \frac{r}{\Phi^2}\frac{d \Phi}{dt}\frac{\partial}{\partial \zeta}B_t(\zeta, \theta) = 0
\end{equation}
It follows that
\begin{equation}
B_t(r, \theta, t) = \frac{1}{\Phi}Q(\zeta, \theta)
\end{equation}
where $Q$ is any arbitrary function. Equation \ref{motion_phi} dictates that $B_t$ as a function of $r$ and $\theta$ can be transformed into a strict function of $B_p$. Thus we can further restrict $B_t$ as :
\begin{equation}
B_t(r, \theta, t) = \frac{1}{\Phi}Q(\zeta, \theta) = \frac{1}{\Phi}Q(B_p[\zeta, \theta])
\end{equation}

Without loss of generality, we can transform any function of $r$, $\theta$ and $t$ into a function of $B_p$ and $\zeta$ \citep{1975ApJ...197..251L}. If we transform the pressure, equation \ref{motion_theta} becomes,
\begin{equation}\label{motion_zeta1}\begin{split}
\frac{\partial^2 B_p}{\partial \zeta^2} + \frac{sin \theta}{\zeta^2} \frac{\partial}{\partial \theta} \left(\frac{1}{sin \theta} \frac{\partial B_p}{\partial \theta}\right) + Q(B_p)\frac{dQ(B_p)}{dB_p} \\ + {4\pi \zeta^2 \Phi^4 sin^2 \theta \frac{\partial p(\zeta, B_p, t)}{\partial B_p}} = 0  
\end{split}\end{equation}
The last term of the equation is the only one explicitly dependent on time. To remove the explicit time dependency from equation \ref{motion_zeta1}, we define :
\begin{equation}\label{p_no_t}
p(\zeta, B_p, t) = \frac{1}{\Phi^4} P(\zeta, B_p)
\end{equation}
Making the same variable transform in equation \ref{motion_r}, we have :
\begin{equation}\begin{split}
&\rho \zeta \left( \frac{\partial^2 \Phi}{\partial t^2} \right) = -\frac{1}{\Phi^5}\frac{P}{\zeta} - \frac{GM\rho}{\zeta^2 \Phi^2}\\
&-\frac{1}{4\pi \zeta^2 \Phi^2 sin^2 \theta}\frac{\partial B_p}{\partial \zeta} 
\left[\frac{\partial^2 B_p}{\partial \zeta^2} + \frac{sin \theta}{\zeta^2} \frac{\partial}{\partial \theta} \left(\frac{1}{sin \theta} \frac{\partial B_p}{\partial \theta}\right) + Q(B_p)\frac{dQ(B_p)}{dB_p} \right].
\end{split}\end{equation}
Using equation \ref{motion_zeta1} to simplify this equation we have :
\begin{equation}\begin{split}
&\rho \zeta \left( \frac{\partial^2 \Phi}{\partial t^2} \right) = -\frac{1}{\Phi^5}\frac{P}{\zeta} - \frac{GM\rho}{\zeta^2 \Phi^2}\\
&-\frac{1}{4\pi \zeta^2 \Phi^2 sin^2 \theta}\frac{\partial B_p}{\partial \zeta} 
\left[{-4\pi \zeta^2 \Phi^4 sin^2 \theta \frac{\partial p(\zeta, B_p, t)}{\partial B_p}} \right]
\end{split}\end{equation}
thus we are left with :
\begin{equation}
\rho \zeta \left( \frac{\partial^2 \Phi}{\partial t^2} \right) = - \frac{GM\rho}{\zeta^2 \Phi^2}
\end{equation}

To match our other variable, we want $\rho$ to be dependent on time only through the variable $B_p$. Thus we require :
\begin{equation}
 \frac{\partial^2 \Phi}{\partial t^2} = -\frac{\alpha}{\Phi^2}
\end{equation}
integrating once 
\begin{equation}
\left(\frac{\partial \Phi}{\partial t}\right)^2 = \frac{\eta\Phi-2\alpha}{\Phi}
\end{equation}
The Lagrangian velocity of a particle in the plasma will be :
\begin{equation}
\frac{dr}{dt} = \zeta_o \frac{d\Phi}{dt}
\end{equation}
where $\zeta_o$ is a constant. We take $\zeta_o$ to the the constant value of $\zeta$ associated with the leading edge. Thus :
\begin{equation}
\frac{dr}{dt} = \zeta_o \left(\frac{\eta\Phi-2\alpha}{\Phi}\right)^{1/2}
\end{equation}
writing the equation with respect to $r$ we have :
\begin{equation}
\frac{dr}{dt} = \zeta_o \left(\frac{\eta r-2\alpha\zeta_o}{r}\right)^{1/2}
\end{equation}

\begin{figure}
\begin{center}
\includegraphics[width=80mm]{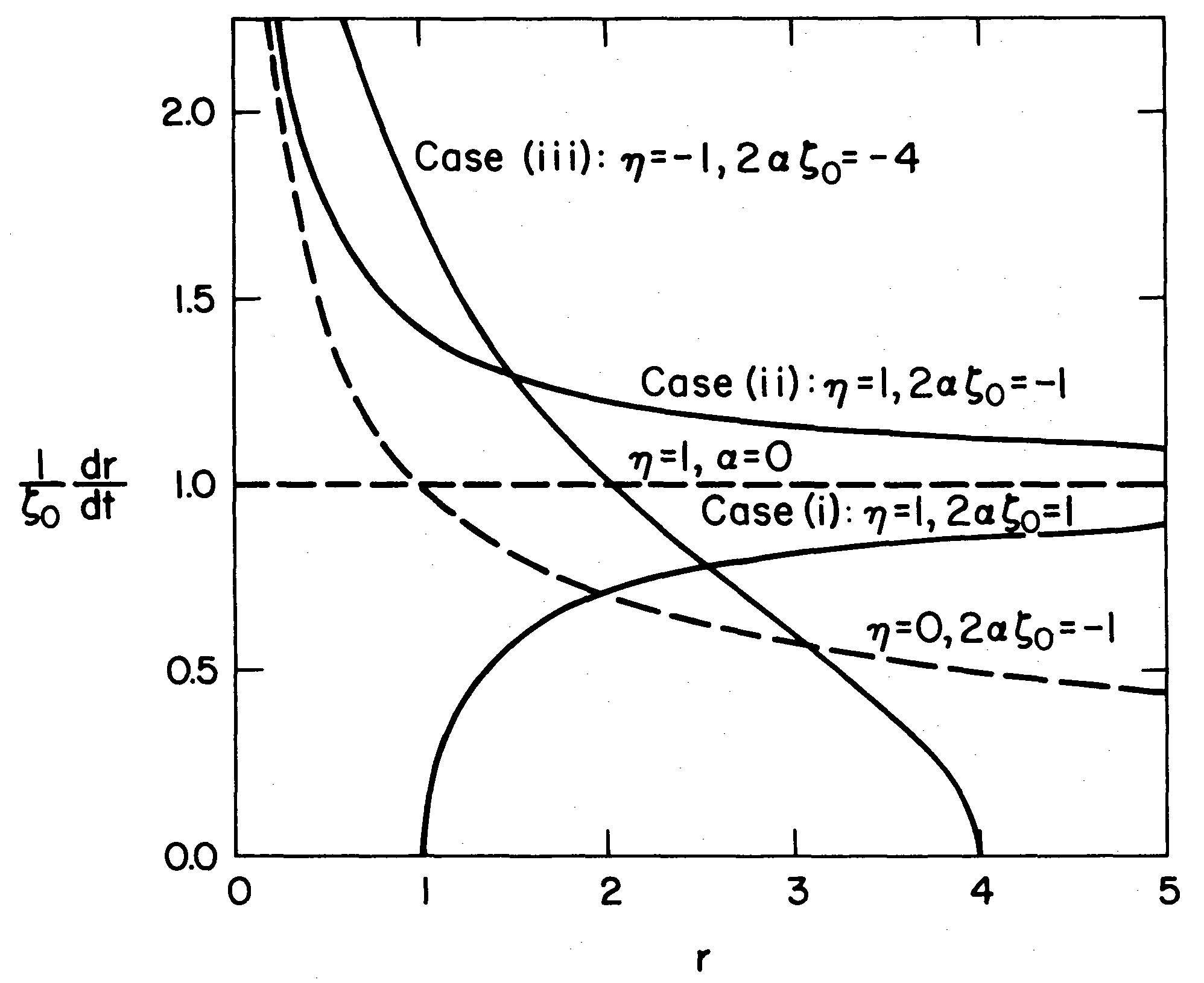}
\caption{Velocity profile for various values of $\eta$ and $\alpha$ \citet{1982ApJ...254..796L}.}
\label{low_fig1}
\end{center}
\end{figure} 
 
Figure \ref{low_fig1} shows that variety of solutions from setting the integration constants to various values. However, for the $\alpha = 0$ case, the solution simplifies to the free-expansion velocity. The general solution of the self-similar MHD model, while more laborious, offers a larger array of solutions that are due to the introduction of the gas pressure and gravity.


\subsection{Flux Rope Model}

\begin{figure}
\begin{center}
\includegraphics[width=70mm]{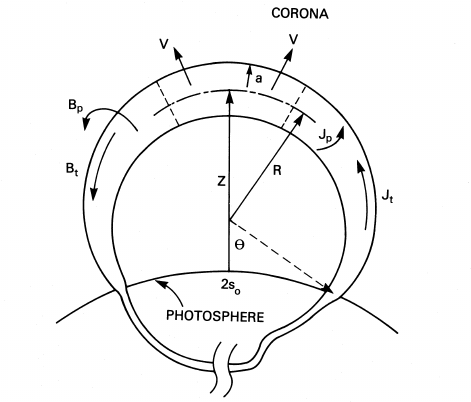}
\caption{Geometry of flux rope \citet{1989ApJ...338..453C}.}
\label{chen_fig1}
\end{center}
\end{figure}

\citet{1989ApJ...338..453C} explored the toroidal forces as applied to a current loop or flux rope embedded in a background plasma such as the solar corona. In this paper, \citet{1989ApJ...338..453C} only considered the MHD aspects of the Lorentz and pressure forces acting on the flux rope. The flux rope consists of a loop wound with a twisted magnetic field. The geometry of the flux rope, figure \ref{chen_fig1}, is defined by the following parameters: $Z$ the height of the center of the loop at the apex above the photosphere, $R$ the major radius of the loop, and $a$ the minor radius. The flux rope is assumed to extend below the photosphere, though the physics of this region are not addressed. The foot-points of the flux rope are separated by the distance $2s_o$ and are assumed to be fixed in the photosphere. The model is only considered for aspect ratios, $R/a$, within the range of 5-10. This range of the aspect ratio allows for simplifications of the model with respect to the current in the photosphere. 

The geometry of the torus magnetic field gives rise to the specific properties in the plasma which do not appear in other magnetic configurations. "Toroidal forces" refers to the properties, which may be approximated as a section of a torus. The toroidal forces have been fully developed for laboratory plasmas. However, laboratory and solar plasma have different boundary conditions. Laboratory plasmas are typically surrounded by a vacuum enclosed in a rigid metal wall. Also, magnetic fields are applied externally to balance the toroidal force, trapping the plasma. Solar plasma is surrounded by ambient plasma and the magnetic field can extend to infinity.

The flux rope consists of a toroidal $B_t$ and poloidal $B_p$ magnetic field with a poloidal and toroidal current density components $J_p$ and $J_t$, respectively.  The curvature of the magnetic field lines creates a tension force that points inward along the minor radius everywhere. In a symmetrically straight cylinder, these forces would cancel out when integrated along the cylinder. However, for the torus the top and the bottom field are not symmetric. To account for the toroid configuration of the magnetic field, it is necessary to integrate the Lorentz force over a toroidal section. The Lorentz force is 
\begin{equation}
{\bf J \times B} =(J_p B_t- J_t B_p) \hat{r}
\end{equation}
where $\hat{r}$ is the unit vector in the outward minor radial direction. The total force due to $(-J_t B_p)$ is $F_p$ and points in the major radially outward direction, and $(J_p B_t)$ gives $F_t$ in the major radially inward direction. Thus the direction of the Lorentz force depends on the difference of the toroidal and poloidal forces. In the solar corona, the pressure force also acts on the current loop in the major radial direction. 

The local force acting on a plasma element is given by 
\begin{equation}
{\bf f} = \frac{1}{4 \pi} ({\bf \nabla \times B}) {\bf \times B} - \nabla p.
\end{equation}
However, to determine the motion of the center of mass of a section of the loop the force must be integrated over a given section. To integrate the force over a section of the loop the external conditions must be specified: the loop is surrounded by a plasma of pressure $p_a$ and the magnetic field vanishes at infinity. The integration over the toroidal loop has been fully developed by \citet{1966RvPP....2..103S}. Here we give the equation of motion for the center of motion of the top section of the flux rope:  
\begin{equation}\label{chen_2a}
F_R = \frac{I_t^2}{c^2 R} \left[ {ln \left( \frac{8R}{a} \right)} +{\frac{1}{2}\beta_p} -{\frac{1}{2}\frac{B_t^2}{B_p^2}} +{2 \left(\frac{R}{a}\right) \frac{B_s}{B_p}} -1 + \frac{\xi_i}{2}    \right] + F_g + F_d.
\end{equation}
The force, $F_R = M {d^2 Z}/{dt^2}$, is the major radial force per unit length of the loop with a mass, $M$. The magnetic fields, $B_{t}$ and $B_s$, are the model ambient fields in the toroidal direction and normal to the plane of the loop, respectively. The quantity $\beta_p$ is defined by 
\begin{equation}\label{chen_pressure}
\beta_p = \frac{\bar{p} - p_a}{B_p^2/8\pi}
\end{equation}
where $\bar{p}$ is the average internal pressure of the loop, $p_a$ is the ambient pressure, and $B_p = B_p(a)$ is the poloidal magnetic field at the outer edge of the minor radius $(r=a)$. The quantity $\xi_i$ is the internal induction, characterizing the minor radial current distribution, given by 
\begin{equation}
\xi_i = \frac{2}{a^2 B_p^2(a)} \int^a_0 {r^2 B_p^2(r)}dr.
\end{equation}
The values of $\xi_i$ vary from $0$ for a surface current distribution to $1/2$ for a uniform current distribution.  It is interesting to note that the form of equation \ref{chen_2a} was derived assuming a circular minor radius of the loop. Also two of the terms of equation \ref{chen_2a} directly depend on the aspect ratio $R/a$ of the toroid.

In equation \ref{chen_2a}, the terms $(ln(\frac{8R}{a}) -1 + \frac{\xi_i}{2})$ come from the $J_t \times B_p$ force which pushed the loop outward. The term  $({\frac{1}{2}\frac{B_t^2}{B_p^2}})$  is from the $J_p \times B_t$ force which pushes the loop downward. The term $({2 \left(\frac{R}{a}\right) \frac{B_s}{B_p}})$ is from the external magnetic field which is holding down the loop. The direction of the force from the gas pressure $\beta_p$ can be in either direction.

The evolution of the minor radius $a(t)$ is given by 
\begin{equation}\label{chen_2b}
\frac{d^2a}{dt^2} = \frac{I^2_t}{Mc^2a} \left( \frac{B_t^2-B_{et}^2}{B^2_p} -1 + \beta_p  \right).
\end{equation}
It is informative to look at the proportionality of equation \ref{chen_2b} which is
\begin{equation}\label{chen_2c}
\frac{d^2a}{dt^2} \propto \left(B_t^2-B_{et}^2 -{B^2_p} +8\pi(\bar{p} -p_a) \right).
\end{equation}
Thus, $B_t^2$ and $\bar{p}$ cause the minor radius to expand and $B_{et}^2$, ${B^2_p}$ and $p_a$ cause it to contract. The effect of the various forces is what we would have expected intuitively. 

If the apex height of a flux rope in equilibrium is perturbed, we can analytically solve for the rate of the minor radius expansion in terms of the major radius expansion. \citet{1996JGR...10127499C} gives this proportionality as 
\begin{equation}\label{chen_20}
\left | \frac{da}{dt} \right |  \propto \left ( \frac{a}{R} \right) \left | \frac{dZ}{dt} \right |
\end{equation}
Thus for the range of aspect ratios considered in the model, the minor radius expansion speed is 10 to 20 percent that of the apex expansion speed.

\begin{figure}
\centering
\subfloat{\includegraphics[width=65mm]{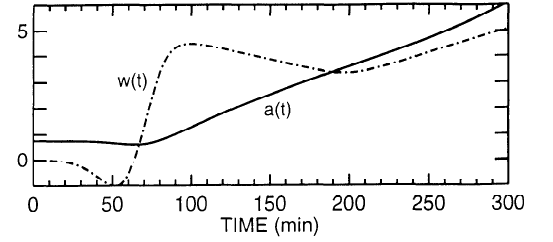}}      
\subfloat{\includegraphics[width=65mm]{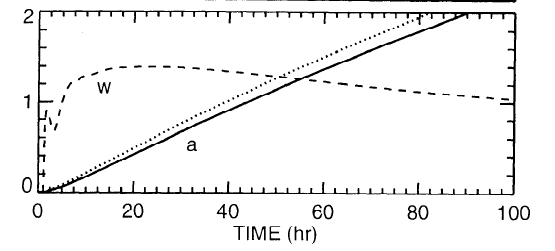}}
\caption{Velocity and radius profile \citet{1996JGR...10127499C}.}
\label{chen_fig2}
\end{figure}

Figure \ref{chen_fig2} shows the velocity and radius for nominal model values given in \citet{1996JGR...10127499C}. As the flux rope increases in height away from the solar surface, the ambient pressure and external toroidal magnetic field will decrease. Applying conservation of flux to the poloidal and toroidal magnetic fields, as the minor radius increases, $B_t$ and $B_p$ will decrease. Thus as the flux rope moves away from the Sun, the pressure term, $8\pi(\bar{p} -p_a)$ becomes increasingly important.

\subsection{Summary of Theoretical Models}

\begin{table}
\begin{center}
\begin{tabular}{|l l p{5cm} p{4cm}|} \hline
\multicolumn{4}{|l|}{ }\\
\multicolumn{2}{|c}{MODELS} & EXPANSION & OBSERVABLES  \\ \hline
\multicolumn{4}{|l|}{\textbf{Kinetic}}\\ 
  & Melon-seed & $r = \left( \frac{8 M v_s^2}{\phi} \right ) \frac{R^{3/2}}{B_o R^2_\odot}$ & $r, R, M, \phi, B_o$ \\
  &  & $r = r_o \left ( \frac{R}{R_\odot} \right )$ & $r, R$\\
  & MSOE & $r(t) \frac{d^2 r(t)}{dt^2} = - \frac{3}{2} \left( \frac{d r(t)}{dt}\right)^2 + \left ( \frac{r_o}{r(t)}\right)^4 \frac{(A_{ho}^2 - A_{ao}^2)}{2} \frac{\rho_o}{\rho_a}$ & $r, r_o, v, a$\\ 

\multicolumn{4}{|l|}{\textbf{Self-Similar MHD}} \\ 
  & Free-expansion & $v = U +\frac{(r_o + Ut)}{(t-t_o)}$ & $v, U, r_o, t_o$\\
  & General solution & $\frac{dr}{dt} = \zeta_o \left(\frac{\eta r-2\alpha\zeta_o}{r}\right)^{1/2}$ & $r, v, \zeta_o$\\
\multicolumn{4}{|l|}{\textbf{Flux rope}} \\ 
  & & $\left | \frac{da}{dt} \right |  \propto \left ( \frac{a}{R} \right) \left | \frac{dZ}{dt} \right |$ & $a, R, \frac{da}{dt}, \frac{dZ}{dt}$\\
\multicolumn{4}{|l|}{ }\\\hline
\end{tabular}
\caption{Summary of all models reviewed. The observables column lists all parameters that can be measured in coronagraphic data.}
\label{summary}
\end{center}
\end{table}

The theoretical models describing the propagation and expansion of CMEs all have advantages and disadvantages. The kinetic models are the most intuitive and the easiest to understand. However, the melon-seed model seems to overly simplify the problem of CME propagation and addresses the expansion as an aside. The melon-seed model offers very little insight to the physical properties of the CME itself since only the external forces are considered. However, as mentioned before, if this simple model can be used to find the propagation and expansion of the CME then perhaps these values could be used in a more complex model to find other physical parameters of the CME. 

The melon-seed overpressure expansion model uses the simplified  geometry of the melon-seed model with the addition of several other terms. Clearly, the melon-seed model can be improved by some consideration of the physical processes internal to the CME. However, it is not clear that the overpressure expansion equation correctly describes these processes.  The replacement of the adiabatic gas pressure with the magnetic pressure requires more explanation and analysis. The addition of the overpressure expansion term which relies on the height of CME creates a set of coupled ordinary differential equations which is an additional complication to the model since the propagation depends on a large number of parameters.

The self-similar MHD model is far more complicated than the kinetic models. The general  solution for the self-similar MHD model incorporates all physical processes of the plasma and reduces to the simple solution for the case of free expansion. However, it is possible that the added complexity of the self-similar MHD model might make the equations too complex to solve. Also the model is still limited by the assumed axially symmetric geometry and the force-free magnetic field.

Last, the flux rope model has many advantages over the other two models. From the current STEREO coronagraph images, the geometry that best fits that of the observed CMEs is a flux rope. Thus both the geometry and the magnetic field configuration of the flux rope model seem to match the CME closely. The flux rope model has the added complication of expansion in both the minor and major radial directions.  In the data these radial directions are difficult to separate and effect the model in very different ways.  Also the integration of the force over the flux rope assumes a circular minor radius. As the CME moves out into the interplanetary medium the minor radius in no longer circular. How this distortion of integration geometry  effects the equation of motion has not been studied.

\section[Previous Studies]{Previous Studies}

The expansion speed of CMEs has been measured in coronagraphic images using an image processing technique called {\it optical flow}. \citet{2006ApJ...652.1747C} used an optical flow algorithm to find the expansion speeds of several CMEs observed in SOHO-LASCO. The optical flow algorithm was applied to images taken with SOHO-LASCO C2. The authors were restricted to working with data from special observing sequences when the image cadence was  between 5 to 12 minutes due to the limitations of optical flow technique. Also they were limited to CMEs with height-time speeds less then 300 km s$^{-1}$. Finally, they were only able to measure expansion speed for eight CMEs with large, well-defined structure. They found that the sum of the bulk and expansion speeds are all very close to the height-time (HT) speed, ranging from within 3\% to 18\% of the HT speed with an average of 9\%. These results indicate that the bulk and expansion speeds of CMEs are being separated accurately by the optical flow method.

\citet{2008AnGeo..26.1919L} used in situ magnetic field and plasma data from the WIND satellite to calculate the expansion speed of magnetic clouds at 1 AU. \citet{2008AnGeo..26.1919L} defined the expansion speed as the component of the velocity perpendicular to the MC's axes. Using a method derived from the force-free, self-similar expansion of an idealized MHD magnetic cloud, the authors calculated the expansion speed of 53 magnetic clouds. They found an average speed of 49 km s$^{-1}$ with a standard deviation of 27 km s$^{-1}$. The estimated error on these speeds is between 15 to 20 km s$^{-1}$ depending on the MC speed.

The expansion speed results from \citet{2006ApJ...652.1747C} near the Sun have an average of 101 km s$^{-1}$  with a standard deviation of 45 km  s$^{-1}$. These speeds are approximately double the expansion speed results from  \citet{2008AnGeo..26.1919L}. It seems that measurements alone do not complete the picture of CME expansion. Both of these methods of measuring CME expansion speeds are limited by geometrical effects. In the coronagraph images, all the velocities are projected onto the plan of the sky and are less then the {\it true} three dimensional velocities. In the derivation of the expansion velocity from the in situ data the cross-section of the MC is assumed to be circular even though this geometry is unlikely \citep{2008AnGeo..26.1919L}.

\section{Program of Study}
We begin by selecting CMEs that were well observed in both the remote sensing and in situ data between January 2010 and June 2011.  In Chapter \ref{eventselection}, we discuss the criteria we used for selecting CMEs for this study. Of the more than 2,000 entries in the SOHO LASCO CME Catalog during this time period, we found only nine CMEs that met our criteria of being well observed in remote sensing and in situ data. Also in the Chapter \ref{eventselection}, we describe some the characteristics of the studied CMEs as observed in the remote sensing and in situ data. In Chapter \ref{analysisremotesensing}, we use the Elliptical and GCS models described in sections  \ref{elliptical_model} and \ref{gcs_model} to analyze the remote sensing data. From the GCS model fit, we find the trajectory and dimensions of the CME at a range of heights between 3.0 and 211 R$_\odot$.  The Ellipse model allows us to parameterize any possible elliptical distortion in the cross-section of the CME. From the Ellipse model we find the center height and the radius of the CME parallel and perpendicular to the CME propagation. At the end of Chapter \ref{analysisremotesensing}, we compare the results of the GCS and Ellipse models. In Chapter \ref{expansonpropagation}, we study the propagation and expansion of the CME in two ways. First in section \ref{empirical}, we analyze the results of Chapter  \ref{analysisremotesensing} to derived the kinematics of the CME. We used the derived kinematics of the CME front to predict the arrival time and velocity of the CME at Earth. We compare our predicted values with those detected in situ.  We also derive the expansion speeds of the CME from the Ellipse model results. In section \ref{theory}, we used the kinematics derived in the first half of the chapter to calculate the forces on the CME as predicted by the theoretical models introduced in section \ref{intro:theory}. In Chapter \ref{remoteinsitu}, we use the model described in section \ref{EFR} to derive a size and orientation of the CMEs from the in situ data. We compare the results from the in situ model with the results of the GCS model projected to 1 AU. Finally in Chapter \ref{summaryanddiscussion}, we summarize and discuss our results and outline any future work that would lead from this study.

\chapter{Studied CMEs}\label{eventselection}
In this study, we analyze nine events chosen from all Earth-directed CMEs observed between January 2010 and June 2011. The main qualification for the selection of an event was that it was well observed in both remote sensing data and in situ data at Earth. In the next sections, we will describe the criteria for well observed in each data set. The qualification that the CME impact Earth strictly limits the number of possible CMEs. During the period of selection, 2,062 CMEs were cataloged in the SOHO LASCO CME Catalog from the remote sensing data \citep{2004JGRA..10907105Y}. During the same time, only 61 possible CMEs were identified in the WIND in situ data at Earth \citep{talktowu}. 

Thus, we began selecting events by attempting to match possible CMEs detected in the WIND data to CMEs observed near the Sun in remote sensing data. We did this by assuming a three-four day transit time from Sun to Earth and identifying Earth-directed CMEs which left the Sun during this time. Matching CMEs seen in remote sensing data to CMEs detected in situ is not an easy task even with continuous observation from the Sun to 1 AU. Due to projection effects, a CME front may appear to reach a height of 1 AU several hours before it impacts Earth. Thus if there is more than one CME it can be difficult to determine which CME caused the in situ observation. This task is somewhat simplified by the relatively low solar activity during the selection time period. Due to the low solar activity there are few CMEs during the three-four day window before the in situ detections.  Also most of the CMEs are isolated in the data. Thus fewer CMEs are obscured by another CME and there is less CME-CME interaction. However, the low solar activity during the rising phase of the solar cycle also reduces the intensity and size of the events during this time limiting the number of CMEs that can be continuously observed in the images out to 1 AU.

\section{Selection Criteria for Studied CMEs}
\subsection{Remote Sensing Criteria}
To select CMEs in the remote sensing data, we first had to identify those that were Earth-directed.  An Earth-directed CME in the LASCO data will appear as a partial or full halo. In the SOHO LASCO CME Catalog a partial halo is defined as having an apparent angular width of more then 120$^o$. With only the LASCO view, determining if a CME is Earth-directed is difficult because CMEs that are directed towards and away from Earth have the same appearance. Before STEREO, observers had to rely on EUV signatures on the solar disk to determine if a CME was directed towards Earth. However, not all CMEs have strong or even any EUV signature \citep{2009ApJ...701..283R}. 

With the addition of the STEREO views, it is much easier to determine if the CME is Earth-directed. During the period between January 2010 and June 2011, the STEREO spacecraft were separated from each other by 132$^o$ to 180$^o$. The spacecraft reached opposition on 6 February 2011 and afterwards began moving closer to each other on the far side of the solar system. On 1 January 2010, the spacecraft were -68$^o$(B) and 64$^o$(A) from Earth. On 30 June 2011, they were -93$^o$(B) and 99$^o$(A) from Earth. Given the configuration of the STEREO spacecraft during the selection period, an Earth-directed CME would appear as a West limb event in STEREO B and an East limb event in STEREO A. Thus, an observer can quickly determine if the CME is in the same quadrant of space as the Earth but it is still impossible to tell if a CME will produce a measurable signal in the in situ data from the remote sensing data. Two of the CMEs we studied were not even identified in the  SOHO LASCO CME Catalog as partial halos (see table \ref{categories}).

The first criterion for a CME to be well observed in the remote sensing data is that it had to be observed in the 8 telescopes used in this study without a significant period ($< $ 1 hour) of missing data. Given the nature of space-born observations and the approximately three day duration of these CMEs, this first criteria eliminated several events. In all the remote sensing  data, the CME had to exhibit a bright easily visible structure. Due to the effects of Thomson scattering, the dimmest observations are from the LASCO viewpoint for Earth-directed CMEs. Thus the visibility of the CME in the LASCO coronagraphs is usually the limiting case for event selection.  The CME had to be easily tracked between instruments, thus the structure  of the CME had to be visible out to nearly the edge of the field of view. Also the CME could not expand outside the upper and lower edges of the HI-1 instruments field of view.  

\subsection{In Situ Criteria}
To identify potential CMEs in the in situ data, we used the automatic detection technique of \citet{2005AnGeo..23.2687L}. The technique was developed to detect potential magnetic clouds in the data based on the definition from \citet{1981JGR....86.6673B}. The technique can also identify CMEs in the data. The detection requirements for a magnetic cloud are higher then for a CME detection. Thus, we looked at all detections which met the the minimum search criteria to find CMEs for this study. The minimum requirements from detection are; the proton plasma beta must be low, the field directions must change smoothly and these two conditions must persist continuously for a minimum of eight hours. 

Specifically, the CME search was performed on in situ data with a one minute running average time step. Periods of low proton plasma beta were identified in the data where the average over 25 minutes was  $<\beta_p>  \leq 0.3$. A smooth change in the direction of the magnetic field was evaluated by a quadratic fit of the latitude of the magnetic field, $\theta_B$, for a running interval of 25 minutes. If the chi-squared of the fit is $\chi^2 \leq 500$, the magnetic field is considered smoothly changing. For a minimum detection, both of these conditions must be met continuously for 8 hours. \citet{2005AnGeo..23.2687L} has a higher criteria for potential magnetic cloud detection. A period of data must meet the minimum criteria as well as have a relatively high average magnetic field strength, a low proton thermal velocity and the change in the magnetic field latitude must be greater then some lower limit-value.  

We ran the detection software on the WIND data during the search period and found 61 potential CMEs using the lower search criteria. We searched for CMEs in the remote sensing data that corresponded with these potential in situ detections. We identified nine CMEs in the data that match our selection criterion. Of the nine CMEs  selected for this study, seven also meet the stricter criteria for possible magnetic cloud detections. 

\section{Description of Studied CMEs}
Throughout the study, we will often refer to the studied CMEs by the date of first appearance in the LASCO C2 data. We have also assigned a number in chronological order to each event. For brevity, we will also refer to the CMEs by these numbers. The date and time of the first appearance and event number of each CME are listed in table \ref{projected_values}.  

\subsection{Remote Sensing Description}\label{remotesensingevents}

To give a basic description of the studied CMEs, we have chosen some qualitative and quantitative characteristics of the CMEs as observed in the remote sensing data. Samples of the remote sensing data can be found in Appendix \ref{appremotesensing}. In table \ref{projected_values}, we have listed the widths of each event as seen in the LASCO C2, SECCHI COR2 A and B images. The bottom line of table \ref{projected_values} lists the average and standard deviation of each column.  All the events have a larger width in C2 and similar widths in both COR2 images as we expect for Earth-directed CMEs. In table \ref{projected_values}, we also give the speed of the CME front from a linear fit of the height-time measurements made for each data sequence. These speeds are projected values of the true three-dimensional speed of the CME. We have included the projected linear speed for the sake of comparison of these events with previous CME studies when only one view point was available. Due to projection effects the same CME can have very different speeds from each view point. Several of these events have strong accelerations and hence a linear fit is not the best. We will examine the kinematics of the CMEs in depth in Chapter \ref{expansonpropagation}.  
\begin{table} 
\begin{threeparttable}
\caption{Summary Projected CME Parameters}\label{projected_values}
\begin{tabular}{c c c | *{3}{c} | *{3}{c}} \hline  
&  &  & \multicolumn{3}{c}{Width (deg)} & \multicolumn{3}{c}{Speed (km s$^{-1}$)}\\
CME & Date & Time & C2\tnote{a} & COR2 A & COR2 B & C2\tnote{a}  &COR2 A &COR2 B\\ \hline
1&19-Mar-2010& 10:30&101&65&49&186&202&180\\
2&03-Apr-2010& 10:33&360&51&51&668&809&800\\
3&08-Apr-2010& 01:31&156&49&50&227&421&425\\
4&16-Jun-2010& 14:54&153&41&54&236&184&238\\
5&11-Sep-2010& 02:00&236&70&78&469&392&406\\
6&26-Oct-2010& 01:36&83&71&73&214&189&260\\
7&15-Feb-2011& 02:24&360&90&91&669&751&773\\
8&25-Mar-2011& 01:25&85&50&48&45&100&142\\
 &25-Mar-2011\tnote{b} & 14:36&$>$191& & &119& & \\
9&01-Jun-2011& 18:36&189&60&51&361&479&516\\ \hline       
  & & & 191$\pm$108&61$\pm$15&61$\pm$16&	319$\pm$218&392$\pm$254&416$\pm$243 \\
\end{tabular}
\begin{tablenotes}
	\item [a] values are taken from the SOHO LASCO CME Catalog
     	\item [b] The CME was listed as two events in the SOHO LASCO CME Catalog.
\end{tablenotes}
\end{threeparttable}
\end{table}

From the LASCO C2 data, the projected speeds of our CMEs range between 45-669 km $s^{-1}$ with an average and standard deviation of 319 km s$^{-1}$ $\pm$217. The average speed of our CMEs as seen in LASCO C2 is less then the average speed of 437 km s$^{-1}$ for all halo and partial halo CMEs during any year between 1996 and 2002 in LASCO \citep{2004JGRA..10907105Y}. For the same phase of the last solar cycle, 1998, the average speed of all CMEs in LASCO was 412 km s$^{-1}$  \citep{2004JGRA..10907105Y}. This speed is similar to the average speeds as seen in the STEREO COR2 views of our CMEs. Thus the selected CMEs are slow for halo and partial halo CMEs but average for all CMEs during this phase of the last solar cycle. 

The average width of the CMEs as seen in STEREO COR2 is 61$^o\pm16$. The qualification that the CME not expand out of the HI-1 field of view puts a restriction on the angular width of the CME from the SECCHI view point. However, smaller CMEs are usually not visible far out into the heliosphere. Thus all the selected CMEs have an angular width between 41 and 78 degrees when viewed as limb events. The selected CMEs are larger then the average CME as seen in the SOHO LASCO Catalog which detects many small CMEs near the Sun \citep{2004JGRA..10907105Y}.  

In table \ref{categories}, we have placed the events into several descriptive categories. The most subjective of theses categories is whether the CME has a flux rope like structure. A flux rope like structure is defined as fitting the three part structure of an idealized CME. Thus the CME must have a rounded leading edge followed by a cavity and bright core. The structure of the CME is evaluated in the COR2 images where it is seen as a limb event. Halo events are not seen to have three part structures. From the SOHO LASCO CME Catalog, we list whether the CME was observed as a halo or partial halo. Next, we indicate whether a white light shock is seen in the data. Shocks can obscure the outline of the CME, effecting the model fitting. Shocks are also detected in the in situ data. It is of interest to compare shock detection in the two data sets.

We also evaluated in the COR2 images the symmetry of the CME between the STEREO A and B views.  The symmetry gives us an estimate of how close the CME is to the Sun-Earth line. The more symmetric a CME is in the STEREO A and B view then presumably the closer the CME is to the Sun-Earth line. The symmetry of the CME is also important for fitting the structure models to the remote sensing data. If the STEREO A and B views are nearly identical then there is little gained from using both view points. However, if the asymmetry is large, this might indicate that the CME is distorted. If the CME is distorted, it is difficult to fit the idealized shape of the CME models to the data.  
\begin{table}
\begin{threeparttable}
\caption{Categorization of CMEs}\label{categories}
\begin{tabular}{*{5}{c} l}\hline
CME	&	Flux-rope	&	Symmetry	&	Halo	&	Shock	&	Comment	\\ \hline
1	&	X	&  		&		&		&	Very circular structure	\\
2	&	X	&	X	&	Halo	&	X	&	Bright asymmetric core material	\\
3	&		&	X	&	PH	&	X	&	Diffuse front 	\\
4	&	X	&		&	PH	&		&	Rotating\tnote{b}	\\
5	&	X	&		&	PH	&	X	&	Face-on flux rope structure	\\
6	&	X	&	X	&		&		&	Face-on flux rope structure; diffuse front	\\
7	&		&	X	&	Halo	&	X	&	Preceding CME; distorted front 	\\
8	&	X	&	X	&	PH\tnote{a}&		&	Very slow rising; two events in catalog	\\
9	&		&	X	&	PH	&	X	&	Preceding CME	\\ \hline
\end{tabular}
\begin{tablenotes}
	\item [a] Second detection	
	\item [b] \citet{2011ApJ...733L..23V}	 and \citet{2010AGUFMSH23B1841N}
\end{tablenotes}
\end{threeparttable}
\end{table}

The CMEs chosen through our selection qualifications have a variety of characteristics. The projected speeds of the CME are averages for all CMEs. It does not seem that the selection qualifications are related to any characteristic other then the size of the events.

\subsection{In Situ Description}

In table \ref{table:insitu}, we have listed some parameters of the studied CMEs as observed in situ. The in situ data for each CME are plotted in Appendix \ref{appinsitu}. For each CME, we have listed its arrival time at the WIND spacecraft. This arrival time corresponds to the arrival of the jump in density and plasma temperature. The CME arrival time is earlier than the period of smoothly varying magnetic field which is used as the arrival time of magnetic clouds.  Similarly, the end of the CME includes any increase in density following the low density cavity of the CME. The duration of the CME in days, $\Delta t$, is given in table \ref{table:insitu}. We list which component of the magnetic field shows a strong rotation and if the CME includes a shock. We also list a type for each CME detection. A type 1 CME only meets the minimum criteria from \citet{2005AnGeo..23.2687L} and a type 2 CME met the higher criteria. A type 3 CME is characterized as a MC based on the quality of the flux rope model from section \ref{EFR}. Finally, we list the minimum Dst during the CME passage to indicate the geo-effectiveness of each CME. CME 2 with the highest Dst of all the events caused a geo-magnetic storm.
\begin{table}
\begin{center}
\caption{Description of CME In Situ}\label{table:insitu} 
\begin{tabular}{*{9}{c}}\hline
CME	&	Date			&	Time		&	$\Delta$t	&	Rotation	&	Shock	&	Type	&	Dst	\\ \hline
1	&	23-Mar-2010	&	23:02	&	0.67	&		&		&	1	&	1	\\
2	&	05-Apr-2010	&	06:43	&	1.47	&		&	X	&	2	&	-76	\\
3	&	11-Apr-2010	&	11:59	&	1.30	&	By	&	X	&	2	&	-65	\\
4	&	20-Jun-2010	&	23:59	&	1.70	&	Bx	&	X	&	3	&	-6	\\
5	&	14-Sep-2010	&	14:24	&	1.80	&		&		&	2	&	-22	\\
6	&	31-Oct-2010	&	04:48	&	1.70	&	By	&		&	2	&	-7	\\
7	&	18-Feb-2011	&	00:00	&	3.52	&		&	X	&	2	&	-25	\\
8	&	29-Mar-2011	&	14:38	&	3.04	&	Bz	&	X	&	3	&	-6	\\
9	&	04-Jun-2011	&	19:30	&	1.70	&	By	&	X	&	3	&	-45	\\ \hline
\end{tabular}
\end{center}
\end{table}

Of the nine CMEs chosen for this study, three have been studied in previous papers.  CME 2 on 3-April-2010 generated the first significant geomagnetic storm of solar cycle 24 and is possibly the most studied CME of the solar cycle to date. \citet{2011ApJ...729...70W} used a reconstruction technique to model the CME  from Sun to Earth to study the kinematics and morphology of the CME. \citet{2010GeoRL..3724103M} studied the in situ signature of the CME. \citet{2012SoPh..276..293R} and \citet{2011ApJ...743..101T} studied the kinematics and interaction with the solar wind of this CME and two others.  Several other papers have been written about other aspects of the CME not related to this study \citep{2012JGRA..11703225M, 2011ApJ...734...84L, 2012A&A...537A..28Z}. These works found that the CME is a complex event. \citet{2011ApJ...729...70W} found that the orientation of the CME is ambiguous from the remote sensing data. \citet{2010GeoRL..3724103M} concluded that the flank of the CME was detected at Earth which kept the magnetic cloud from being modeled using  standard approaches.  CME 4 on 16-June-2010 was studied in \citet{2011ApJ...733L..23V} and \citet{2010AGUFMSH23B1841N}. \citet{2011ApJ...733L..23V} found that in the middle corona the CME is rapidly rotating. \citet{2010AGUFMSH23B1841N} studied the CME in the heliosphere and showed that it continues rotating using in situ data from MESSENGER and WIND. In \citet{2011SpWea...901005D},various methods were used to predict the arrival time of CME 3 on 8-April-2010 .

To date, this is the largest study of Earth impacting CMEs using the continuous STEREO observation and in situ data. Several studies like \citet{2011ApJ...729...70W, 2010GeoRL..3724103M, 2010AGUFMSH23B1841N} have studied a single CME from Sun to Earth using both remote sensing and in situ data. \citet{2011JASTP..73.1201R} reviewed these studies which incorporated remote sensing and in situ data using a variety of techniques. In this study, we apply the same analysis to nine CMEs to derive those parameters which can be compared between the remote sensing and in situ data sets.

\chapter{Analysis of Remote Sensing Data}\label{analysisremotesensing}
To analyze the remote sensing data, we used the Elliptical Model and the GCS Model described in sections  \ref{elliptical_model} and \ref{gcs_model}. With these models, we are able to parameterize the position and dimension of the CMEs over time from the remote sensing data. In this chapter, we present the results of each modeling method. We will further analyze these results in Chapter \ref{expansonpropagation} to calculate the kinematics of the CMEs. Also from the remote sensing data, we obtained the mass of the CME. We will use the CME mass later to calculate the net force on the CME.

\section{GCS Model Fit of Remote Sensing Data}

The GCS model is fitted to the remote sensing data by an observer adjusting six parameters of the model. The geometry of the GCS model is presented in figure \ref{Fig:GCS}. The parameters of the model are longitude, latitude, rotation, ratio, half angle and height. Longitude, latitude and rotation are the Euler angles which relate the axes of the model to the Heliocentric Earth Ecliptic (HEE) coordinate system. The HEE coordinate system is defined by the x-axis along the Sun-Earth line, z-axis is the north pole for the ecliptic of date and the y-axis completes the right-handed orthogonal set. The half angle is the angle between the model axis and the centers of the legs of the model. The ratio controls the radius of the model. The height is the apex height of the front of the model. In table \ref{raytrace}, we have listed the model parameters for each CME when the model height is $\sim$10 R$_\odot$.  Also in table \ref{raytrace}, we have given the width (W) and radius (a) of the model. The geometry of these parameters are shown in figure \ref{Fig:GCS}. The width and radius are derived from the fit parameters using the equations given in  \citet{2011ApJS..194...33T}.

The figures in Appendix \ref{appremotesensing} show the simultaneously obtained images from the three view points as well as the fitted GCS model. In each image, the model is projected onto the plane of the image using a grid of points over the surface of the model.  For each CME, we show the model fit at three heights. The parameters in table \ref{raytrace} roughly correspond  to the first figure for each CME in Appendix \ref{appremotesensing}. 

\begin{table}
\begin{center}
\caption{GCS Model Parameters at 10 R$_\odot$}\label{raytrace}
\begin{tabular}{*{8}{c}} \hline  
CME & 	Lon 	&	Lat	& 	Rot	& 	Ratio&Half-Angle& Width &	 Radius \\ 
	&	(deg)& 	(deg)&	(deg)&		& 	(deg)&(R$_\odot$) &(R$_\odot$)\\ \hline
1	&	27	&	-6	&	-35	&	0.28	&	11	&	3.2	&	2.2	\\
2	&	6	&	-19	&	12	&	0.37	&	25	&	4.2	&	2.6	\\
3	&	-2	&	-3	&	-29	&	0.19	&	33	&	4.3	&	1.7	\\
4	&	-18	&	4	&	-33	&	0.26	&	10	&	2.9	&	2.0	\\
5	&	-21	&	14	&	-43	&	0.41	&	18	&	4.0	&	2.8	\\
6	&	22	&	-28	&	-55	&	0.27	&	30	&	4.3	&	2.1	\\
7	&	2	&	-2	&	27	&	0.47	&	41	&	5.1	&	3.2	\\
8	&	-27	&	3	&	-12	&	0.31	&	41	&	4.8	&	2.3	\\
9	&	-22	&	-5	&	-16	&	0.39	&	30	&	4.7	&	2.8	\\  \hline
\end{tabular}
\end{center}
\end{table}

When the CME is visible in the LASCO data, we use all three view points to make the fit. To uniquely fit the GCS model to the remote sensing data, the LASCO view  point is essential. Since the CMEs are Earth-directed, the geometry of the CMEs as viewed from STEREO A and B are similar (table \ref{categories}). The symmetry of the STEREO data set is why the LASCO view is of great importance to fit the GCS model. The LASCO view point gives us essential information about the orientation and dimensions of the CME which is ambiguous from the SECCHI data for Earth-directed CMEs.

However, given the effects of Thompson scattering and the LASCO field of view, the CMEs are not visible in LASCO past $\sim$ 25 R$_\odot$ depending on the density of the front. Thus, we must make assumptions about the propagation of the CME to continue fitting the model after the CME is no longer visible in the LASCO view. We assume that at large heights the CME is expanding self-similarly. This assumption is simple to implement since the model expands self-similarly when all parameters except height are held constant. Thus after $\sim$ 25 R$_\odot$,  we hold the rotation, ratio and half angle of the model constant. We adjust the longitude and latitude to fit any changes in the propagation direction. For most CMEs the model parameters have small variations when fitted using the LASCO view. A notable exception is the CME on 16-June-2010 which has a rapid change in the rotation angle in the LASCO field of view \citep{2011ApJ...733L..23V}. The effects of the rotation on the GCS model fit to this CME are discussed in \citet{2010AGUFMSH23B1841N}.

We fit the GCS Model at a maximum height of 211 R$_\odot$ (0.98 AU) for the 3-April-2010 CME. The average maximum height for all the studied CMEs is 179 R$_\odot$ (0.83 AU). The front height from the GCS model is plotted in figure \ref{raytrace_velocity} for each event.

For all the CMEs, the longitude and latitude is within $\pm$30$^o$ of the Sun-Earth line. In this study we have not associated the CMEs with a solar surface source region. Assuming the source regions are similar to the CME trajectory, the locations of the studied CMEs are consistent with previous studies of geo-effective CMEs \citep{2007JGRA..11212103Z}. Thus the derived three-dimensional trajectory of the CMEs is consistent with all of these CMEs being detected in situ at Earth.

We also note from table \ref{raytrace}, that the width and radius of the CMEs at 10 R$_\odot$ are similar. The range of widths is 2.9-5.1  R$_\odot$ and radii is 1.7-3.2 R$_\odot$. As we discussed in section \ref{remotesensingevents}, there was a selection bias for the projected angular width of the CME. This selection bias persists for the three-dimensional size of the CMEs as well.

\section{Elliptic Model Fit of Remote Sensing Data}
\begin{figure}
\begin{center}
\includegraphics[width=.7\textwidth]{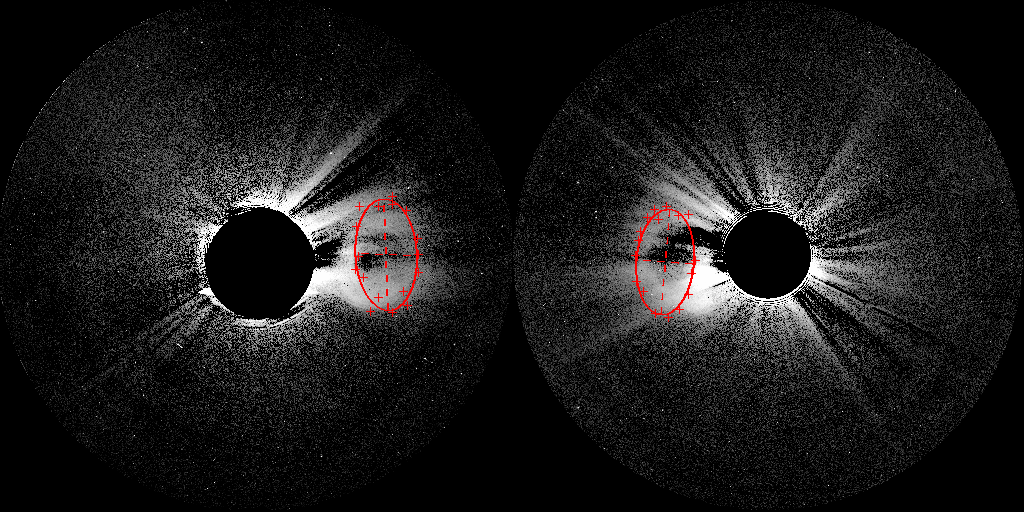}
\caption{Ellipse fit on the SECCHI COR2 A (right) and B (left) images from19-March-2010 17:54 UT.  The direct measurements (plus signs), ellipse fit (solid line) and major and minor axes (dashed lines) have been plotted on the images.}
\label{ellipse_image}
\end{center}
\end{figure}

As we discussed in section \ref{Structures Models}, the Ellipse Model allows us to parameterize any possible elliptical distortion in the cross-section of the CME. 
We fit an ellipse to each CME by making direct measurements of the CME envelope in the SECCHI remote sensing data. Figure \ref{ellipse_image} shows an example of an ellipse fit to SECCHI COR2 A and B data for the 19-March-2010 CME. The direct measurements made by the observer are plotted with plus signs. An ellipse fit is plotted over the direct measurements with a solid line. For each fit, we get the center of the ellipse, $(x_c, y_c)$, the semi-major and semi-minor axis, ($a, b$) and the tilt angle ($\phi$).  The major and minor axes of the ellipse fit are plotted with dashed lines, in figure \ref{ellipse_image}. We define the semi-minor axis as the parallel radius and the semi-major axis as the perpendicular radius where parallel and perpendicular refer to the direction of the CME propagation. 

The results from the ellipse fit at a CME front of 10 R$_\odot$ from the GCS Model are given in table \ref{table:ellipse} in units of R$_\odot$. These parameters are from the same images as the GCS parameters in table \ref{raytrace}. To convert the measurements in pixels to distance in units of R$_\odot$, we must know the distance between the measured feature and the observer. Without this information, we can only convert the pixel measurements to elongation in degrees. For the parameters in table \ref{table:ellipse}, we used the CME trajectory from the GCS Model. In table \ref{table:ellipse}, we list the conversion factor from R$_\odot$ to degrees for each spacecraft. The first two parameters in table \ref{table:ellipse} are the height of the center of the ellipse fit. 

Even with the correction for the CME trajectory, the ellipse fits in STEREO A and B data are sometimes different by as much as 10 R$_\odot$ for the extreme case of CME 1. There are several possible reasons for the differences between the STEREO A and B parameters. One possible reason is that the STEREO A and B planes of the sky are different. Thus from each view point, we are looking at different projections of the CME.  \citet{2011JASTP..73.1173L} found a $<$10$^o$ error in the longitude of the leading edge for tie-pointing and triangulation technique. They attribute this error to different apparent leading edges. However, for our CMEs, there is no obvious correlation between the differences in the ellipse parameters and the spacecraft separation. Also there is no correlation between any of the GCS parameters for the orientation of the CME and difference in the ellipse parameters.  We will investigate the difference in the Ellipse measurements more in the next section by comparing the Ellipse and GCS model results.
\begin{table}
\begin{center}
\caption{Ellipse Model Parameters at 10 R$_\odot$}\label{table:ellipse}
\begin{tabular}{*{9}{c}} \hline  
CME	&\multicolumn{2}{c}{Center (R$_\odot$)} 	&\multicolumn{2}{c}{Semi-Major (R$_\odot$)} 	&\multicolumn{2}{c}{Semi-Minor (R$_\odot$)}&	\multicolumn{2}{c}{ deg/R$_\odot$} 		\\
	&	A	&	B	&	A	&	B	&	A	&	B	&	A	&	B	\\ \hline
1	&	9.3	&	8.0	&	5.1	&	3.6	&	2.8	&	2.0	&	0.177	&	0.263	\\
2	&	9.0	&	8.3	&	3.5	&	3.5	&	1.6	&	1.8	&	0.245	&	0.259	\\
3	&	9.7	&	10.6	&	4.5	&	3.8	&	1.6	&	1.6	&	0.261	&	0.249	\\
4	&	7.6	&	9.8	&	2.9	&	4.0	&	2.4	&	2.3	&	0.278	&	0.205	\\
5	&	7.1	&	9.1	&	4.4	&	6.6	&	2.1	&	3.1	&	0.269	&	0.199	\\
6	&	7.9	&	7.5	&	5.4	&	4.8	&	2.8	&	3.0	&	0.243	&	0.239	\\
7	&	7.5	&	7.7	&	4.3	&	4.6	&	2.1	&	2.0	&	0.276	&	0.258	\\
8	&	9.6	&	9.0	&	4.1	&	3.6	&	2.6	&	2.3	&	0.252	&	0.246	\\
9	&	8.0	&	8.4	&	3.6	&	3.0	&	2.5	&	2.5	&	0.250	&	0.251	\\ \hline
\end{tabular}
\end{center}
\end{table}

We were able to make ellipse fits to the remote sensing data out to a maximum CME front heights of 69 R$_\odot$ for the 11-Sep-2010 CME. The average maximum front height for all the CMEs is 60 R$_\odot$. The outer edge of the SECCHI HI1 field of view (FOV) is 84 $R_\odot$ for the plane of the sky. Thus we were not able to make ellipse fits over the entire HI1 FOV nor into the HI2 FOV despite the fronts being visible throughout the HI1 and HI2 FOVs. This limitation is because the top and bottom edges of the CME diffuse faster than the CME front and it becomes difficult to estimate the extent of the CME beyond a certain height.

\section[Comparison of the GCS Model Fit to the Ellipse Model Fit]{Comparison of the GCS Model Fit to the Ellipse \\ Model Fit}\label{compareGCSE}

\begin{figure}
\begin{center}
\includegraphics[height=.95\textwidth]{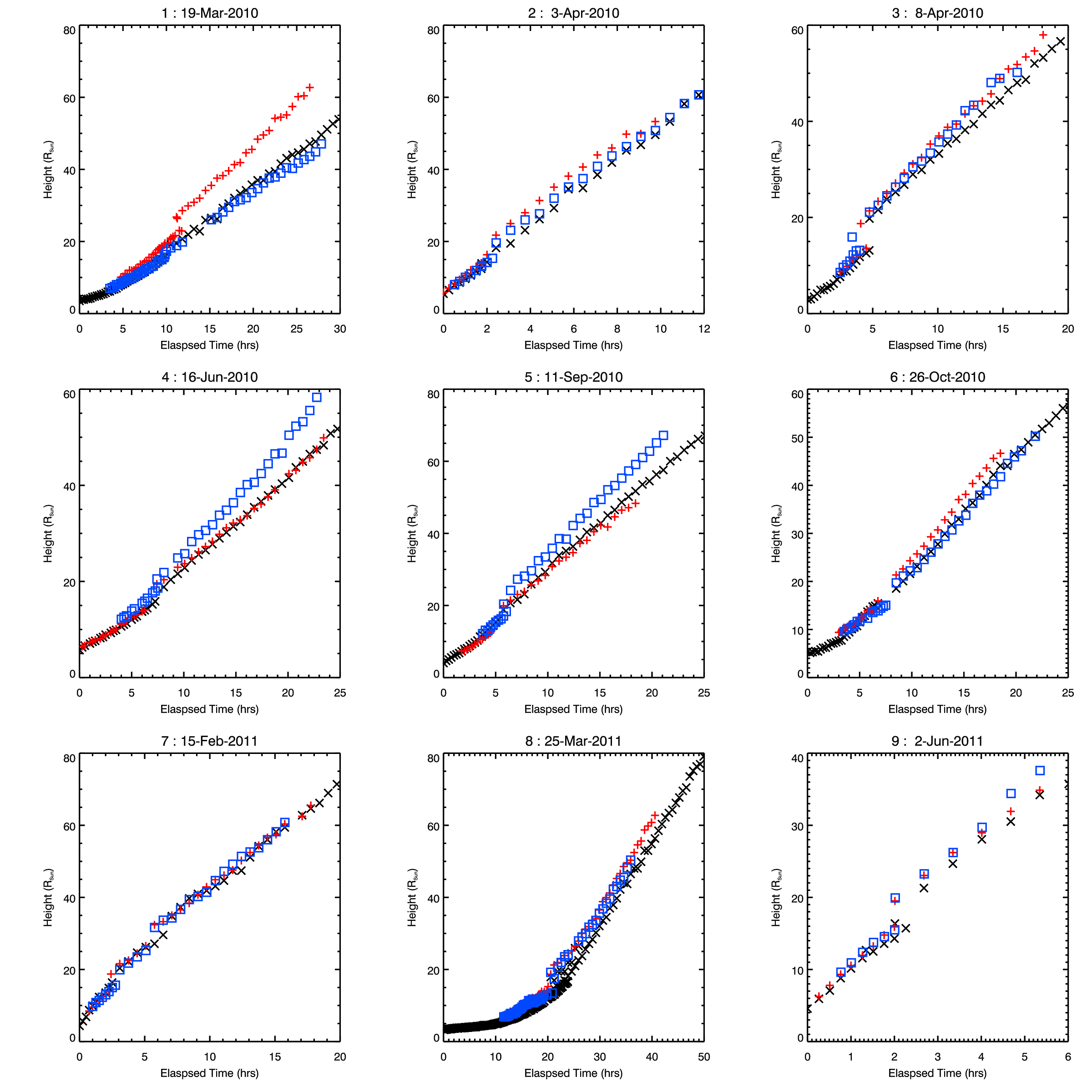}
\caption{Comparison of the GSC front height with the sum of the Ellipse center height and parallel radius.  From the GCS Model the front heights are plotted with black Xs. From the Ellipse Model, we have plotted the center height plus the parallel radius. STEREO A and B values are plotted with red plus signs and blue squares, respectively.}
\label{ellipse_raytrace}
\end{center}
\end{figure}

We begin by comparing the height of the GCS front with the sum of the parallel radius and center height from the Ellipse model.  In figure \ref{ellipse_raytrace}, the GCS data is plotted with black Xs and the Ellipse data for STEREO A and B are plotted with red plus signs and blue squares, respectively. We see that the two values are similar. However, we find that for some CMEs ( 1,4 and 5) the observations from from one view point diverges from the other two view pints. In table \ref{categories}, these CMEs were identified as being asymmetric between the STEREO A and B views. Since the direct measurements from one of the STEREO views match the GCS fit, we assume that the GCS model was preferentially fitted to this view. The preference many not be user error (though that is always a possibility) but that one  STEREO viewpoint showed a larger deviation from the idealized geometry of the model. For example, in figure \ref{2010-03-19_3}, the CME extends further north in STEREO A than in the STEREO B view. The northern extension in the STEREO A data is not fitted by the model. Thus the user preferentially fitted the model to the STEREO B data. 

We believe that the discrepancies among the CME measurements is due to intrinsic differences in the CME when viewed from different sides. A difference in the CME envelope between the STEREO A and B view could be caused by a distortion of the CME by the ambient corona. Such a distortion could be caused by a pile up of on one side of the CME as it propagates through an non-uniform solar wind. Another possibility is that the asymmetry is not a feature of the CME but the projection of another coronal feature such as a streamer which  only effects one view point. 

\begin{figure}
\begin{center}
\includegraphics[height=.95\textwidth]{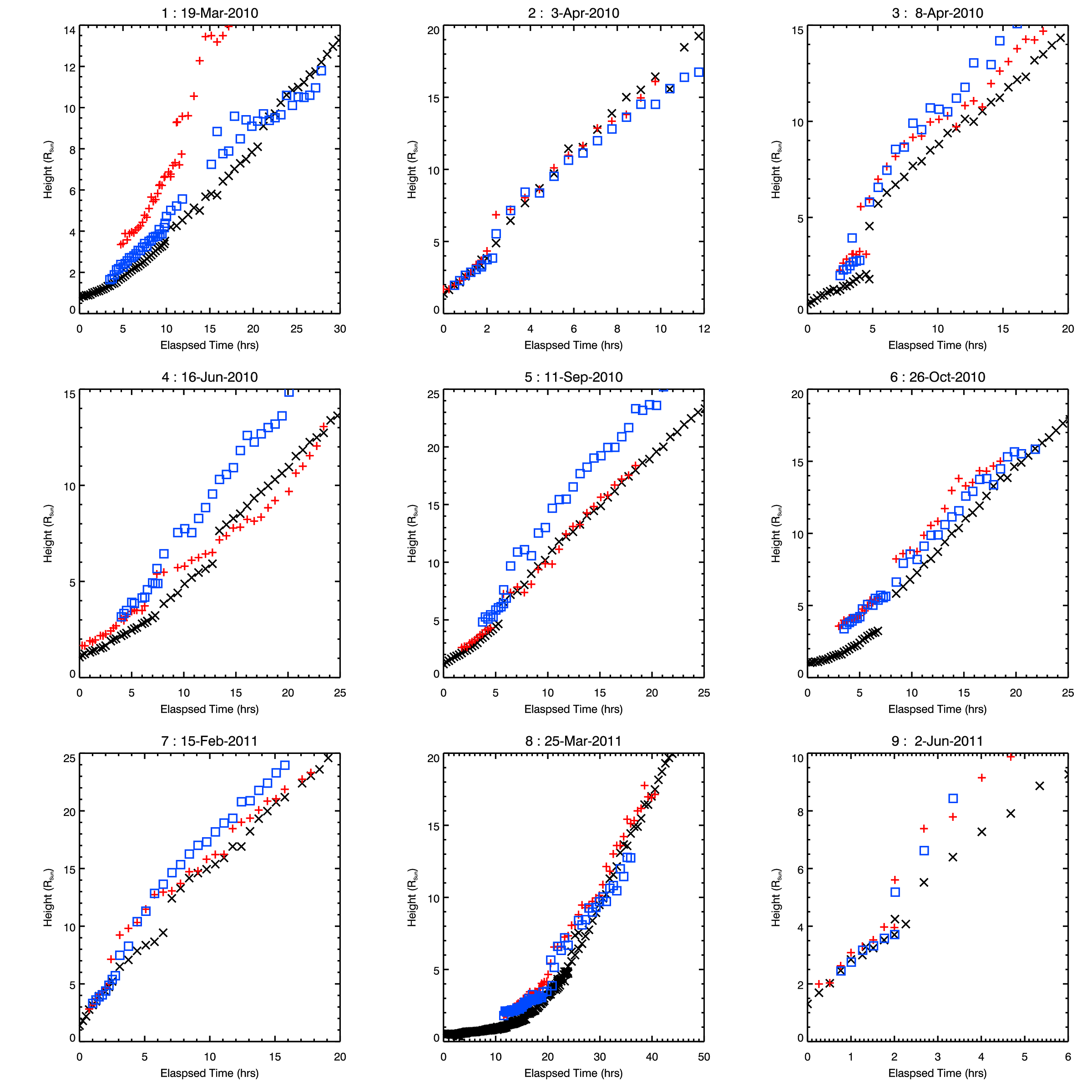}
\caption{Comparison of GCS radius with the Ellipse radii. From the GCS Model the radius are plotted in black X. From the Ellipse Model, we have plotted the mean of the parallel and perpendicular radii.  STEREO A and B values are plotted with red plus signs and blue squares, respectively.}
\label{ellipse_raytrace_radius}
\end{center}
\end{figure}

We also compare the radius of the CME derived from the two models. In figure \ref{ellipse_raytrace_radius}, we have plotted the mean of the parallel and perpendicular radii and the radius from the GCS model using the same convention as figure \ref{ellipse_raytrace}. The radii from the GCS model are similar to the mean of the ellipse radii but show the same discrepancies as the front heights.  Thus the area of cross-section of the two models is also similar. For the theoretical models in section \ref{theory}, we will mainly use the GCS values since we have measured the CME with the GCS model at greater heights and all the theoretical models assume a circular cross section. We use the Ellipse model values when necessary and informative.

\section{Measurement of CME Mass}

To calculate the mass of the CME, we used the Thomson scattering geometry from \citet{1966gtsc.book.....B} which relates the observed brightness to coronal electron density. This technique has been used to measure the mass of CMEs in \citet{1981SoPh...69..169P}, \citet{2000ApJ...534..456V}, \citet{2002ESASP.506...91V}, \citet{2007A&A...467..685S},  and \citet{2009ApJ...698..852C}.  To convert the number of electrons to mass, we assume that the ejected material comprises a mixture of completely ionized hydrogen and 10\% helium. To correctly calculate the mass of a CME from the Thomson scattering geometry, we must know the three dimensional distribution of the CME mass. A first order estimate can be made by assuming all the mass is concentrated to the plane of the CME propagation. We used the longitude from the GCS Model fit to calculate mass of the CME.

\begin{figure}
\begin{center}
\includegraphics[width=1\textwidth]{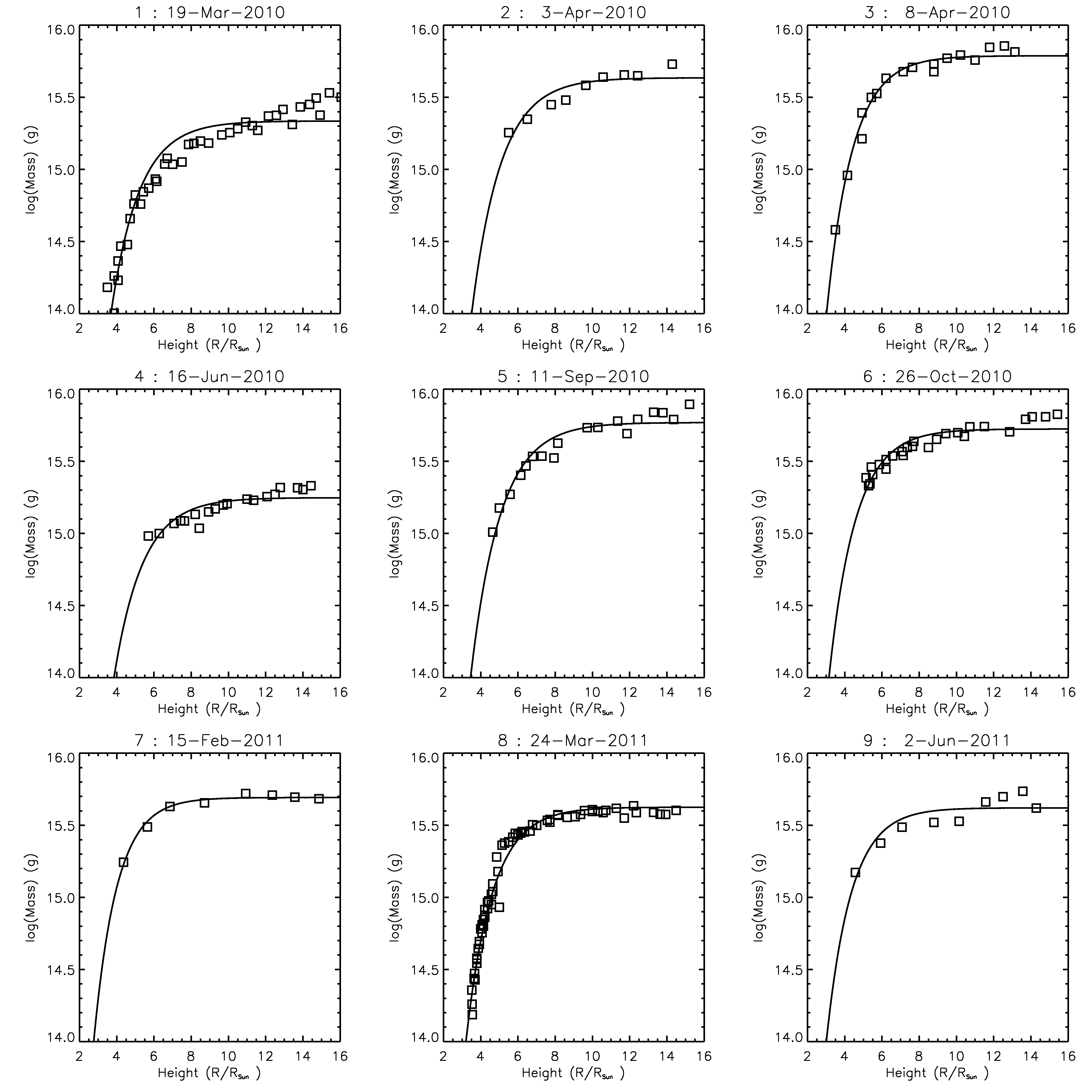}
\caption{Mass of the studied CMEs as calculated using the Thomson scattering geometry and the longitude from the GCS Model fit plotted with squares. The fit of equation \label{mass} to the data is plotted with a solid line.}
\label{mass_plot}
\end{center}
\end{figure}

Figure \ref{mass_plot} shows the total mass of the studied CMEs versus the height of the GCS model front.
 \citet{2009ApJ...698..852C} found that the CME mass increases with time and height until it reaches a constant value above about 10R$_\odot$. To find the final mass of the CME, they used the function :
\begin{equation}
M(h) = M_c (1-e^{-h/h_c})
\label{eq:mass}
\end{equation}
where M$_c$ is the final total mass of the event and h$_c$ is the height where the mass reaches 63\% of its final mass. We have over plotted the fit of this function to the data in figure \ref{mass_plot}. The results of the fit are given in table \ref{table:mass}. The final masses are similar to those found in \citet{2009ApJ...698..852C}. However, the h$_c$ values ($<$h$_c>$=1.4) are lower then the average $<$h$_c>$=2.1 found in \citet{2009ApJ...698..852C}. The difference in determining deprojected the CME height might account for the difference in h$_c$. Here, we are using the front height found from the GCS Model fit. \citet{2009ApJ...698..852C} used the heights deprojected by the longitude found using their mass technique. 
\begin{table}
\begin{center}
\caption{CME Mass}\label{table:mass}
\begin{tabular}{*{4}{c}} \hline  
CME	&	h$_c$	&	$\log$(M$_c$)	&	M$_c$		\\ 
	&	R$_\odot$&            (g)		& (g)\\ \hline
1	&	1.51	&	15.3	&	2.16	$\cdot 10^{15}$	\\
2	&	1.56	&	15.6	&	4.31	$\cdot 10^{15}$	\\
3	&	1.39	&	15.8	&	6.13	$\cdot 10^{15}$	\\
4	&	1.54	&	15.2	&	1.77	$\cdot 10^{15}$	\\
5	&	1.58	&	15.8	&	5.87	$\cdot 10^{15}$	\\
6	&	1.42	&	15.7	&	5.30	$\cdot 10^{15}$	\\
7	&	1.24	&	15.7	&	4.92	$\cdot 10^{15}$	\\
8	&	1.42	&	15.6	&	4.22	$\cdot 10^{15}$	\\
9	&	1.33	&	15.6	&	4.18	$\cdot 10^{15}$	\\ \hline
\end{tabular}
\end{center}
\end{table}

The results from equation \ref{eq:mass} suggest that CMEs are not completely formed in the low corona ($<$1.5 R$_\odot$). Based on the values of h$_c$, the CMEs continue to gain 33\% of their mass between heights of 1.5 and 10 R$_\odot$. The increase in mass cannot be completely explained by the CME emerging into the SECCHI COR2 field of view from behind the occulter. This increase in mass may be caused by a continued injection of mass from the low corona or a pileup of  the ambient corona at the CME front. \citet{1979SoPh...61...95A} found an in flow of mass from lower in the corona for a CME observed in Skylab. Whatever process increases the mass, it does not continue past 10 R$_\odot$. The change in CME mass with height will be used when we look at the forces driving the propagation of the CME.

\chapter[CME Expansion and Propagation in the Inner Heliosphere]{CME Expansion and Propagation \\ in the Inner Heliosphere}\label{expansonpropagation}

\section{Empirical Results : Elliptical and GCS Model}\label{empirical}

In this section we derive the kinematics of the CME based on the measurements made with the GCS and Elliptical models. From the GCS model height and time (HT) data, we derive the kinematics of the CME front. We use these values to estimate the arrival time and velocity of the CME at Earth and compare these results with the in situ data. We also use the Ellipse results to find the kinematics of the CME center and the expansion speeds parallel and perpendicular to the propagation. 

\subsection{Propagation}
\begin{figure}
\begin{center}
\includegraphics[width=1\textwidth]{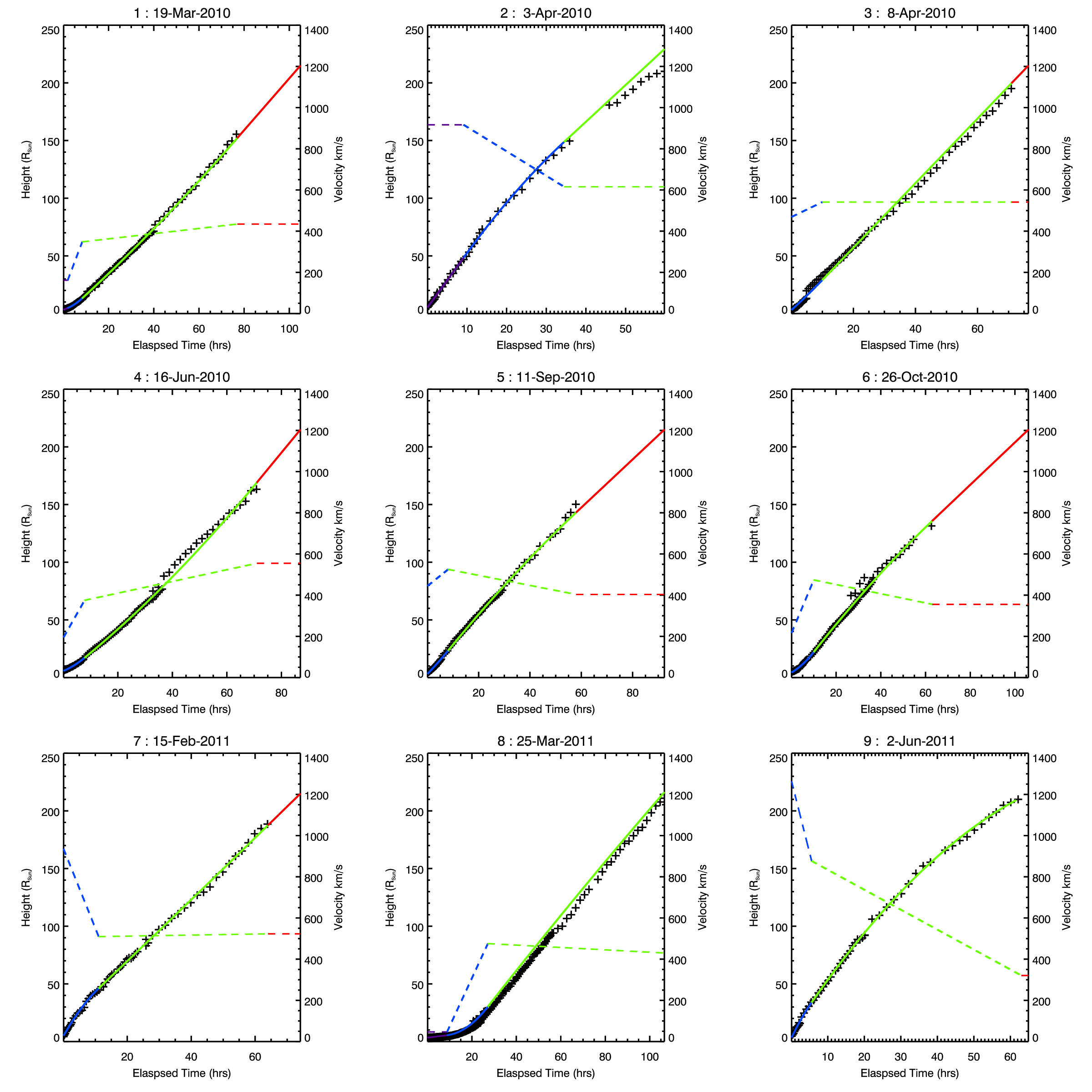}
\caption{Multi-function fit of time and height data from the GCS Model fit of the studied CMEs. The height and time data for each model fit are plotted with plus signs. The multi-function fits are plotted in solid lines. Each phase is plotted in a different color; purple, blue, green and red, respectively. The velocities of the CME fronts are plotted with dashed lines using the same color scheme. }
\label{raytrace_velocity}
\end{center} 
\end{figure}

We used the heights from the GCS Model to determine the three-dimensional kinematics of the CME front. The apex of the GCS Model front in R$_\odot$ is plotted versus elapsed time with plus signs in figure \ref{raytrace_velocity}. The elapsed time begins when the CME is visible in LASCO C2, COR2 A and B and the first model fit is made. To analyze the kinematics of the CME from the HT data, we must fit an analytical function to these data. We began by fitting the data with a first and second order polynomial function. However, we found that a single polynomial function could not describe all the HT data.  Thus, we have chosen to fit the data with multiple polynomial functions. The HT data was fitted using equations \ref{equ:htmod_fit1} - \ref{equ:htmod_fit2}. The height and velocity of the polynomial equations are continuous by imposing the restraints of equations \ref{equ:htmod_fit3} - \ref{equ:htmod_fit4}. This multi-function polynomial fit is used by \citet{2009ApJ...694..707W} and \citet{2009SoPh..259..163W} to fit the kinematics of two CMEs observed in STEREO from the solar surface to 1 AU.
\begin{align}
\label{equ:htmod_fit1}
&v_0 t + h_0 = h  &0<t<t_1\\ 
&0.5 a_1 t^2 + f_1 t + H_1 = h &t_1<t<t_2\\
&0.5  a_2 t^2 + f_2 t + H_2 = h &t_2<t<t_3\\
&v_3 t + H_3 = h  &t_3<t<\infty
\label{equ:htmod_fit2}
\end{align}
where
\begin{align}
\label{equ:htmod_fit3}
H_1 &=  v_0 t_1 + h_0 				&\text{and} \qquad  	&v_1(t=t_1) = v_o  \\
H_2 &= 0.5 a_1 t_2^2 + f_1 t_2 + H_1	&\text{and} \qquad 	&v_2(t=t_2) = a_1 t_2 + f_1 \\
H_3 &=  0.5  a_2 t_3^2 + f_2 t_3 + H_2 	&\text{and} \qquad 	&v_3 = a_2 t_3 + f_2 
\label{equ:htmod_fit4}
\end{align}

The parameters of the multi-function fit are given in table \ref{table:htmod_fit}. For the fit to converge some of the parameters must be held constant by the user. The constant parameters for each fit are listed in boldface in table  \ref{table:htmod_fit}. For all the CMEs the final time, t$_3$, is held constant at the time of the last measurement. When fixed by the user, the initial height, h$_0$ is set at the height of the first measurement. Five of the CMEs do not have an initial linear phase at the heights measured. For these CMEs the duration of the initial phase is set equal to zero (t$_1$ = 0). The total chi-squared of the fit is also given in \ref{table:htmod_fit} assuming a $\pm$1 pixel error in the fits for each instrument.
\begin{table}
\begin{center}
\caption{Parameters of Multi-Function Fit}\label{table:htmod_fit} 
\begin{tabular}{c *{8}{r}}\hline
CME	&	h$_0$	&	v$_0$	&	t$_1$	&	a$_1$	&	t$_2$	&	a$_2$	&	t$_3$	&	$\chi^2$	\\
	&	(R$_\odot$)	&	(km s$^-1$)	&	(hrs)	&	(m s$^-2$)	&	(hrs)	&	(m s$^-2$)	&	(hrs)	&		\\ \hline
1	&	\bf{3.2}	&	161.8	&	1.9		&	8.05	&	8.3		&	0.35	&	\bf{77}	&	2182	\\
2	&	\bf{5.0}	&	916.5	&	9.2		&	-3.30&	60.0		&	\bf{0.0}	&	\bf{60}	&	1921	\\
3	&	\bf{3.0}	&	468.5	&	\bf{0.0}	&	2.02	&	\bf{10.0}	&	\bf{0.0}	&	\bf{71}	&	23411	\\
4	&	\bf{5.7}	&	193.4	&	\bf{0.0}	&	6.47	&	7.8		&	0.79	&	\bf{71}	&	3852	\\
5	&	\bf{3.1}	&	444.3	&	\bf{0.0}	&	2.77	&	8.1		&	-0.68	&	\bf{58}	&	7653	\\
6	&	4.3		&	214.7	&	\bf{0.0}	&	7.20	&	\bf{10.0}	&	-0.62	&	\bf{63}	&	4780	\\
7	&	4.7		&	938.5	&	\bf{0.0}	&	-10.77&	11.1		&	0.07	&	\bf{64}	&	2962	\\
8	&	3.3		&	46.7		&	8.8		&	6.49	&	27.2		&	-0.16	&	\bf{107}	&	14706	\\
9	&	\bf{3.0}	&	1269.5	&	0.1		&	-19.55&	5.6		&	-2.69	&	\bf{63}	&	4312	\\ \hline
\end{tabular}
\end{center}
\end{table}

In figure \ref{raytrace_velocity}, we have plotted the results for each multi-function fit with the height and time data from the GCS Model. Each phase of the fit is plotted in a different color; purple, blue, green and red, respectively. The instantaneous velocity is also plotted with a dashed line for each CME using the same color scheme. The right axis of the plot gives the scale for the velocity data.

From the kinematic data, we can immediately see a trend that has been well observed in CME propagation \citep{2006SSRv..124..145G}. The CME velocities are converging as the CME propagates to 1 AU. The range of initial CME velocities is 47-1270 km s$^{-1}$ with a standard deviation of 426  km s$^{-1}$ while the range of range of final velocities is 321-615 km s$^{-1}$ with a standard deviation of 99 km s$^{-1}$. The final velocities are listed in table \ref{table:earth_values}. \citet{2010ApJ...717L.159P} found similar results. By comparing the velocities in table \ref{table:htmod_fit} with the projected velocities in table \ref{projected_values}, we see that the initial three-dimensional velocities are larger and have a greater range than the projected values. For the LASCO view the initial CME velocities range from 45 to 669 km s$^{-1}$ with a standard deviation of 217 km s$^{-1}$. Thus the convergence of CME velocities from the inner corona to 1 AU is larger by a factor of three when the three-dimensional velocities are used. 

Specifically, we see in table \ref{table:htmod_fit} that the CMEs with the lowest initial velocities (CMEs 1, 4 and 8) have the largest accelerations in phase 1. While CMEs 2, 7 and 9 have the highest initial velocities and are the only ones that decelerate in phase 1. For all the CMEs, the acceleration in phase 2 is small, with the exception of CME 9 which has the highest initial velocity and decelerated considerably before reaching Earth. CME 9 also has the lowest final velocity.  As noted in table \ref{categories} CME 9 is preceded by another CME which many account for the unusual kinematics of the event. 

Another extreme case is CME 8. This CME has an initial velocity which is an order of magnitude smaller than the mean of the other CMEs initial velocities. This slow initial phase persists for 8 hours before the CME begins to accelerate. By the time the CME reaches Earth the finial velocity is similar to the other CMEs.

Thus all of the CMEs decelerate or accelerate between the Sun and Earth. Only CME 3 does not have some acceleration in the final phase of the multi-function fits. CME 3 also has the largest error for the multi-function fit.  The other eight CMEs have a small acceleration in the final phase of the fit which persists over an average of 60 hours. An acceleration of the CME requires an imbalance of the net force on the CME. We will look at the forces on the CME in section \ref{theory}.

\subsection{Predicted CME Values at Earth}\label{earth}
Using the analytical fits to the CME HT data, we can predict the arrival time of the CME at Earth. By comparing the predicted arrival time with the in situ data, we can assess the accuracy of our three-dimensional trajectories and propagation fits. We can also verify the association between the CME observed remotely and in situ in the case of multiple events. To estimate the arrival time of the CME, we assume that the CME propagates at a constant velocity after the last height measurement, equation \ref{equ:htmod_fit2}. Table \ref{table:earth_values} gives the predicted arrival times, $\Delta$T, the difference between the predicted  and in situ arrival times (table \ref{table:insitu}), and normalized arrival time error, the ratio of the transit time of the CME and $\Delta$T. The average transit time for these CMEs is 3.6 days which is slower then then 2.7 day average transit time for CMEs which caused major geomagnetic storms \citep{2003ApJ...582..520Z} but faster then the 4.3 day arrival time for slow CMEs \citep{2001JGR...10629207G}.

For our sample, the maximum discrepancy in the arrival times is -17.6 hours or $\pm$20 \% of the transit time. This is the largest sample date of arrival times for CMEs using STEREO data. \citet{2011SpWea...901005D} presented the predicted arrival time of CME 3 on 8-April-2010 from eight techniques. These eight techniques used a range of various lead times and had an error of $\pm$12 hours. Our arrival time predictions and those from \citet{2011SpWea...901005D} are not a significant improvement over those made with only the LASCO data.  \citet{2001JGR...10629207G} used an empirical model to predict the arrival times of 47 CMEs observed in LASCO and found an error range of $\pm$15 hours for 72\% of the CMEs. We find that even with nearly continuous observations of the CME in the heliosphere and the best available models, it is still difficult to predict the arrival time of a CME in situ.

\begin{table}
\begin{center}
\caption{CME Arrival Times and Speeds}\label{table:earth_values} 
\begin{tabular}{*{8}{c}}\hline
CME	&\multicolumn{2}{c}{Arrival}		&	$\Delta$T	&	$\Delta$T/Tran	&	v$_3$	&	$<$v$_{in situ}>$&	$\Delta$v	\\
	&	Date			&	Time		&	(hrs)	&	(\%)	&	(km s$^{-1}$)	&	(km s$^{-1}$)	&	(km s$^{-1}$)	\\ \hline
1	&	23-Mar-2010	&	20:32	&	-2.5	&	-2	&	435	&	289	&	146	\\
2	&	05-Apr-2010	&	17:40	&	11.0	&	20	&	615	&	639	&	-24	\\
3	&	11-Apr-2010	&	07:46	&	-4.2	&	-6	&	541	&	408	&	133	\\
4	&	20-Jun-2010	&	06:21	&	-17.6&		-20	&	555	&	365	&	190	\\
5	&	14-Sep-2010	&	22:46	&	8.4	&	9	&	403	&	366	&	37	\\
6	&	30-Oct-2010	&	17:39	&	-11.2&		-11	&	356	&	360	&	-4	\\
7	&	18-Feb-2011	&	04:32	&	4.5	&	6	&	523	&	476	&	47	\\
8	&	29-Mar-2011	&	14:00	&	-0.6	&	-1	&	430	&	342	&	88	\\
9	&	05-Jun-2011	&	01:04	&	5.6	&	9	&	321	&	452	&	-131	\\ \hline
\end{tabular}
\end{center}
\end{table}

\begin{figure}
\begin{center}
\includegraphics[width=4in]{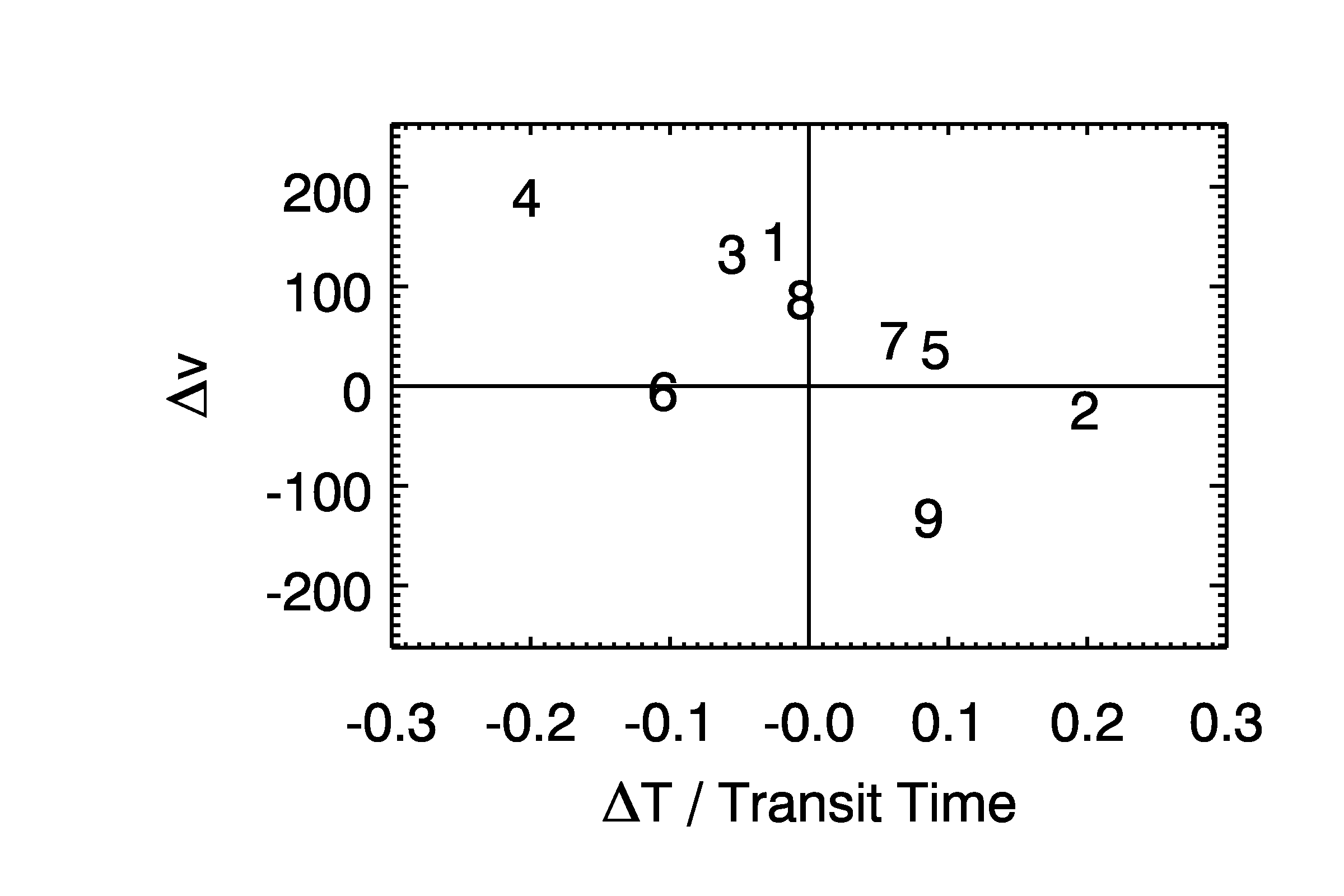}
\caption{The ratio of the transit time and the error in the arrival time plotted versus the difference in the final velocity at Earth.}
\label{arrival_speed}
\end{center} 
\end{figure}

The only other parameter which can be compared between the remote sensing and in situ data without further use of models is the velocity at Earth. In table \ref{table:earth_values}, we give final predicted velocity, v$_3$ and the velocity in situ at the time of the CME detection. The difference between the final and in situ velocity, v$_3$-v$_{in situ}$ = $\Delta$v, ranges from -131 to 190 km s$^{-1}$ with a standard deviation of 99 km s$^{-1}$. The $\Delta$v is less then 10 \% of the in situ velocity for only three CMEs. To see how $\Delta$v effects the arrival time, in figure \ref{arrival_speed}, we have plotted $\Delta$v versus the normalized arrival time error, $\Delta$T/Transit. From this plot, we can see that the difference in the arrival time and the difference in the final velocity are correlated. If the predicted velocity is too fast then the predicted CME arrival time in too early and the opposite is also true.  

\begin{figure}
\begin{center}
\includegraphics[width=4in]{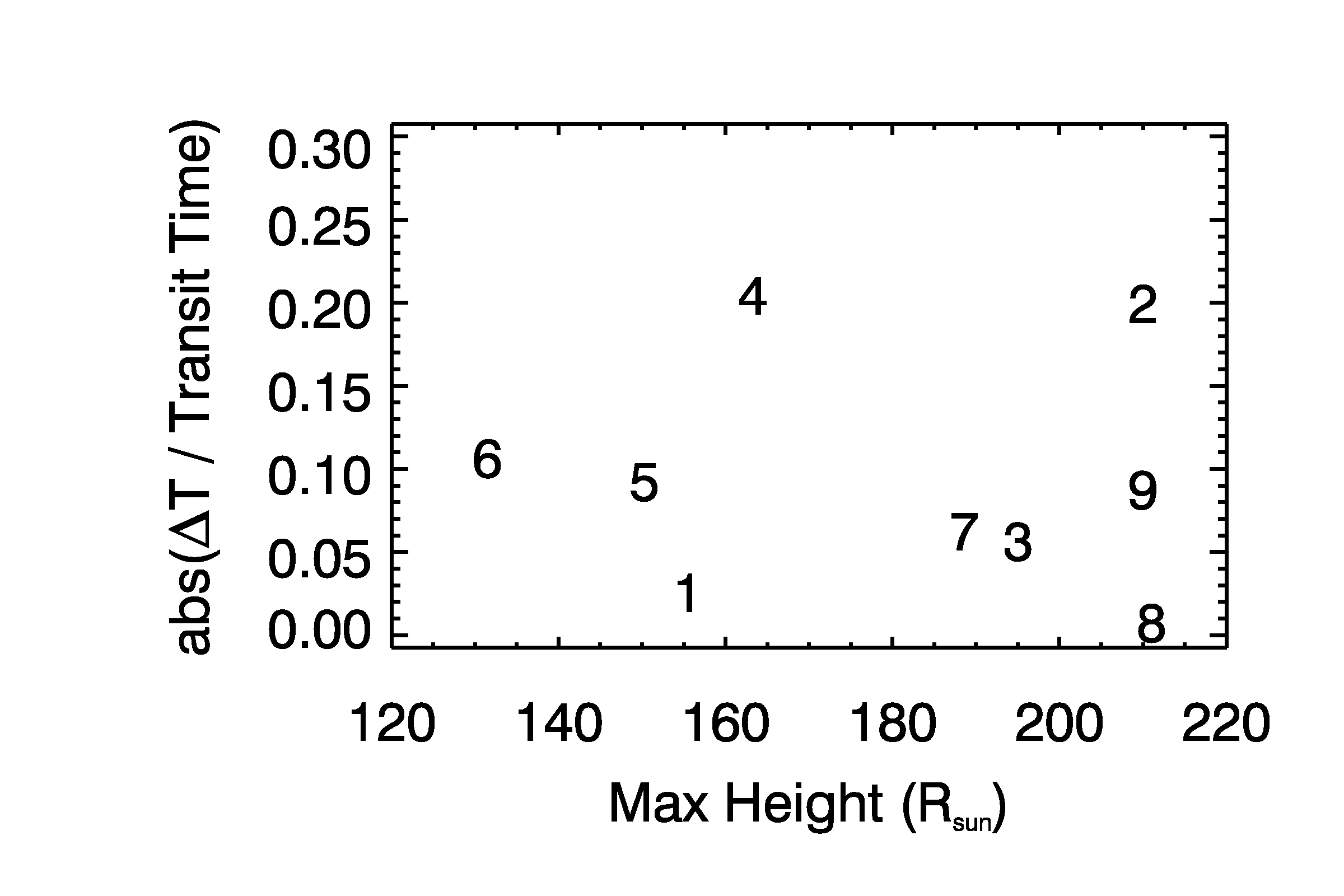}
\caption{The maximum height of the GCS Model fit plotted versus the ratio of the transit time and the normalized error in the arrival time.}
\label{height_arrival}
\end{center} 
\end{figure}

Based on the results from figure \ref{arrival_speed}, we want to test our assumption that the CME propagates at a constant velocity after the last height measurement.  If the extrapolation of the final velocity is inaccurate, we expect the normalized arrival time error to be greater with lower maximum heights.  In other words, if the extrapolation of a linear final velocity is causing the error in the arrival time, we would expect the arrival time error to be small for CMEs that are measured close to Earth. In figure \ref{height_arrival}, we have plotted the maximum height of the GCS Model fit plotted versus normalized arrival time error. The plot shows that error in the arrival time is not correlated with the maximum height. Thus our assumption of a linear final velocity is not directly effecting the arrival times.

There are several possible sources of error in the predicted arrival times other than an error in the fit of the analytical function to the HT data. One possible source of the error is that the predicted arrival times are based on the height of the CME apex. For example, if the CME does not impact Earth at the apex the arrival time in situ would be delayed assuming a convex front. \citet{2012arXiv1202.1299M} found that for an assumed circular CME front the arrival time delay can be as large as 2 days. However, this effect would only produce an early predicted arrival time ($\Delta$T $<$ 0). Our arrival time errors are both positive and negative. Thus this source of error could not apply to all our CMEs.

Another possible source of error is the fit of the GCS model to the remote sensing data. If the longitude of the model fit is incorrect the heights from the fit would also be incorrect. In the middle of the HI2 field of view, a 1${^o}$ change in longitude can cause a change in height of $\sim$2 R$_\odot$ for the same point in the image. The exact error depends on the several parameters of the observation.  An error in the longitude can make the predicted arrival time early or late. Given the distribution of our arrival time error and the sensitively of the fit, we assume that this is the most likely source of error. The error in fitting the GCS model has not been studied and is beyond the scope of this work.

\subsection{Expansion}
In addition to the kinematics of the CME front, we want to analyze the expansion of the CME. To isolate the expansion, we decompose the front velocity of the CME into two components; the bulk and parallel expansion velocities \citep{2006ApJ...652.1747C}. The sum of these two velocities is equal to the front velocity. We also want to analyze the expansion of the CME perpendicular to the propagation trajectory. We calculate these three velocities (bulk, parallel and perpendicular) from the parameters of the Elliptical Model. The motion of the center of the ellipse is the bulk velocity. The change in the semi-major and semi-minor axes are the perpendicular and parallel velocities, respectively. 

The GCS model has a circular cross-section and thus has only one expansion velocity. The height of the center of the model, R, (figure \ref{Fig:GCS} point C), is related to the radius, r, by 
\begin{equation}
r = \kappa R
\label{eq:gcs_expan}
\end{equation}
where $\kappa$ is the user defined GCS model parameter listed in table \ref{raytrace} as ratio. If $\kappa$ is constant, the model expands in a self-similar way with height.  The LASCO field of view is needed for $\kappa$ to be well contained. After the CME is no longer visible in the LASCO view, we keep $\kappa$ constant. To assess any changes in the bulk and expansion velocities at larger heights, we must use the Ellipse Model.

\begin{figure} 
\begin{center}
\includegraphics[height=.9\textheight]{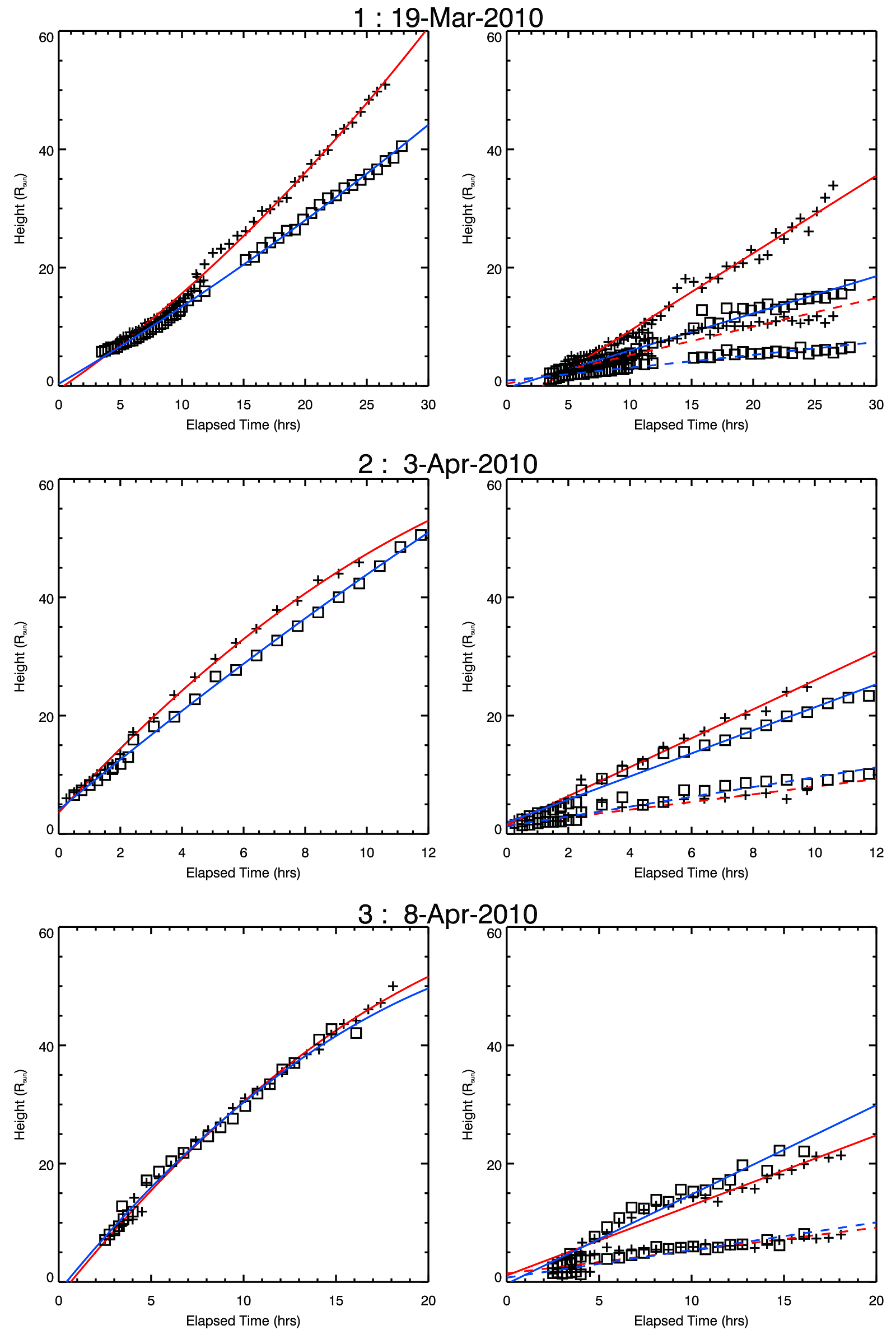}
\caption{The height of the ellipse center (right) and the parallel and perpendicular radii from the Ellipse model. The STEREO-A and B values are plotted with plus signs and squares, respectively. We have fitted the data with quadratic functions. These fits are plotted with solid lines in red and blue for STEREO A and B, respectively. All height values have been calculated using the longitude found from the GCS Model.}
\label{ellipse_plot_1}
\end{center}
\end{figure}

\begin{figure}
\begin{center}
\includegraphics[height=.9\textheight]{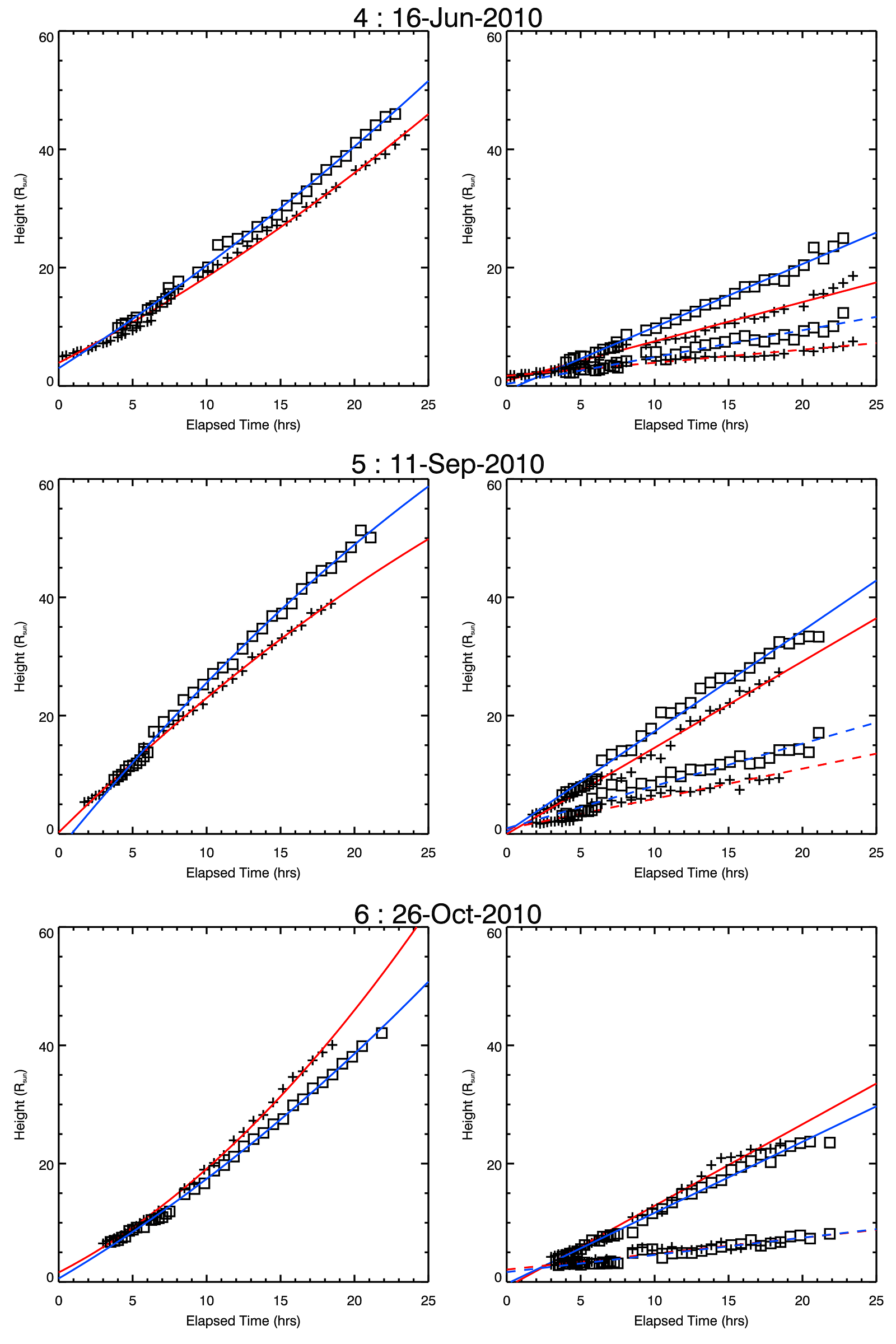}
\caption{The height of the ellipse center (right) and the parallel and perpendicular radii from the Ellipse model. The STEREO-A and B values are plotted with plus signs and squares, respectively. We have fitted the data with quadratic functions. These fits are plotted with solid lines in red and blue for STEREO A and B, respectively. All height values have been calculated using the longitude found from the GCS Model.}
\label{ellipse_plot_2}
\end{center}
\end{figure}

\begin{figure}
\begin{center}
\includegraphics[height=.9\textheight]{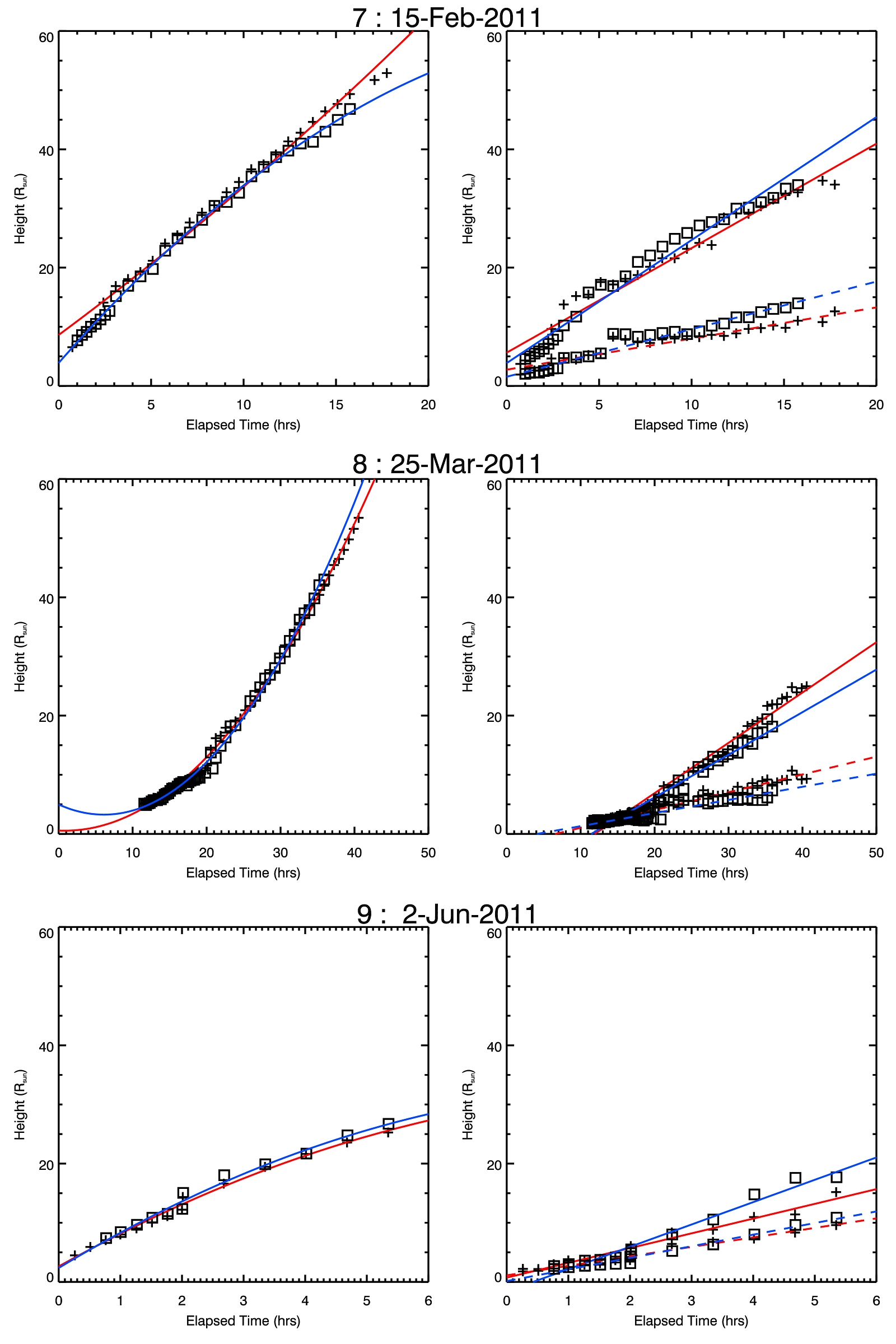}
\caption{The height of the ellipse center (right) and the parallel and perpendicular radii from the Ellipse model. The STEREO-A and B values are plotted with plus signs and squares, respectively. We have fitted the data with quadratic functions. These fits are plotted with solid lines in red and blue for STEREO A and B, respectively. All height values have been calculated using the longitude found from the GCS Model.}
\label{ellipse_plot_3}
\end{center}
\end{figure}

The results of the ellipse fits are shown in figures \ref{ellipse_plot_1}-\ref{ellipse_plot_3}. We have plotted the height of the ellipse center versus elapsed time (left column). We have also plotted the parallel and perpendicular radii versus elapsed time (right column). The STEREO-A and B values are plotted with plus signs and squares, respectively. We have fitted the data with polynomial functions. These fits are plotted with solid lines in red and blue for STEREO A and B, respectively. The elapsed time along the x-axis is the same as in figure \ref{raytrace_velocity}.

To obtain the bulk velocity, we fit the center height versus time with a second order polynomial. Since the ellipse measurements do not cover the range of height that the GCS fits do, we found that we need only a single polynomial to fit the data. Table \ref{table:bulk_motion} gives the maximum height of the CME for each ellipse fit, the acceleration and the velocity at the maximum height.

 The expansion speeds  were calculated by fitting the radii with a first order polynomial. The speeds from these fits are give in table \ref{table:expansion}. Also in table \ref{table:expansion}, we list the ratio of the parallel and perpendicular speeds. For all the CMEs, the parallel speed is less then half the perpendicular velocity with the exception of CME 9 which is about 50 \%.  These speeds  confirm what we can see in the images, the CMEs become more elliptical as they propagate towards 1AU. 

\begin{table}
\begin{center}
\caption{Ellipse Model : Bulk Motion}\label{table:bulk_motion} 
\begin{tabular}{c *{6}{r}}\hline
CME	& \multicolumn{2}{c}{Max Height}& \multicolumn{2}{c}{Accel.}& \multicolumn{2}{c}{Velocity}\\
	& \multicolumn{2}{c}{(R$_\odot$)} & \multicolumn{2}{c}{(m s $^{-2}$)}&\multicolumn{2}{c}{(km s$^{-1}$)}\\
	&	A	&	B	&	A	&	B	&	A	&	B	\\ \hline
1	&	50.9	&	40.5	&	2.2	&	0.8	&	487	&	321	\\
2	&	45.9	&	50.5	&	-13.9	&	-3.7	&	608	&	679	\\
3	&	50.0	&	42.8	&	-6.3	&	-6.8	&	339	&	379	\\
4	&	42.4	&	46.0	&	1.7	&	1.5	&	390	&	430	\\
5	&	38.9	&	51.3	&	-2.1	&	-2.6	&	340	&	401	\\
6	&	40.1	&	42.1	&	5.0	&	2.3	&	582	&	466	\\
7	&	52.9	&	46.8	&	2.0	&	-5.9	&	578	&	350	\\
8	&	53.4	&	43.0	&	3.6	&	4.9	&	520	&	530	\\
9	&	25.3	&	26.7	&	-30.4	&	-34.6	&	538	&	545	\\ \hline
\end{tabular}
\end{center}
\end{table}

\begin{table}
\begin{center}
\caption{Ellipse Model : Expansion Velocities}\label{table:expansion} 
\begin{tabular}{c *{6}{c}}\hline
CME	& \multicolumn{2}{c}{Perp. Speed}& \multicolumn{2}{c}{Para. Speed}& \multicolumn{2}{c}{Para./Perp.}\\
	& \multicolumn{2}{c}{(km s$^{-1}$)}&\multicolumn{2}{c}{(km s$^{-1}$)}&\multicolumn{2}{c}{($\%$)}\\
	&	A	&	B	&	A	&	B	&	A	&	B	\\
1	&	254	&	122	&	93	&	42	&	37	&	35	\\
2	&	473	&	377	&	125	&	161	&	26	&	43	\\
3	&	229	&	292	&	75	&	90	&	33	&	31	\\
4	&	128	&	206	&	43	&	88	&	33	&	43	\\
5	&	282	&	329	&	98	&	138	&	35	&	42	\\
6	&	266	&	232	&	52	&	56	&	20	&	24	\\
7	&	342	&	402	&	102	&	156	&	30	&	39	\\
8	&	165	&	139	&	58	&	43	&	35	&	31	\\
9	&	482	&	730	&	310	&	379	&	64	&	52	\\ \hline
\end{tabular}
\end{center}
\end{table}

\begin{figure}
\begin{center}
\includegraphics[width=4in]{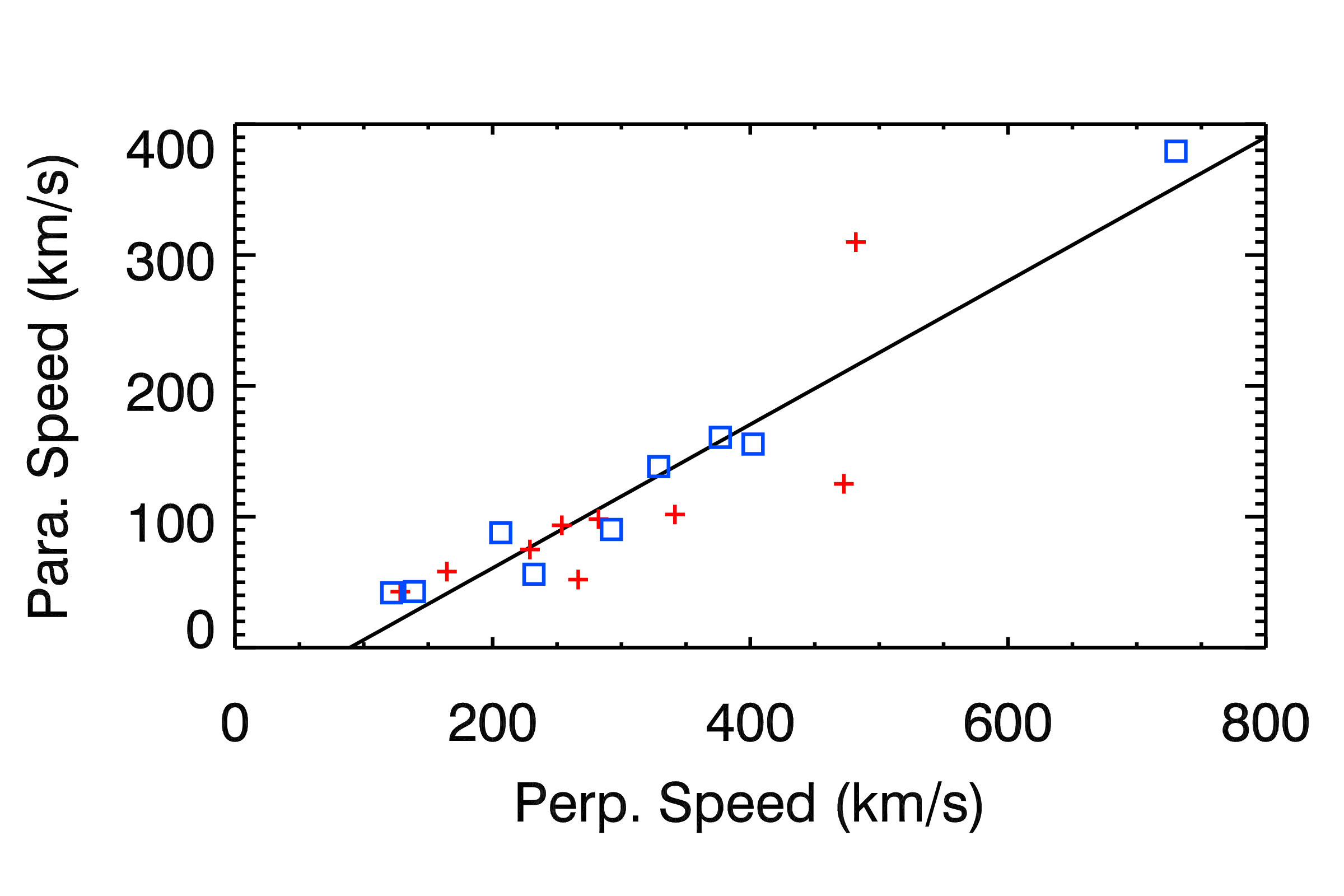}
\caption{From the Ellipse Model, the perpendicular expansion speeds plotted versus the parallel expansion speeds. STEREO A values are plotted with red plus signs. STEREO-B values are plotted with blue squares.}
\label{Fig:perp_para}
\end{center}
\end{figure}

\begin{figure}
  \centering
  \subfloat{\includegraphics[width=0.5\textwidth]{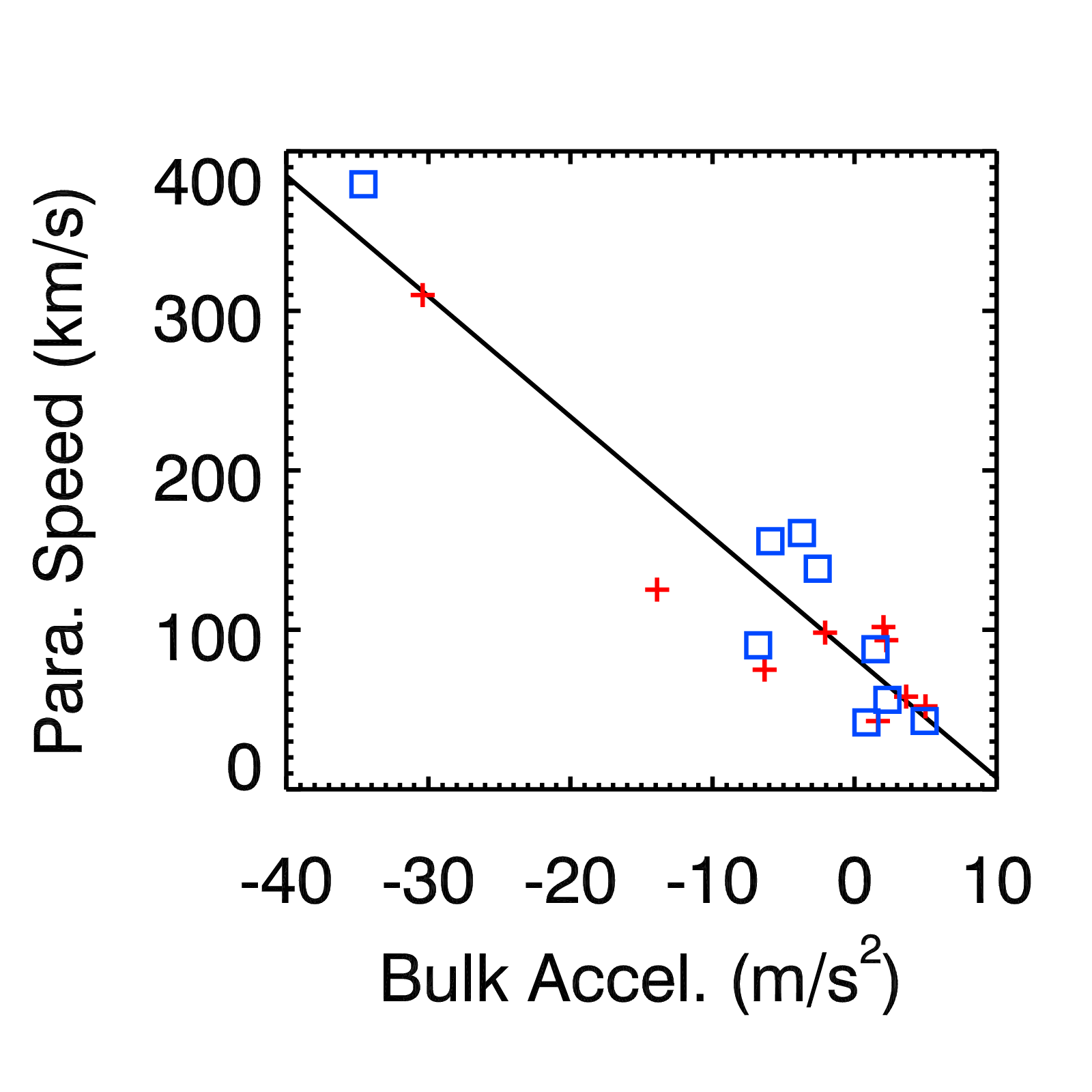}}                
  \subfloat{\includegraphics[width=0.5\textwidth]{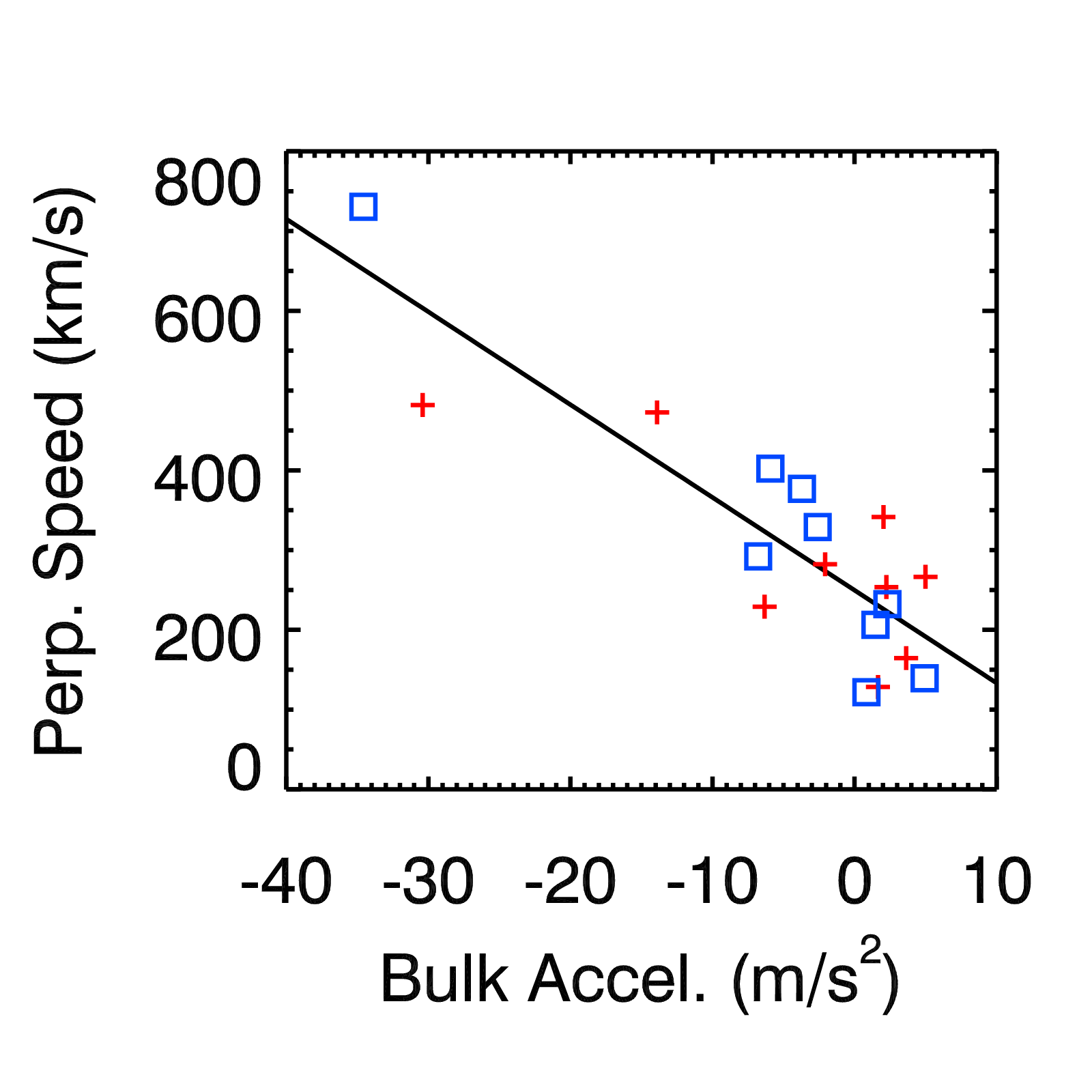}}     
  \caption{From the Ellipse Model, the parallel and perpendicular expansion speeds plotted versus the bulk acceleration. STEREO A values are plotted with red plus signs. STEREO-B values are plotted with blue squares. }
  \label{Fig:acc_para_perp}
\end{figure}

To analyze the elongation of the CME, in figure \ref{Fig:perp_para}, we have plotted the perpendicular speeds versus the parallel speeds. STEREO A and B values are plotted with red plus signs and blue squares, respectively. We have found that the parallel and perpendicular speeds are related by
\begin{equation}
V_{para} = 0.5 \times V_{perp} - 50.
\label{eq:para_perp}
\end{equation}
This result would indicate that the parallel and perpendicular  expansions are not independent but are dependent on another parameter of the CME. In figure \ref{Fig:acc_para_perp}, we have plotted the parallel and perpendicular speeds versus the bulk acceleration of the CME.  Again, STEREO A and B values are plotted with red plus signs and blue squares, respectively. The parallel and perpendicular speeds are related to the bulk acceleration by
\begin{align}
V_{para} &= -7.5 \times A_{bulk} - 83 \\
V_{perp} &= -11.6 \times A_{bulk} - 250.
\label{acc}
\end{align}
Thus for these range of heights where the Ellipse Model was applied, approximately 5 to 50 R$_\odot$, the expansion speeds are constant, the parallel and perpendicular expansion speeds are linearly related, and both expansion speeds are linearly related to the the bulk acceleration. Since there is no acceleration in the expansion speeds, a result of the ellipse fits is the absence of a net force in the interior of the CMEs. The bulk acceleration requires that a net force is applied to the exterior of the CME. The result that expansion speeds and the bulk acceleration are related, implies that the expansion speeds are also dependent on the net external force on the CME.

\section{Theoretical Models of  CME Propagation and Expansion}\label{theory}
 
 In this section, we use the dimensions and kinematics of the CMEs found from observational data with the theoretical models of CME propagation and expansion introduced in section \ref{intro:theory} to derive other physical properties of the CMEs. With the theoretical models, we will explore the net forces on the CME which drive propagation and expansion. 
 
\subsection{Kinematic Model}

There are two implementations of the kinematic model, the melon-seed model and melon-seed overpressure expansion (MSOE) model, introduced in section \ref{intro:kinematic}. We will use the melon-seed model to analyze the expansion of the CME and the MSOE model to analyze the external forces on the CME.

\subsubsection{Melon-Seed Model} 

The melon-seed model predicts two regimes for the expansion of the CME; the pressure-dominated and magnetic-dominated case \citep{1984SoPh...94..387P}. The pressured-dominated case (equation \ref{eq:pressure_case}) predicts that the CME is expanding faster than the center propagates assuming the sound speed is constant (adiabatic expansion). Our measurements (tables \ref{table:bulk_motion} and \ref{table:expansion}) show that the pressure dominated case does not apply at these heights in the corona. Thus we focus on the magnetic dominated case, which prescribes self-similar expansion
\begin{equation}
r = r_o \left ( \frac{R}{R_\odot} \right ),
\label{eq:magnetic_case}
\end{equation}
where the radius, r, is dependent on the center height of the CME, R, by some arbitrary function r$_o$. The GCS model expands self-similarly, where r$_o$ is the user defined parameter $\kappa$, equation \ref{eq:gcs_expan}. The parameter $\kappa$ is kept constant after the CME is no longer visible in LASCO.

\begin{figure}
\begin{center}
\includegraphics[height=.95\textwidth]{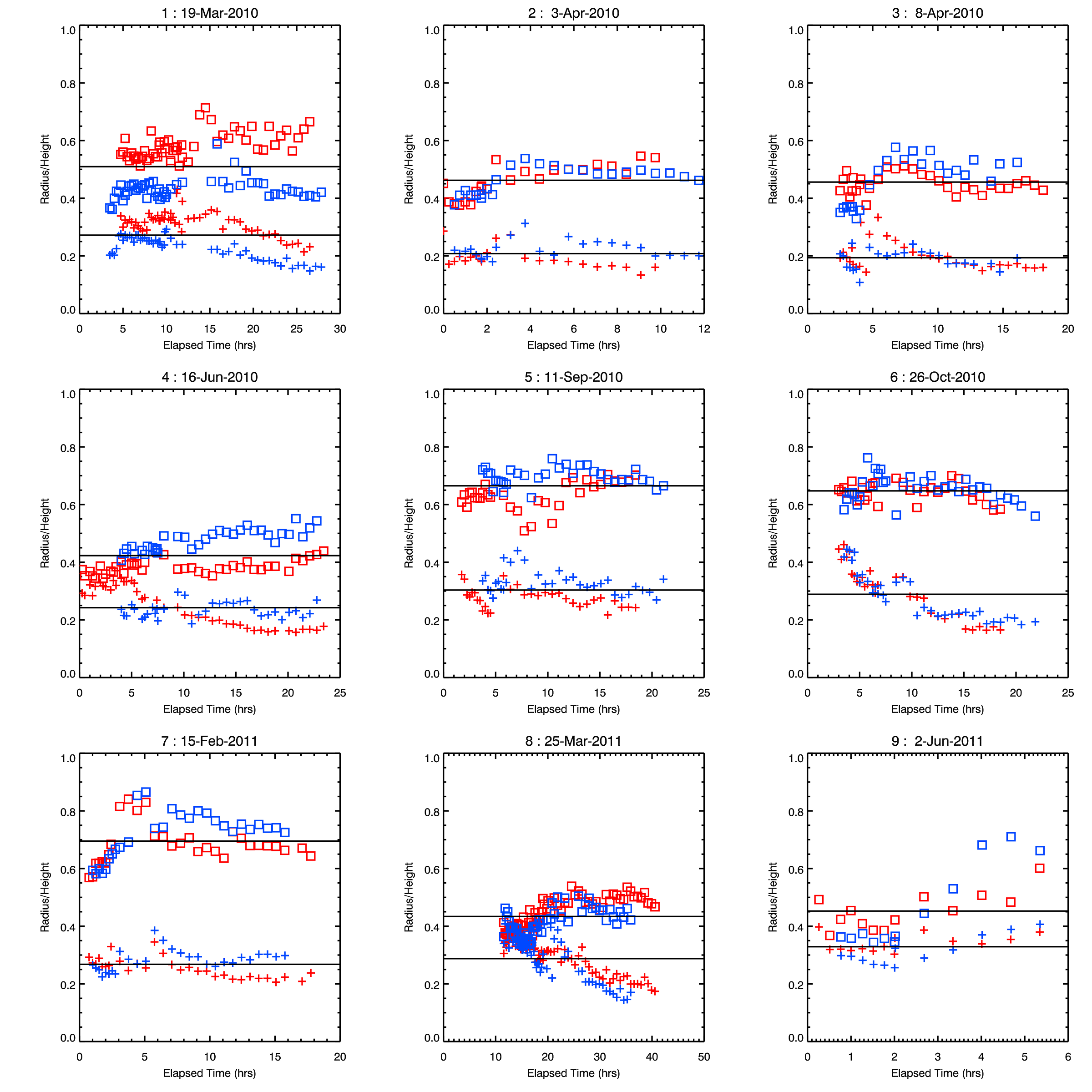}
\caption{The ratio of the parallel and perpendicular radii and center height from the Ellipse Model. The perpendicular radius is plotted with squares and the parallel radius is plotted with plus signs. The STEREO A and B data are plotted in red a blue, respectively. We have over plotted the mean of each data set in black lines.}
\label{melon_seed_magnetic}
\end{center}
\end{figure}

From the Ellipse Model, we have plotted the ratio of the radii and center height in figure \ref{melon_seed_magnetic}.  The perpendicular data is plotted with squares and the parallel data is plotted with plus signs. The STEREO A and B data are plotted in red and blue, respectively. We have over plotted the mean of each data set with black lines. 
The overall trend is that CMEs are expanding self-similarly where the ratio of the radii and the center heights are constant. Some events, most notably CME 8, seem to show a variation in the parallel and perpendicular ratio early in the CME propagation that then becomes constant. We investigated this apparent trend and found that variation in the ratio occurs between 10 and 15 R$_\odot$, the boundary between the COR2 and HI1 instruments. These two instruments have different photometeric sensitivities. If the CME has a diffuse front, the outer envelop of the CME will appear to increase as it moves into the more sensitive HI1 field of view. A similar effect has not been observed in limb CMEs as viewed from LASCO C3 where the CME is observed in the same instrument until 32 R$_\odot$. Thus it is possible that the change in the ratios is due to observational effects and is not physical. Other CMEs, such as CME 6 and CME 9, have variable ratios at later times. However, the variation in the CME 6 data is only in the parallel radius and in the CME 9 data is only in the STEREO B data for the perpendicular radius.  Without a larger sample of CMEs with varying ratios, we cannot form any conclusions. Thus we conclude that, most CMEs expand in a self-similar fashion in both the parallel and perpendicular directions between 5 and 50 R$_\odot$.%

In table \ref{table:ratio_expan}, we have listed the mean of the parallel and perpendicular ratios to center height and the $\kappa$ parameter from the GCS Model. The mean and standard derivation of each data set is listed in the final line. The perpendicular ratio varies more than the parallel ratio. We looked at several parameters of the CME to see if there was any correlation with the value of the perpendicular ratio. We did not see any clear trend between the perpendicular ratio and the mass, latitude, rotation of the CME  or magnetic field of source region. The perpendicular ratio may be dependent on structure of the ambient corona.   Also the $\kappa$ is near the mean of the parallel and perpendicular radii for most of the CMEs. This is expected from the comparison of the Ellipse and the GCS fits. 
\begin{table}
\begin{center}
\caption{Ratio of Radius and Center Height}\label{table:ratio_expan} 
\begin{tabular}{c c c c}\hline
CME	&	Para.	&	Perp.	&	GCS	\\ \hline
1	&	0.27	&	0.51	&	0.33	\\
2	&	0.21	&	0.46	&	0.50	\\
3	&	0.19	&	0.46	&	0.36	\\
4	&	0.24	&	0.42	&	0.36	\\
5	&	0.30	&	0.66	&	0.53	\\
6	&	0.29	&	0.65	&	0.46	\\
7	&	0.27	&	0.70	&	0.55	\\
8	&	0.29	&	0.43	&	0.60	\\
9	&	0.33	&	0.45	&	0.40	\\ \hline
        &  0.27$\pm$0.04 &0.53$\pm$0.11&0.45$\pm$0.10\\
\end{tabular}
\end{center}
\end{table}

\subsubsection{Melon-Seed Overpressure Expansion Model}

The melon-seed overpressure expansion (MSOE) model \citep{2006SoPh..239..293S} prescribes the forces external to the CME. From the MSOE model there are three forces that combine to create the net force on the CME and drive its propagation, 
\begin{equation}\label{siscoe_equ5}
F_{net} = M \frac{dv}{dt} = F_m - F_g - F_d. 
\end{equation} 
where F$_m$ is the magnetic force, F$_g$ is the gravitation force, F$_d$ is the drag force, M is the mass and $\frac{dv}{dt}$ is the acceleration of the CME front.

The gravitational force, F$_g$ equation \ref{eq:Fg}, depends on the center height and mass of the CME, both measured quantities. The magnetic force, F$_m$ equation \ref{eq:Fm}, depends on the volume of the CME and the ambient magnetic field. We calculate the volume of the CME assuming an ellipsoid shape,
\begin{equation}
Vol = \frac{4}{3} \pi r^2 \frac{W}{2}
\end{equation}
where the radius, r, and width, W, are from the GCS model. To calculate the magnetic force, we first need to estimate the ambient magnetic field. We assume that the ambient magnetic field is described by the function 
\begin{equation}\label{Bsw}
B_a(R) = B_o  \left ( \frac{R_\odot}{R} \right )^2.
\end{equation}
for heights R$>$5R$_\odot$ \citep{1984SoPh...94..387P}. To find B$_o$, we solve equation \ref{Bsw} where B$_a$(R) is the average upstream magnetic field from the in situ data and R is 1AU from the in situ data. The values of the upstream magnetic field and B$_o$ are listed in table \ref{table:upstream}. The drag force, Fd equation \ref{eq:Fd}, is dependent on the area of the CME, mass density and drag coefficient of the solar wind and difference between the propagation velocity and the solar wind velocity. To estimate the density of the solar wind, we use the function from \citet{1996JGR...10127499C},
\begin{equation}\label{nr}
n(R) = 4 \left [ 3 \left ( \frac{R_\odot}{R} \right )^{12} + \left ( \frac{R_\odot}{R} \right )^4 \right ] \times 10^8 +2.3 \times 10^5 \left ( \frac{R_\odot}{R} \right )^2
\end{equation}
This function gives a density of 5 cm$^{-3}$ at 1 AU which is consistent with the in situ data. To calculate the solar wind velocity, we use the function from \citet{2006SoPh..239..293S},
\begin{equation}
V_{sw}(R) = \frac{n(1 AU)}{n(R)} V_{SW}(1 AU) \left ( \frac{1 AU}{R} \right )^2
\end{equation}
where we use the average upstream solar wind velocity from the in situ data for  V$_{SW}$(1 AU). The upstream solar wind velocities are listed in table \ref{table:upstream}. We use a drag coefficient of 2.5 \citep{2010PhDT.......182P}. Using these assumed values and estimations of the solar wind, we have all the parameters needed to calculate the forces on the CME. 
\begin{table}
\begin{center}
\caption{Ambient Corona}\label{table:upstream} 
\begin{tabular}{c c c c}\hline
CME	&	Upstream B	&	B$_o$	&	Upstream V	\\
	&	nT	&	G	&	km s$^{-1}$	\\ \hline
1	&	3.12	&	1.44	&	293	\\
2	&	2.40	&	1.11	&	219	\\
3	&	2.31	&	1.07	&	235	\\
4	&	3.00	&	1.39	&	375	\\
5	&	6.63	&	3.07	&	325	\\
6	&	4.01	&	1.86	&	369	\\
7	&	3.15	&	1.46	&	362	\\
8	&	3.69	&	1.70	&	335	\\
9	&	6.14	&	2.84	&	365	\\ \hline
\end{tabular}
\end{center}
\end{table}

\begin{figure}
\begin{center}
\includegraphics[height=.95\textwidth]{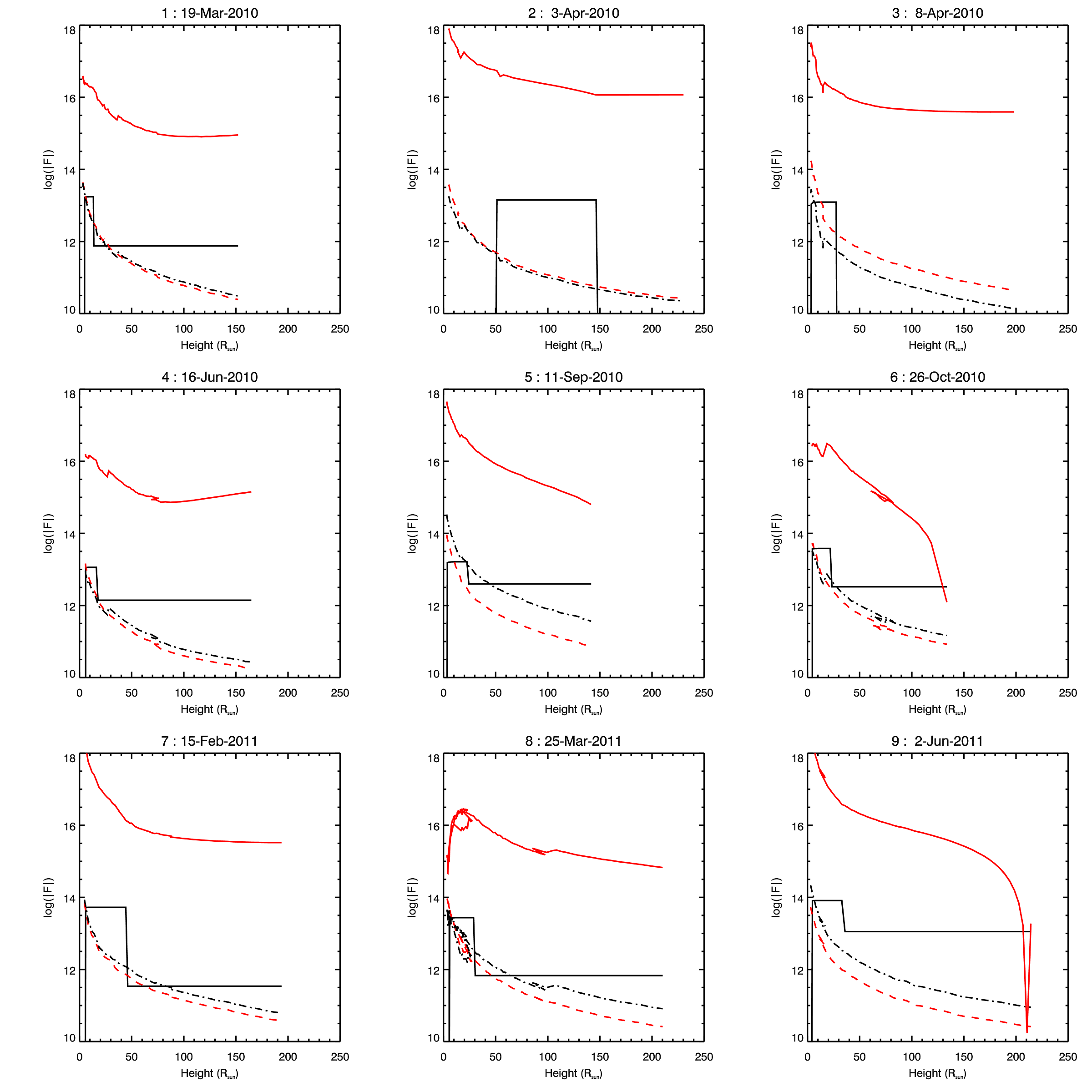}
\caption{The forces from the MSOE model F$_{net}$ mass times the acceleration of the CME (solid black), F$_m$ (dot-dash black), F$_d$ (solid red) and F$_g$ (dash red).}
\label{msoe}
\end{center}
\end{figure}

In figure \ref{msoe}, we plot the logarithm of the absolute value of the forces from the MSOE model. We have plotted F$_{net}$, mass times the acceleration of the CME, (solid black), F$_m$ (dot-dash black), F$_d$ (solid red) and F$_g$ (dash red) versus front height. The forces plotted in red oppose the propagation of the CME while F$_m$ propels the CME. It is clear from figure \ref{msoe} that the drag force completely dominates the other forces by several orders of magnitude. Given the values we have used to calculate the forces from the MSOE model, the CME would not propagate at these heights. For the forces to equal the net force, the ambient magnetic field would need to be 2 to 3 orders of magnitude larger and vary from equation \ref{Bsw} by an order of magnitude. Similarly, if we were to adjust the drag coefficient to balance the forces, it would need to be of the order of 0.001 to 0.01. Thus we must presume that there is a fundamental problem in the formulation of the MSOE model. We assume that the force imbalance is because there is no consideration in the model of the force internal to the CME. The forces internal to the CME could be driving the CME as well as the external force. We conclude that considering only the magnetic field external to CME in not sufficient to propagate the CME using the observed geometry and kinematics from 5 R$_\odot$ to 1 AU.

\subsection{Flux Rope Model}

The flux rope model has the internal magnetic field, gas pressure imbalance and hoop force driving the CME as well as the external magnetic field \citep{1989ApJ...338..453C}. Similar to the MSOE model, we can measure or estimate most of the model values. We will use the model to calculate the internal magnetic field and compare it to the values detected in situ. Using the equation for the force on the center of the CME
\begin{equation}\label{chen_2a}
F_R = M \frac{dv}{dt} = \frac{I_t^2}{c^2 R} \left[ {ln \left( \frac{8R}{a} \right)} +{\frac{1}{2}\beta_p} -{\frac{1}{2}\frac{B_t^2}{B_p^2}} +{2 \left(\frac{R}{a}\right) \frac{B_s}{B_p}} -1 + \frac{\xi_i}{2}    \right] + F_g + F_d,
\end{equation}
the force on of the minor radius
\begin{equation}\label{chen_2b}
F_a = M\frac{dw}{dt} = \frac{I^2_t}{c^2a} \left( \frac{B_t^2}{B^2_p} -1 + \beta_p  \right),
\end{equation}

equation \ref{chen_pressure} and the definition of the total toroidal current, we find that the internal poloidal magnetic field of the CME is
\begin{equation}\label{Bp}
B_p = \frac{-b \pm (b^2 - 4ac)^{\frac{1}{2}}}{2a}
\end{equation}
where
\begin{align}
a &= \ln \left ( \frac{8R}{a} \right ) - \frac{3}{2} + \frac{\xi_i}{2} \\
b &= 2 \left ( \frac{R}{a} \right ) B_s \\
c &=  8\pi(\bar{p}-p_a) + \frac{1}{2}aF_a - \frac{4R}{a^2}(F_R - F_g - F_d).
\end{align}
Once we have solved the poloidal magnetic field, we can solve for the internal toroidal magnetic field using the equation
\begin{equation}
B_t = \left [ a F_a - 8 \pi (\bar{p}-p_a) + B_p^2 \right ]^{0.5}.
\end{equation}

To use the geometric parameters from the GCS model with the flux rope model, we must reconcile the geometry of the two models. The GCS model creates a croissant-like CME where the minor radius varies along the axis. In the flux rope model the minor radius is constant along the axis. Also the major radius of the flux rope model is constant for all angles. For the GCS model the distance between the center of the model and the center of the cavity varies with angle. However, the apex of the two models have the same geometry.  Thus we use the apex radius and center height of the GCS model for the flux rope parameters, a and Z. We use the equation of \citet{1996JGR...10127499C} to define the major radius
\begin{equation}
 R = (Z^2 + s_o^2)/2Z
 \end{equation}
 where s$_o$ is the foot point separation from the GCS model. For all the CMEs the flux rope major radius is larger than the width of the GCS model. Thus using the geometry of the flux rope model, we are modeling a larger magnetic structure which does not represent the size of the density structure seen in the remote sensing data.  
 
To derive the other parameters in equation \ref{Bp}, we begin by solving for the internal pressure. Assuming the equation of state is
\begin{equation}
\frac{d}{dt} \left ( \frac{\bar{p}}{\bar{\rho}^\gamma} \right ) = 0 
\end{equation}
where $\bar{\rho}$ = $\bar{n}$m$_i$  is the average internal mass density and $\gamma$= 4/3 is the adiabatic index and m$_i$ is the mass of a proton.  Using the geometry of the flux rope to find the density, we can also solve for the internal pressure. To find the ambient pressure, we use the equation 
 \begin{equation}
p_a = 2 n_a k T_a
\end{equation}
 where n$_a$ density and T$_a$ temperature of the solar wind. For the ambient temperature, we use the equation 
\begin{equation}
T_a(Z) = T_o \left ( \frac{Z}{R} \right ) ^{-\alpha}
\end{equation}
where T$_o$ = 2 $\times$ 10$^6$ K and $\alpha$ = -0.85 \citep{1989ApJ...338..453C}. We use equation \ref{nr}  for the solar wind density. Thus we can solve for the average internal pressure and the ambient pressure. We use an internal inductance of $\xi_i$ = 1.2, which assumes a constant magnetic field distribution. We use the same external magnetic field  and drag force and gravitational force as for the MSOE model. Thus we have all the parameters needed to solve for the internal poloidal magnetic field.
 
\begin{figure}
\begin{center}
\includegraphics[height=.95\textwidth]{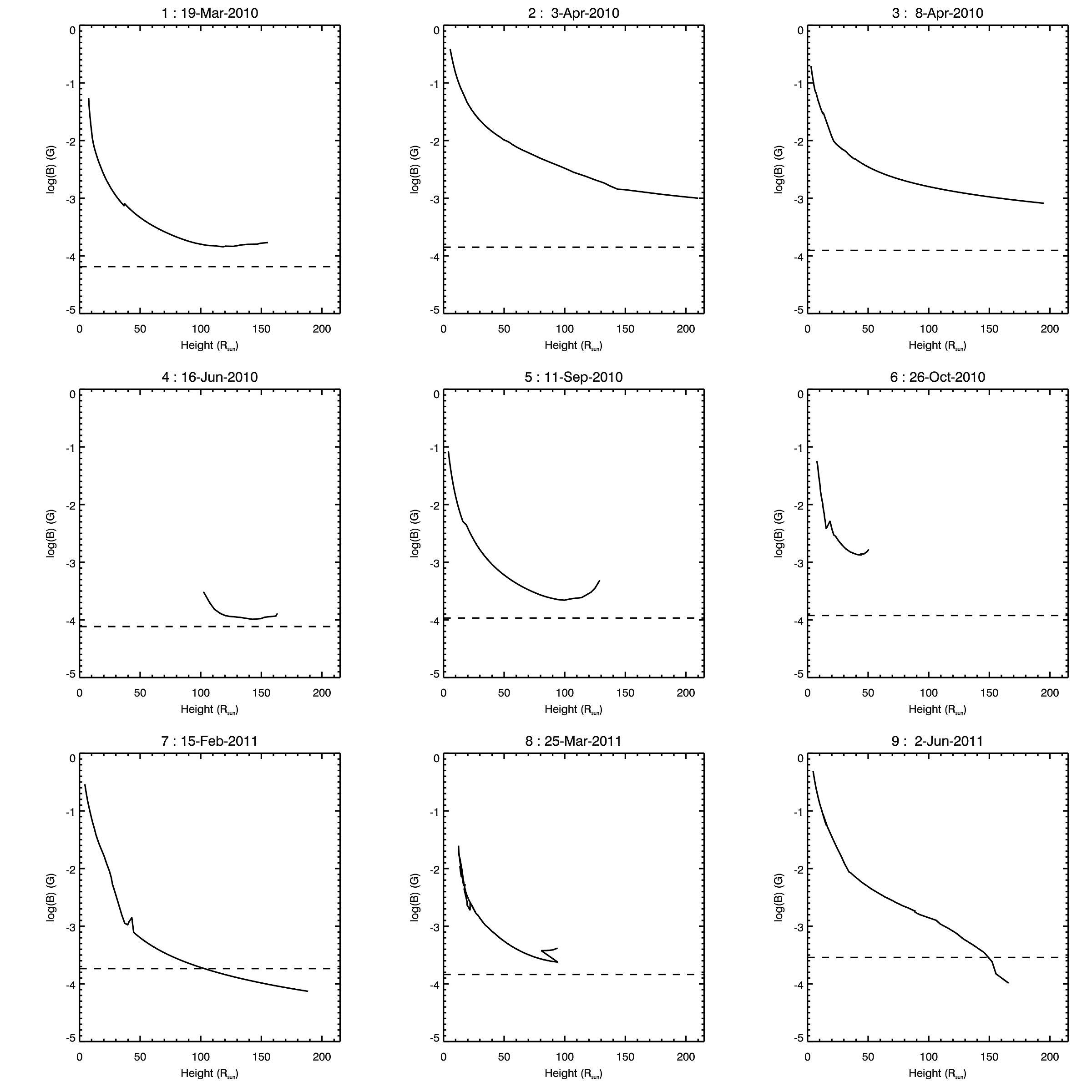}
\caption{The magnitude of the internal magnetic field from the flux rope model and the maximum magnetic field of the CME at earth (dashed).}
\label{flux_rope}
\end{center}
\end{figure}

In figure \ref{flux_rope}, we have plotted the magnitude of the internal magnetic field from the flux rope model. The dashed line at the bottom is at the value of the maximum magnetic field detected at Earth during the CME. For the flux rope model, there are regions where the solution for the poloidal magnetic field is imaginary (the square root term in equation \ref{Bp} is negative). The force balance of the model is very sensitive to the input parameters.

In table \ref{table:fluxrope_B}, we have listed the minimum internal magnetic field from the flux rope model, the height of the minimum field and the maximum magnetic field detected at Earth. The results from the flux rope model are very promising. It is possible that by adjusting the assumed parameter such as adiabatic index, drag force, and the solar wind parameters, we could find values where the flux rope magnetic field would match the measured magnetic field at Earth. \citet{2010ApJ...715L..80K} were able to reproduce the propagation and magnetic field for a single CME by giving the CME an initial energy profile and adjusting the other parameters of the model. \citet{2010PhDT.......182P} was able to reproduce the propagation and expansion speeds of four CMEs in the inner heliosphere by adjusting several parameters of the model. However,  \citet{2010PhDT.......182P} did not have magnetic field measurements for their CMEs to compare with the model results.

Adjusting the parameters of the model may only compensate for the difference in the geometry of the flux rope model and the GCS model. The flux rope geometry creates a larger CME then we find with the GCS model. Before we can draw any conclusion from the flux rope model, we must first investigate how the difference in the observed and flux rope geometries effect the outcome of the model. This investigation could be done by deriving the flux rope model for the GCS geometry. 
\begin{table}
\begin{center}
\caption{Flux Rope Internal Magnetic Field}\label{table:fluxrope_B} 
\begin{tabular}{c c c c}\hline
CME	&	H$_{fr}$	&	B$_{fr}$	&	B$_E$	\\
	&	R$_\odot$	&	nT	&	nT	\\ \hline
1	&	118	&	14	&	7	\\
2	&	210	&	100	&	14	\\
3	&	195	&	82	&	12	\\
4	&	142	&	10	&	8	\\
5	&	100	&	22	&	11	\\
6	&	44	&	133	&	12	\\
7	&	189	&	7	&	18	\\
8	&	94	&	24	&	15	\\
9	&	166	&	10	&	29	\\ \hline

\end{tabular}
\end{center}
\end{table}

\chapter{CME Structure from Remote Sensing and In Situ Modeling}\label{remoteinsitu}
\section{Modeling of In Situ Data}
\subsubsection{Self-Similar Magnetohydrodynamics Model}
 
We fit the in situ data using the non force free Elliptical Flux Rope (EFR) model described in section \ref{EFR}.  The EFR model is similar to the general solution of the self-similar MHD model in section \ref{intro:ssmhd}. The MHD  equations in section \ref{intro:ssmhd} were solved in spherical coordinates. The EFR model solves the MHD equation in elliptical cylindrical coordinates. The relation of elliptical cylindrical coordinates to cartesian is given by equation \ref{elliptic}. Solving the MHD equation in elliptical cylindrical coordinates, allows the model to find solutions from the in situ data where the cross-section of the flux rope is elliptical. 


The EFR model is fitted to the in situ data for the regions of the CME where the magnetic field is varying smoothly and the proton $\beta$ is low. The boundaries of the EFR fit regions are shown in the figures in Appendix \ref{appinsitu} with vertical dashed lines. Only three of the CMEs (4, 8 and 9) show enough rotation in one of the magnetic field directions to be classified as magnetic clouds (table \ref{table:insitu}). We expect the EFR model fit to these CMEs to be better than the other CMEs that show little or no rotation in the magnetic field. The in situ data used in the model was smoothed to four magnetic field measurements per hour.  

The model parameters that we investigate here are those that describe the size and orientation of the CME.  The orientation of the CME is defined by the three Euler angles ($\theta$, $\phi$, $\xi$) that relate the Y, X and Z axis, respectively, of the CME to Geocentric Solar Ecliptic (GSE) coordinates. For the remote sensing data, we have been using the Heliocentric Earth Ecliptic (HEE) coordinates. The GSE coordinates are centered at Earth and mirror the HEE coordinates which are centered at the Sun.  Also the vector, $\bf{Y_o}$ gives the impact parameter, the distance of the spacecraft crossing from the center of the flux rope. The semi-minor and semi-major axes define the cross-section of the CME. The results of the EFR model fit are listed in table \ref{table:efr}. The best results from the EFR model are for those CMEs that have been identified from the in situ data as magnetic clouds, CMEs 4, 8, and 9.  These results are highlighted with boldface text  in table \ref{table:efr}.

\begin{table}
\begin{center}
\caption{EFR Model Parameters}\label{table:efr} 
\begin{tabular}{*{7}{c}}\hline
CME	&	Lon $\phi$&	Lat $\theta$&	Rot $\xi$&	Major	&	Minor	&	Y$_o$	\\
	&	(deg)	&	(deg)	&	(deg)	&	(AU)	&	(AU)	&	(AU)	\\ \hline
1	&	206	&	-44	&	108	&	0.35	&	0.33	&	0.34	\\
2	&	119	&	-66	&	12	&	0.30	&	0.29	&	0.23	\\
3	&	358	&	-5	&	28	&	0.02	&	0.01	&	0.02	\\
\bf{4}&	\bf{41}&	\bf{-14}&	\bf{174}&	\bf{0.10}&	\bf{0.09}&	\bf{0.05}	\\
5	&	19	&	-31	&	16	&	0.36	&	0.08	&	-0.27	\\
6	&	179	&	-69	&	145	&	0.20	&	0.14	&	0.03	\\
7	&	63	&	-61	&	384	&	0.37	&	0.36	&	0.14	\\
\bf{8}&	\bf{1}&	\bf{7}&	\bf{89}&	\bf{0.04}&	\bf{0.01}&	\bf{0.01}	\\
\bf{9}&	\bf{273}&	\bf{-18}&	\bf{0}&	\bf{0.09}&	\bf{0.08}&	\bf{-0.08}	\\ \hline
\end{tabular}
\end{center}
\end{table}

\section{Comparison of the GCS Model with the EFR Model}

\begin{figure}
\begin{center}
\includegraphics[height=.95\textwidth]{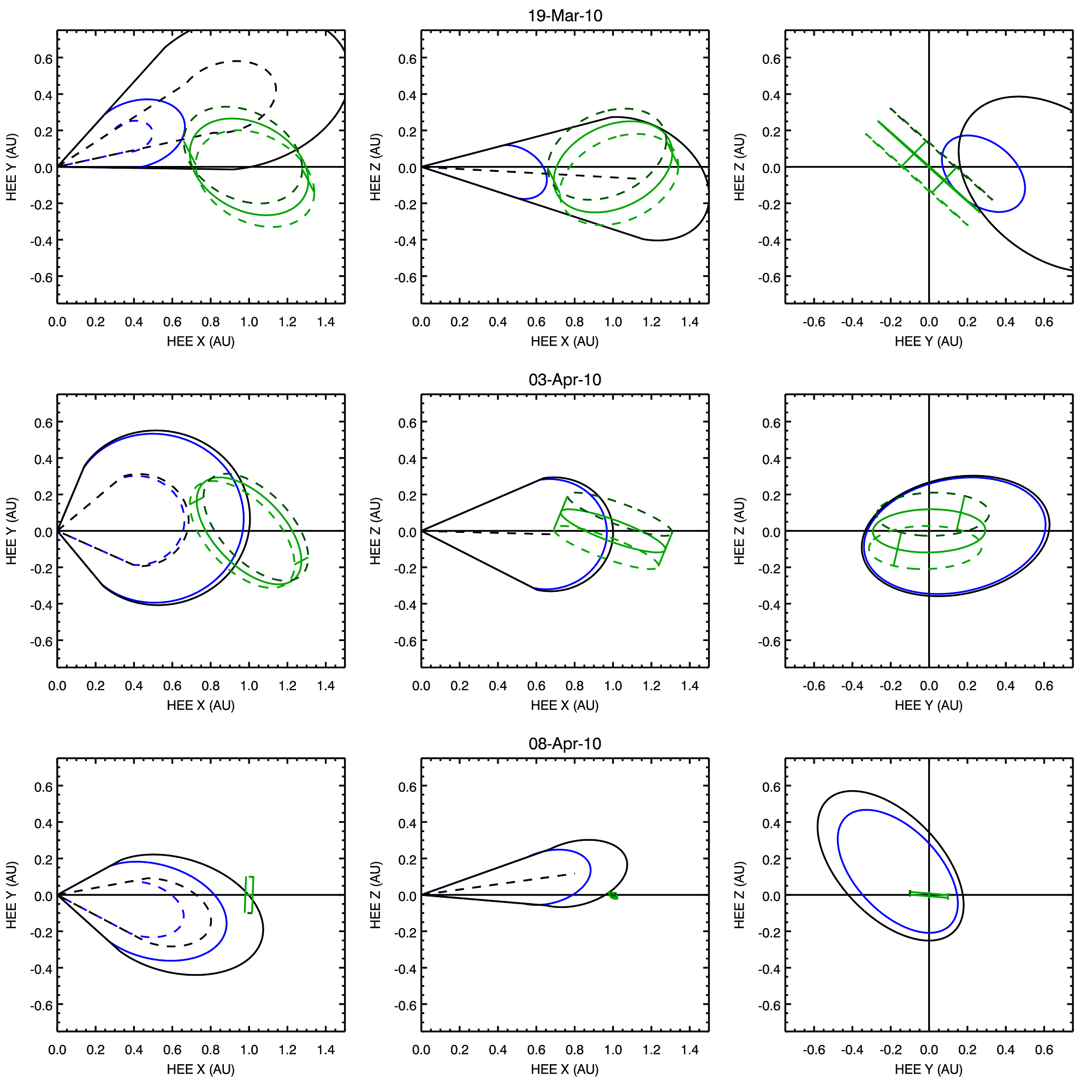}
\caption{Comparisons of the size and orientation of the CMEs 1, 2, and 3 from the GCS Model and the EFR Model. }
\label{raytrace_wind1}
\end{center}
\end{figure}

\begin{figure}
\begin{center}
\includegraphics[height=.95\textwidth]{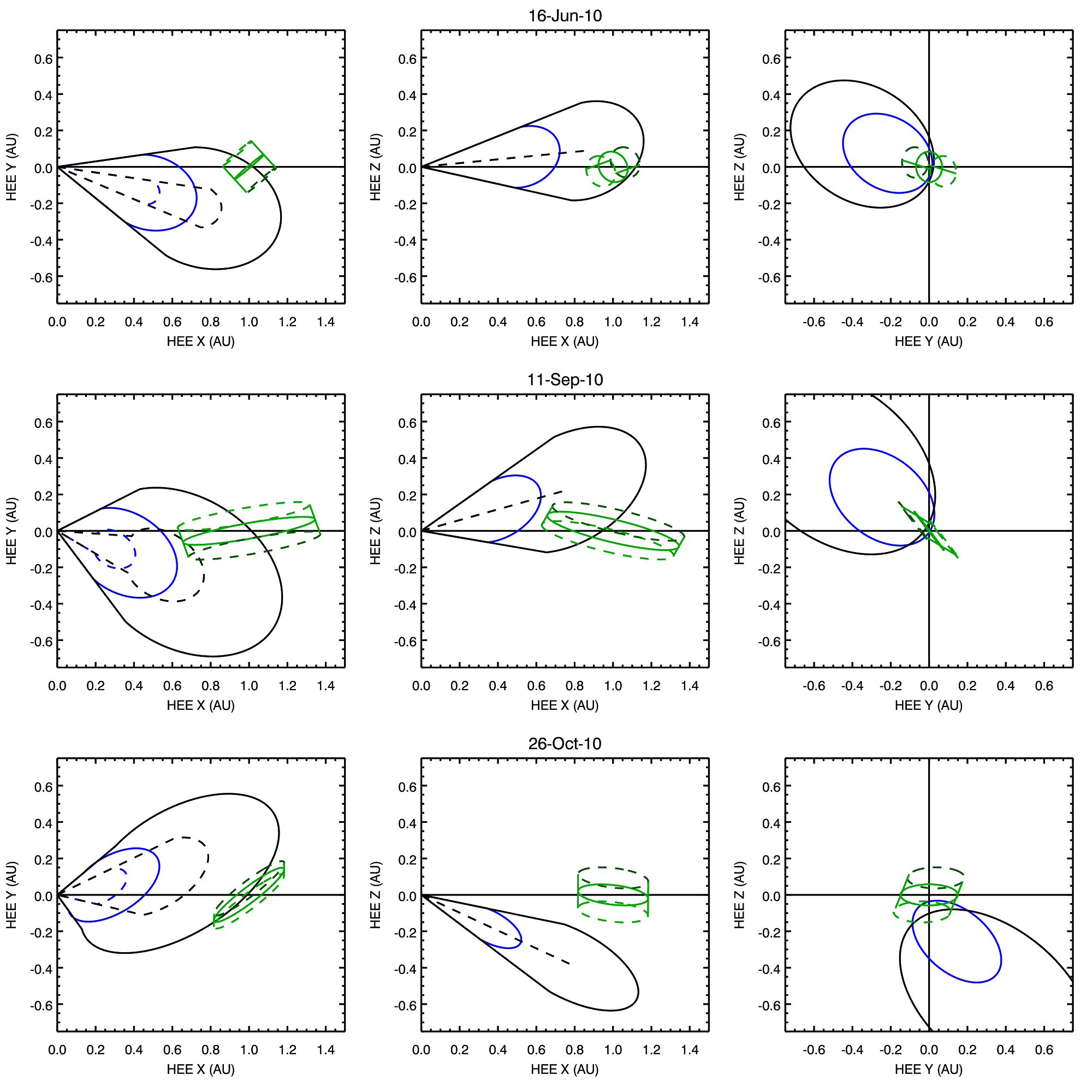}
\caption{Comparisons of the size and orientation of the CMEs 4, 5, and 6 from the GCS Model and the EFR Model.}
\label{raytrace_wind2}
\end{center}
\end{figure}

\begin{figure}
\begin{center}
\includegraphics[height=.95\textwidth]{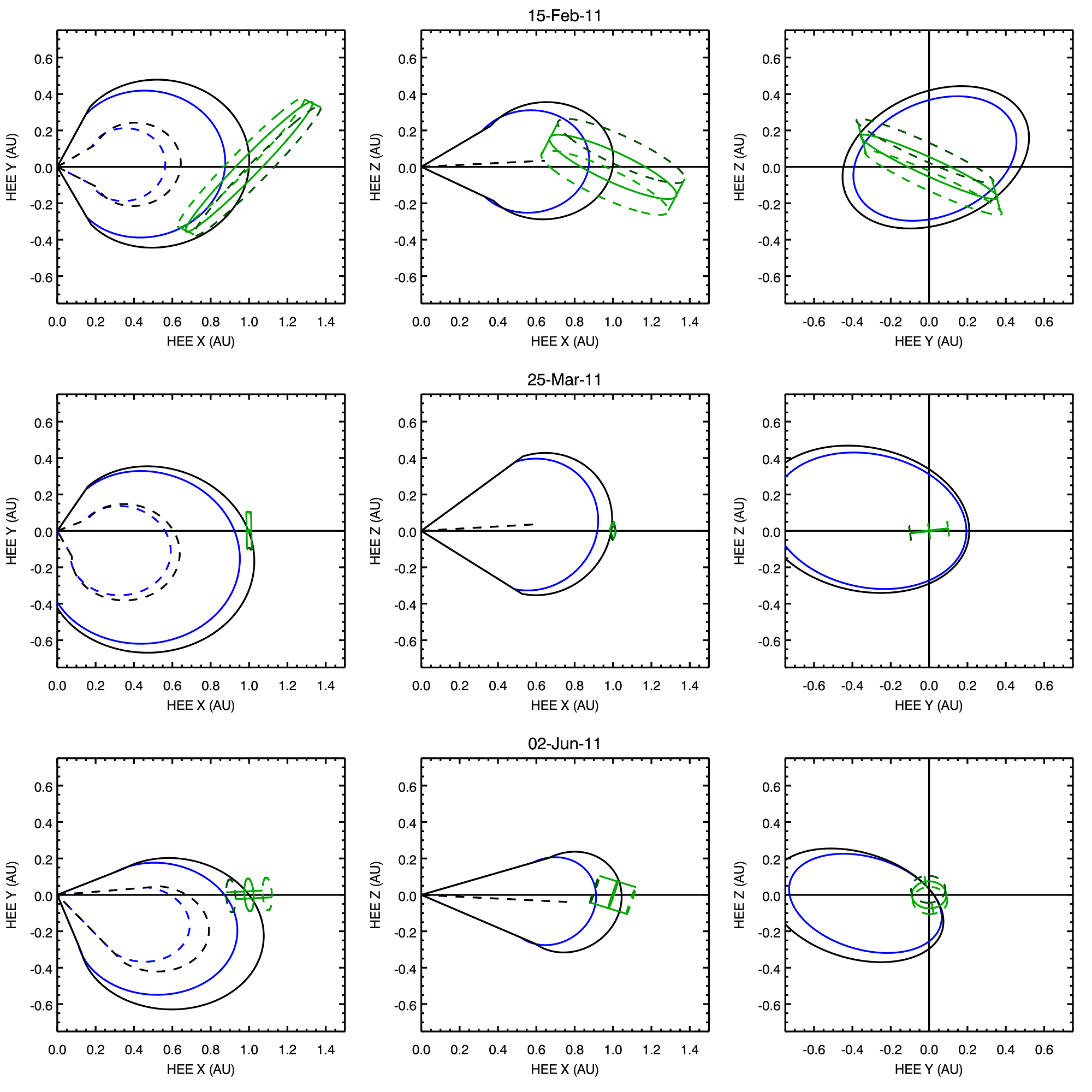}
\caption{Comparisons of the size and orientation of the CMEs 7, 8, and 9 from the GCS Model and these EFR Model.}
\label{raytrace_wind3}
\end{center}
\end{figure}

Given the different coordinates systems and geometry of the GCS and EFR models it is not straight forward to compare the size and orientation results. In figures, \ref{raytrace_wind1}, \ref{raytrace_wind2} and \ref{raytrace_wind3}, we have plotted representations of the two models in three planes. The left column is the elliptic plane (top view) where the Sun is at (0,0) and Earth is at (1,0). The center column is the Sun-Earth line plane (side view) where again the Sun is at (0,0) and Earth is at (1,0). The right column is the Earth observer view (head-on view) where the Sun and Earth are at (0,0).  

In each view, we have plotted the projection of the elliptical cylinder of the EFR model in green. The solid ellipse crosses Earth. The dashed ellipses are $\pm$ 0.1 AU from the center ellipse. The darker ellipse is the same in all three views to give the correct sense of depth. 

For the GCS model, the blue outline is the maximum measured height. The black outline is the model projected to the height when the front would cross Earth. The dashed line is the axis at the center of the cavity. In the left, we have plotted the face-on outline of the GCS model, figure \ref{Fig:GCS}. In the center column, we have plotted the edge-on outline of the model. In the right column, we have plotted a top-view profile where the major axis of the ellipse is the GCS width and the semi-major axes it the GCS radius. In table \ref{table:gcs}, we list the GCS parameters of the model projected to Earth crossing.
\begin{table}
\begin{center}
\caption{GCS Model Parameters}\label{table:gcs} 
\begin{tabular}{*{7}{c}}\hline
CME	&	H	&	Lon	&	Lat	&	Rot	&	Radius	&	Width	\\
	&	(R$_\odot$)	&	(deg)	&	(deg)	&	(deg)	&	(AU)	&	(AU)	\\ \hline
1	&	1.66	&	26	&	-3	&	-35	&	0.41	&	0.60	\\
2	&	1.01	&	8	&	-2	&	12	&	0.32	&	0.50	\\
3	&	1.10	&	-10	&	8	&	-50	&	0.28	&	0.49	\\
4	&	1.22	&	-17	&	6	&	-33	&	0.32	&	0.41	\\
5	&	1.31	&	-21	&	17	&	-43	&	0.46	&	0.59	\\
6	&	1.33	&	13	&	-29	&	-55	&	0.42	&	0.62	\\
7	&	1.00	&	2	&	3	&	26	&	0.36	&	0.51	\\
8	&	1.06	&	-20	&	3	&	-11	&	0.40	&	0.59	\\
9	&	1.12	&	-21	&	-3	&	-16	&	0.29	&	0.50	\\ \hline
\end{tabular}
\end{center}
\end{table}

When comparing the sizes of the CMEs, we expect the results of the EFR model to be smaller then the GCS model. From the in situ data in Appendix \ref{appinsitu}, we can see that the density structures have a larger transit time than the magnetic structures. Assuming a uniform distribution of the density around the magnetic structure and that both structures are traveling at the same speed, we would expect the ratio of the transit times to be the same as the ratio of the sizes of the two structures. 

CME 1 : According to the GCS, fit the CME only hits Earth with a glancing blow. From the head-on view (right column) the tilt of the CME is different by only 9$^o$ comparing the two models. Also the radius of the CMEs from both models are different by 0.08 AU. However, the impact parameter from the EFR model does not match the GCS model. CMEs with large impart parameters are the most difficult to fit with the EFR model since the in situ data shows very little rotation in the magnetic field  to constrain the fit.

CME 2 :  The size and impact parameter of the CME is similar from both models. The difference in the radius from the two model is 0.02 AU. However, the orientation from the two models is very different. From the GCS data, the CME has an East-West orientation while the EFR model result has a North-South orientation. \citet{2011ApJ...729...70W} found the same difference in orientation for this event between the Sun and Earth. Given these two independent results it is possible that the CME has rotated between 25 R$_\odot$ and 1AU.

CME 3 : There is no similarity between the results of the two models for this CME. For this CME, the size of the CME between the two models differs by an order of magnitude.  The tilt of the two models as seen in the head-on view (right column) is different by 45$^o$.

CME 4 : The results of the two model are most similar for this CME. The in situ detection of this CME is classified as a magnetic cloud. The magnetic structure from the EFR model is smaller then the density structure from the GCS model as we would expect. Also the orientation of the EFR model in nearly parallel to the front of the CME in the ecliptic view (left column). The tilt of the CME is different by 19$^o$. 

CME 5 : From the EFR model the cross section of this CME is highly elliptical. From both models, the cross section of the CME is large. However, from the in situ data for this CME the density structure is five times larger then the magnetic structure.  All other parameters are similar between the two models for this CME  with the tilt angle which differs by 12$^o$. 
 
CME 6 : This CME is an another glancing blow at Earth. The orientation of the two model fits are at very different as well as the impact parameter. The EFR structure is orientated North-South while the remote sensing structure is at 55$^o$ from the axis.

CME 7 :  This is the only CME where the radius from the EFR model is larger then the GCS fit. We would not expect the magnetic structure to be larger then then density structure from the in situ data. Also the EFR model intersects Earth at a large angle in the first panel while the GCS model has the CME intersecting Earth almost straight on.

CME 8 :   Again for this CME, the EFR radius is an order of magnitude smaller then the GCS model while density structure in situ is only 1.4 larger then the magnetic structure. Despite the difference in size, the orientation of the CME from both models is similar. The difference in the tilt is 3$^o$. This CME was identified in the in situ data as a magnetic cloud.

CME 9 :   This CME was also identified as a magnetic cloud. The orientation of the two models is different. However, for this CME, we might be detecting the leg of the CME with the EFR model. If we are seeing the leg of the CME then orientation of the EFR model pointing back to the Sun is what we would expect. The orientation of the GCS model agrees with this interpretation of the two model results.

Prior to this study, the comparison of the remote sensing density structure to the magnetic structure from the in situ data has not been attempted. Given the different data sets and the difference in geometry and coordinate systems of the two models, it is difficult to quantify this comparison. The radii from the EFR model vary between 0.01-0.37 AU while the GCS radii only vary between 0.28-0.42 AU. With only the results from these two models, we cannot know if the magnetic structure of CMEs varies more than the density structure or if the results from either model are incorrect. However, looking at the ratio of the transit times of the two structures and the ratio of the radii from the two models, we suspect that the models are incorrect.

For the CMEs that are identified as magnetic clouds (4, 8 and 9 ), the radii from the EFR model are less then 0.1 AU. The EFR model is best fitted for these CMEs. Considering only these CMEs, we would conclude that the density structures seen in remote sensing data are 3-10 times larger then the internal magnetic structure. However, the transit times of the spacecraft through the structures does not support this conclusion. The transit times of the density structure are only 1.5-5.7 times larger then the magnetic structure. 

The disagreement between these two models has large implications to space weather since the orientation of the CME's magnetic field is the other major factor in geo-effectiveness. Until the models from these two data sets can be reconciled there in very little progress that can be made.  Since the in situ data is our only source of direct information of the magnetic field and remote sensing data the only means of prediction, it is essential that the analysis of these two data sets is reconciled.

\chapter{Summary and Discussion}\label{summaryanddiscussion}

In this study, we analyzed nine CMEs in both remote sensing and in situ data from Sun to Earth. By using the remote sensing and in situ data sets, we were able to test many of the parameters derived from the models applied to the remote sensing and in situ data. From the models of the remote sensing data, we tested predicted values such as arrival time, impact velocity and magnetic field at Earth. We also compared the size of the CME derived from the in situ data with the density structures seen in remote sensing. By using both sets of data for each CME, we can more comprehensively evaluate our analysis of the CMEs. We found that there are many avenues for future work based on the results of this study.  

The nine CMEs we selected for this study  occurred between January 2010 and June 2011. Of the more than 2,000 CMEs recorded in the SOHO LASCO CME Catalog during this time, these were the only nine that met our selection criteria of being Earth impacting CMEs that were well observed in the remote sensing data. During the selection period, solar activity was increasing from a remarkably low period, the nadir of which was August 2008. The low level of solar activity is reflected in the number of CMEs we were able to track from Sun to Earth in the remote sensing data and the properties of these events. From the SOHO LASCO CME Catalog, the projected speeds of our CMEs range between 45-669 km s$^{-1}$ with a standard deviation of 217 km s$^{-1}$. The average speed of these events is lower then the average speed of halo and partial halo event as reported in \citep{2004JGRA..10907105Y}. We did find a greater range of speeds when we calculated the true three-dimensional velocities of the CMEs. The angular widths of the CMEs are  between 41 and 78 degrees when viewed as limb events. Our selection criteria limited the maximum size of the CMEs as viewed from STEREO. By limiting the size of the selected CMEs, we may have limited the velocities of the CMEs as well since fast CMEs are often wide \citep{2004JGRA..10907105Y}. Only three of the CMEs where identified as having magnetic cloud structures detected in situ. Overall, the CMEs in this study are unremarkable, average CMEs, with one exception. CME 2 on 03-April-2010, caused the first significant geometric storm of the solar cycle 24.
 
We began our analysis in Chapter \ref{analysisremotesensing} by fitting the remote sensing data with two  geometric models of the CME structure; the GCS and Ellipse model. The GCS model allowed us to estimate the three-dimensional position and size of the CME using the three available views of the CME. However, the GCS model assumes a circular cross-section for the CME. From the observations, CMEs in the inner heliosphere appear to have elliptical cross-sections. To study the cross-section of the CMEs in more detail we used the Ellipse model. The Ellipse model is very simple and does not account for projection effects. But this simple approach to the data allows us to overcome the limitation of a circular cross section of the GCS model. From the Ellipse model, we found the center height and the radius of the CME parallel and perpendicular to the propagation. 

In section 3.3, we compared the results of the two models. We found that the sum of the center height and parallel radius from the Ellipse model is equal to the front height of the GCS model as expected. We also found that the average of the parallel and perpendicular radii is similar to the radius of the GCS model. Thus the cross-sectional area of the CMEs is similar in both models. This is a helpful result for this study because we found it difficult to fit the Ellipse model at heights greater then $\sim$50 R$_\odot$. But we were able to fit the the GCS model up to a maximum height of 211 R$_\odot$. To study the kinematics and forces on the CME, we wish to use the model parameters at the greatest measured heights. Ultimately, our goal is to use the parameters from the remote sensing data in the theoretical models of  CME propagation and expansion. All of these models assume the CME has a circular cross-section. 


We used the height and time measurements from the GCS and Ellipse model in section \ref{empirical}, to derive the kinematics of the CME.  From the GCS model, we found that the kinematics of the CME front could not be well described by a single polynomial equation. Thus we used multiple polynomial functions to fit the data. Based on our fit, the CME has different acceleration values during different times in its propagation. All but two of the CMEs had an acceleration in the final phase of the its propagation. All the acceleration in the final phase are small ($<$ 3 m s$^{-2}$) but persisted over several hours. We know from the in situ data that five of the CMEs do not have the same velocity as the solar wind when they cross Earth. Thus these CMEs are not being carried in the solar wind like a leaf in a stream. Also the range of CME velocities converge from the Sun to Earth. The range of initial CME velocities is 47-1270 km s$^{-1}$ with a standard deviation of 426  km s$^{-1}$ while the range of final velocities at 1 AU is 321-615 km s$^{-1}$ with a standard deviation of 99 km s$^{-1}$.  Thus CMEs must decelerate or accelerate between the Sun and 1 AU \citep{2006SSRv..124..145G}. There does not appear to be a fixed height where the CME reaches a constant velocity. 

An acceleration requires an imbalance of the net forces on the CME. The changes in the acceleration of the CME suggests that the net force on the CME is not constant. There may be regions where the dominance of individual forces change. It is this force imbalance we wish to investigate further with the theoretical models.

In section \ref{earth},we used the derived kinematics of the CME front to predict the arrival time and velocity of the CME at Earth. We compared our predicted values with those detected in situ.  The arrival times have a maximum error of $\pm$20\% of the total transit time and the speeds have an error between -131 to 190 km s$^{-1}$ with a standard deviation of 99 km s$^{-1}$. These arrival times are not better then only using LASCO data \citep{2001JGR...10629207G}. We found that the error in the arrival time is correlated with the error in the final velocity. This result led us to test our use of a constant velocity to project the CME to 1AU after the final measurement. We did not find that this method was causing the error in the arrival times. Delay in the CME arrival time versus the predicted arrival could be caused by the shape of the CME \citep{2012arXiv1202.1299M}. However, our predicted arrival times are both earlier and later then the actual arrival time. An error in the arrival time can also be caused by a miscalculation of the CME longitude. In the the HI2 field of view a 1$^o$ change in the longitude changes the height by $\sim$2 R$_\odot$ for the same position in the image. We attribute the overall error in the arrival times to both the estimate of the CME longitude from the GCS model and the fit of the kinematic profile to the height and time data. 

To explore the expansion of the CME, we used the results of the Ellipse model. From \citet{2011ApJS..194...33T}, we know that the GCS model is proportional to the center height by the factor $\kappa$ which is defined by the user. We fit the parallel and perpendicular radii with a linear function. We found that the parallel and perpendicular speeds are related by equation \ref{eq:para_perp}. Thus the parallel and perpendicular expansions are not independent. This result would preclude that the perpendicular expansion is being driven by the solar wind while the parallel expansion is being driven by internal forces \citep{2004ApJ...600.1035R}. We also found that the expansion speeds are related to the acceleration of the center height of the CME. The acceleration of the center height is related to the net force on the CME. Thus both expansion speeds are related to the net force. However, since the expansion is at a constant velocity, the net force is not driving the expansion.

In section \ref{theory}, we used the kinematics derived from the remote sensing data to calculate the forces on the CME as predicted by the theoretical models introduced in section \ref{intro:theory}. The kinematic melon-seed model predicts that the CME will expand self-similarly for the magnetically dominated case. We found that most CMEs tend to expand self-similarly. There are some deviations from self-similar expansion. However, we cannot determine if these deviations are due to observational error or are physical changes in the CMEs. This study is focused on Earth impacting CMEs and was not designed to explore if CMEs are expanding self-similarly. To properly study self-similar expansion a much larger data set would be needed and the CMEs should be measured lower in the corona. From the melon-seed model, we can conclude that the interior of the CME is dominated by the magnetic field and not gas pressure at heights between 5 and 50 R$_\odot$.

From the kinematic MSOE model, we calculated all the forces on the CME as prescribed by the model. We estimated the solar wind velocity and magnetic field using the values measured in situ upstream of each CME. We found that with the MSOE model, the drag force is several orders of magnitude larger then the other forces. The net force on the CME calculated from the measured mass and acceleration cannot be obtained unless the magnetic field is  2 to 3 orders of magnitude larger then what we estimated. Similarly, the drag coefficient would need to be of the order of 0.001 to 0.01 for the drag force to balance the other forces. From comparing our results with those presented in \citet{2006SoPh..239..293S}, they did not use a drag force for most of their examples. Also \citet{2006SoPh..239..293S} has a maximum radius 0.15 AU for their modeled CMEs at 1 AU where the minimum radius we found in the remote sensing data is 0.28 AU. We conclude that there is a fundamental problem in the formulation of the MSOE model. We assume that the force imbalance is because there is no consideration in the model of the force internal to the CME. 

We then investigated the flux rope model. With the flux rope model, we provide the model with all the input values with the exception of the internal magnetic field. We then allow the model to solve for this value. From the in situ data we have a measurement of the magnetic field to which we can compare the model result. From the flux rope model we got values of the internal magnetic field that are close to those measured in situ (table \ref{table:fluxrope_B}). We are encouraged by the results of the flux rope model and see the potential for future work. 

To continue using the flux rope model with the results of the GCS model we would first need to reconcile the geometry of these two models. If we use the dimensions of the GCS model apex to generate the flux rope model geometry, we got a structure that is larger then what we observe in the remote sensing data. In the future, we would like to derive the equations of the flux rope model for the geometry of the GCS model. We believe that with the correct geometry  we would be able to predict the magnitude of the internal magnetic field of CMEs from the remote sensing data. 

In Chapter \ref{remoteinsitu}, we used the EFR model described in section \ref{EFR} to derive a size and orientation of the CMEs from the in situ data. We compare the results from the in situ EFR model with the results of the GCS model projected to 1 AU.  We found that the results of the two models do not agree for most events.  The EFR model best fits those CMEs that are identified as magnetic clouds. For these CMEs, magnetic structures from the EFR models are much smaller then the density structure in the remote sensing data. When comparing the sizes of the CMEs we expect the results of the EFR model to be smaller then the GCS model. However, we are found that the magnetic structures can be an order of magnitude smaller then the density structure. For the CMEs where the sizes from the EFR and GCS models are similar, the orientations do not match. The differences between the results of these two models is very important. Until the models from these two data sets can be reconciled there in very little progress that can be made.   
 
 In conclusion, we have several new results on the kinematics and forces on CMEs as they propagate between the Sun and Earth. We found that based on our fit of the height and time data, CMEs have different acceleration values during different times in their propagation. Also, there is not a fixed height where the CME reaches a constant velocity. From the kinematic results, we predicted the arrival time of the CME at Earth and compared this arrival time to the detection of the CME in situ. The predicted arrival times have a maximum error of $\pm$20\% of the total transit time and the speeds have an error between -131 to 190 km s$^{-1}$ with a standard deviation of 99 km s$^{-1}$. These arrival times are not better then only using LASCO data \citep{2001JGR...10629207G}. Even with remote sensing observations of the inner heliosphere there is still a large uncertainty in the trajectory of the CME.  We also compared the structure of the CME as predicted from the remote sensing and in situ data at Earth. We compare the size and orientation results from the in situ EFR model with the results of the GCS model projected to 1 AU.  We found that the results of the two models do not agree for most events. We also found the kinematics of the CME expansion parallel and perpendicular to the propagation of the CME. We found that the parallel and perpendicular expansions are not independent. We also found that the expansion speeds are related to the acceleration of the center height of the CME.
 
We then used the derived kinematics of the CME with three theoretical models of CME propagation and expansion. From the melon-seed model, we found that most CMEs tend to expand self-similarly. Also, we can conclude that the interior of the CME is dominated by the magnetic field and not gas pressure at heights between 5 and 50 R$_\odot$. With the MSOE model, we found that the drag force is several orders of magnitude larger then the other external forces. The net force on the CME calculated from the measured mass and acceleration cannot be obtained with this model unless the magnetic field is  2 to 3 orders of magnitude larger then what we estimated or the drag coefficient would need to be of the order of 0.001 to 0.01. Our most promising results are from using the kinematics and dimensions of the CME with the flux rope model. From the flux rope model, we calculated values of the internal magnetic field that are close to those measured in situ (table \ref{table:fluxrope_B}). We believe that with the correct geometry we would be able to predict the magnitude of the internal magnetic field of CMEs from the remote sensing data.

\appendix
\appendixeqnumbering
 \appchapter[Remote Sensing Data]{Remote Sensing Data}\label{appremotesensing}
In this Appendix, we give a sample of the remote sensing data used in the study. For each of the nine CMEs in the of the study, we have included the remote sensing data for three heights during the CME propagation. In the caption for each figure we give the CME number, the time of the remote sensing observation and the height of the GCS model fit.  All of the figures in the appendix have the same layout. The left panel is data from STEREO A. The center panel is data from LASCO. The right panel is data from STEREO B. The data in the top and bottom panel are the same. The images in the bottom panel have been over plotted with the GCS model. The GCS model is represented by a grid of points on the surface of the model. The first figure for each CME has data from STEREO COR2 and LASCO C2. The next two figures have data from STEREO HI1 and LASCO C3.

{\bf NOTE : This appendix has been reduced to one figure per event to reduce the file size.}

\begin{landscape}
\begin{figure}
\begin{center}
\includegraphics[width=20cm]{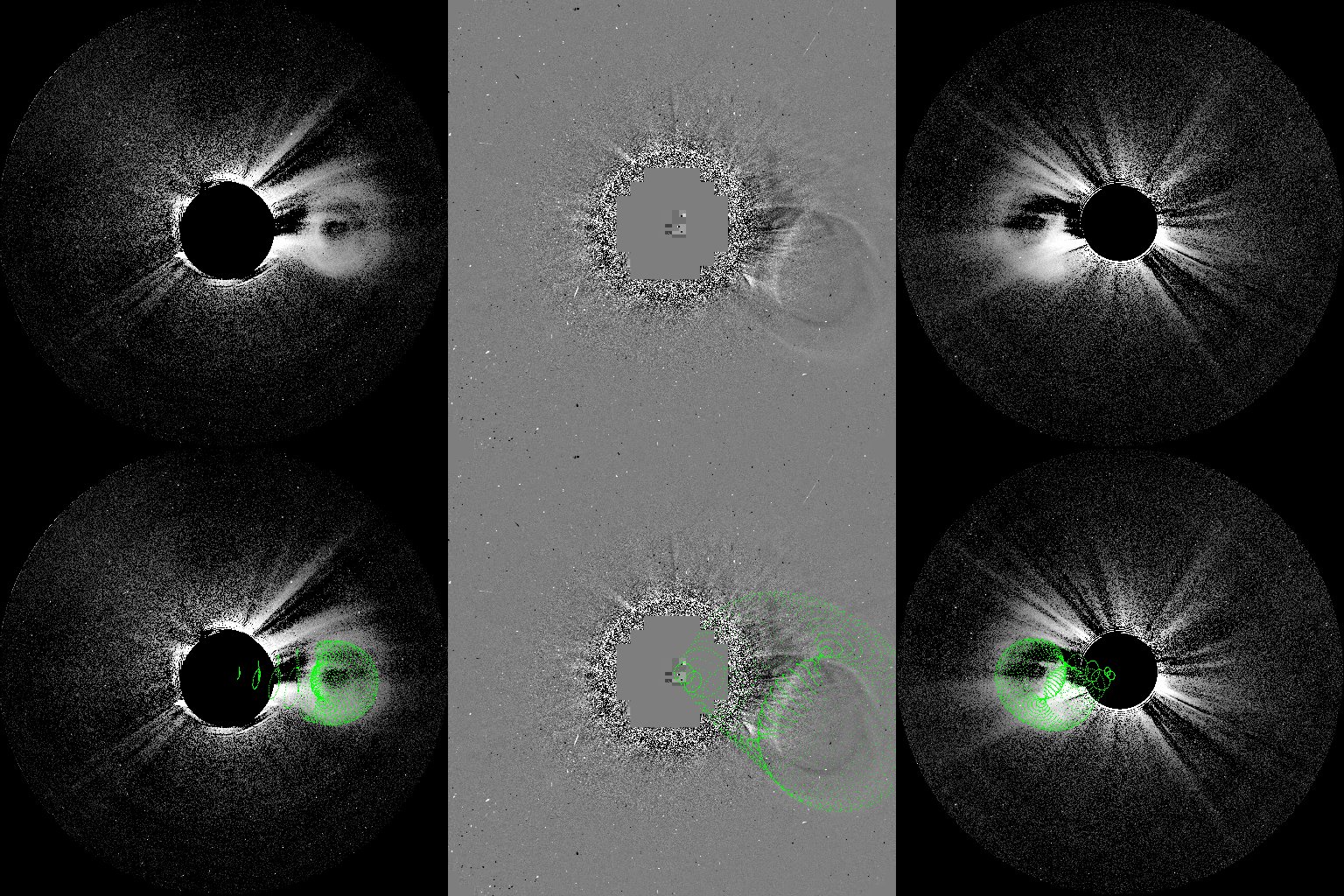}
\caption{ CME 1 19-Mar-2010 18:39 UT height = 10}
\label{2010-03-19_1}
\end{center}
\end{figure} \end{landscape} 



\begin{landscape} \begin{figure}
\begin{center}
\includegraphics[width=20cm]{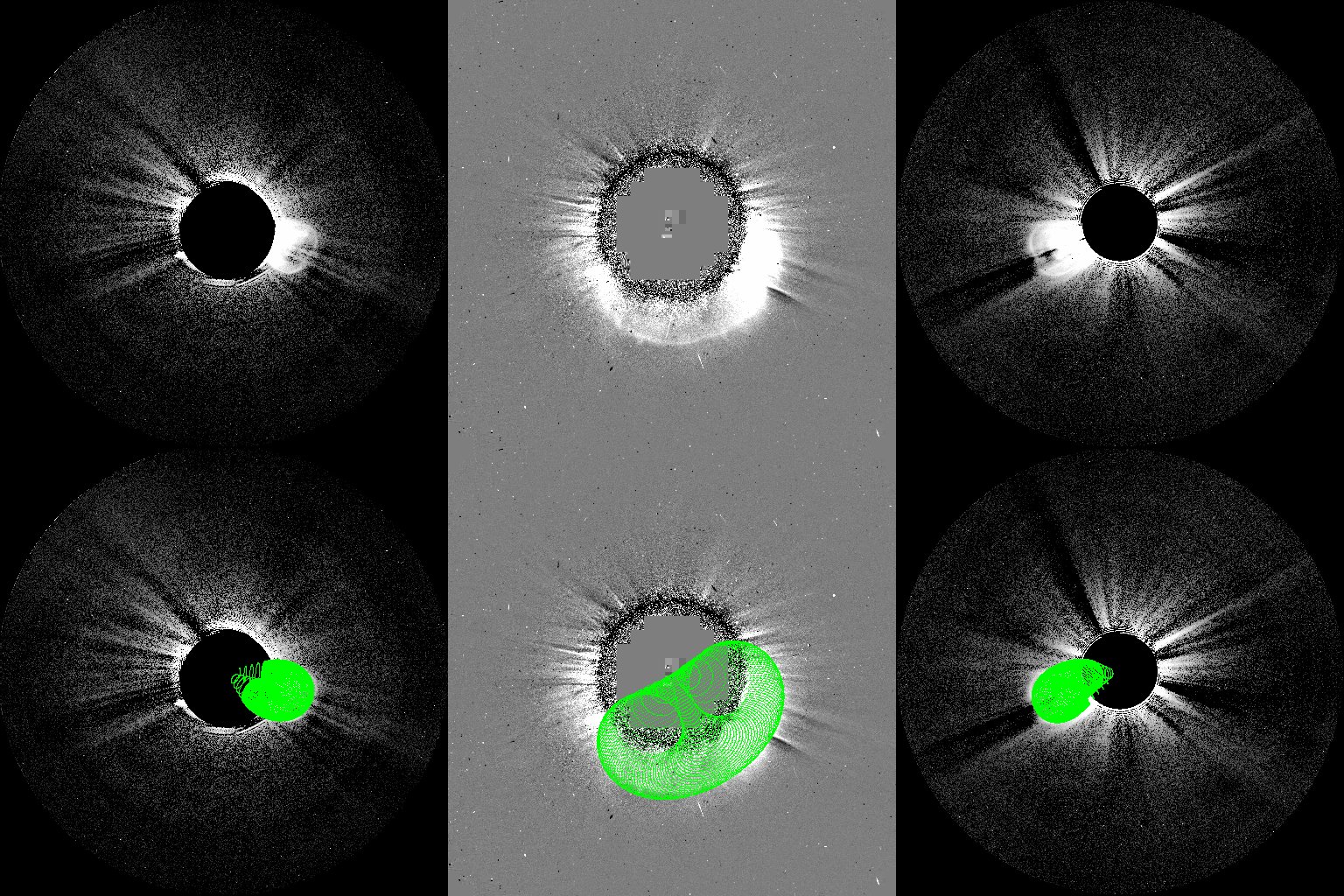}
\caption{CME 2 03-Apr-2010 10:39 UT height = 6.5}
\label{2010-04-03_1}
\end{center}
\end{figure} \end{landscape} 



\begin{landscape} \begin{figure}
\begin{center}
\includegraphics[width=20cm]{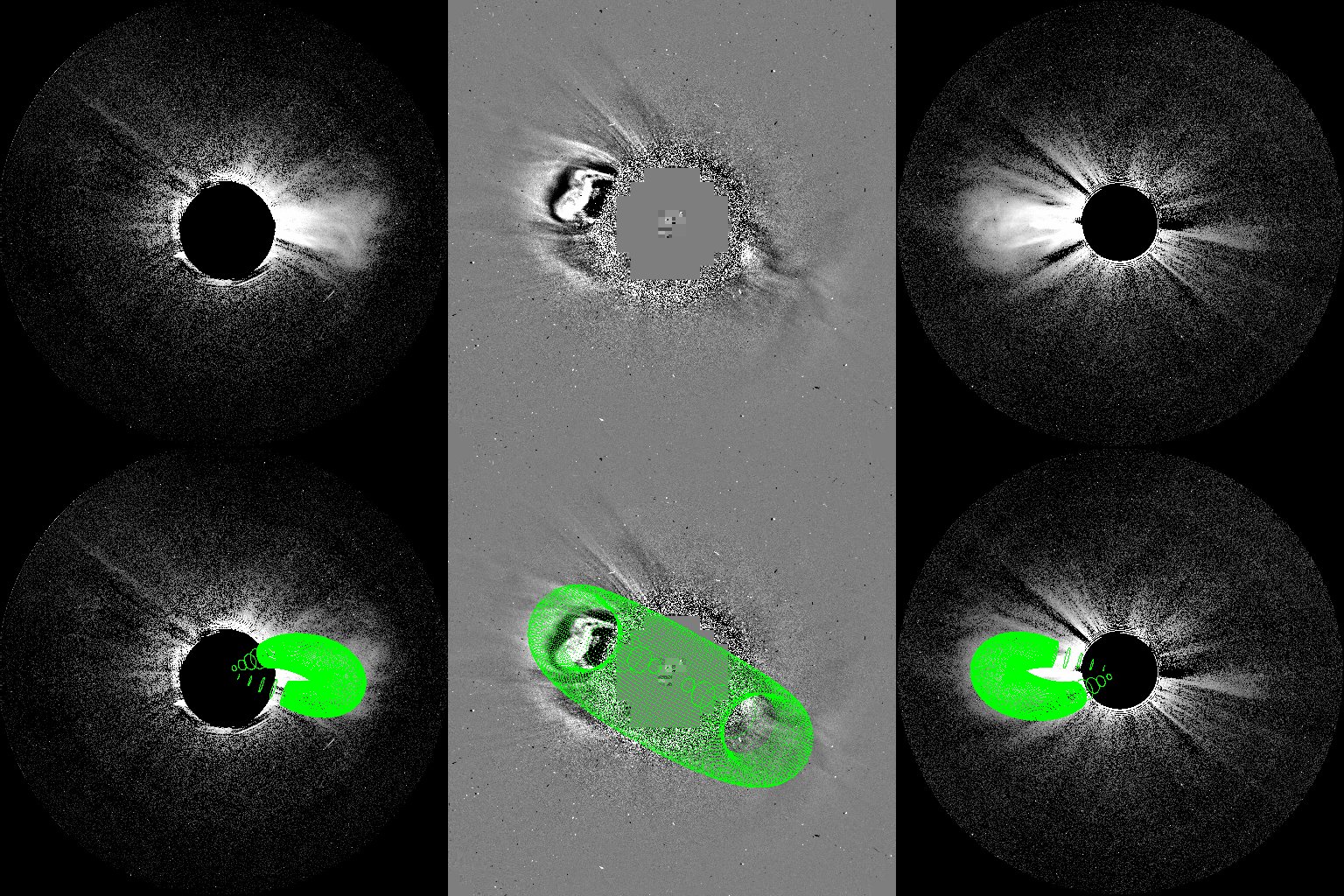}
\caption{CME 3 08-Apr-2010 06:54 UT height = 10}
\label{2010-04-08_1}
\end{center}
\end{figure} \end{landscape} 



\begin{landscape} \begin{figure}
\begin{center}
\includegraphics[width=20cm]{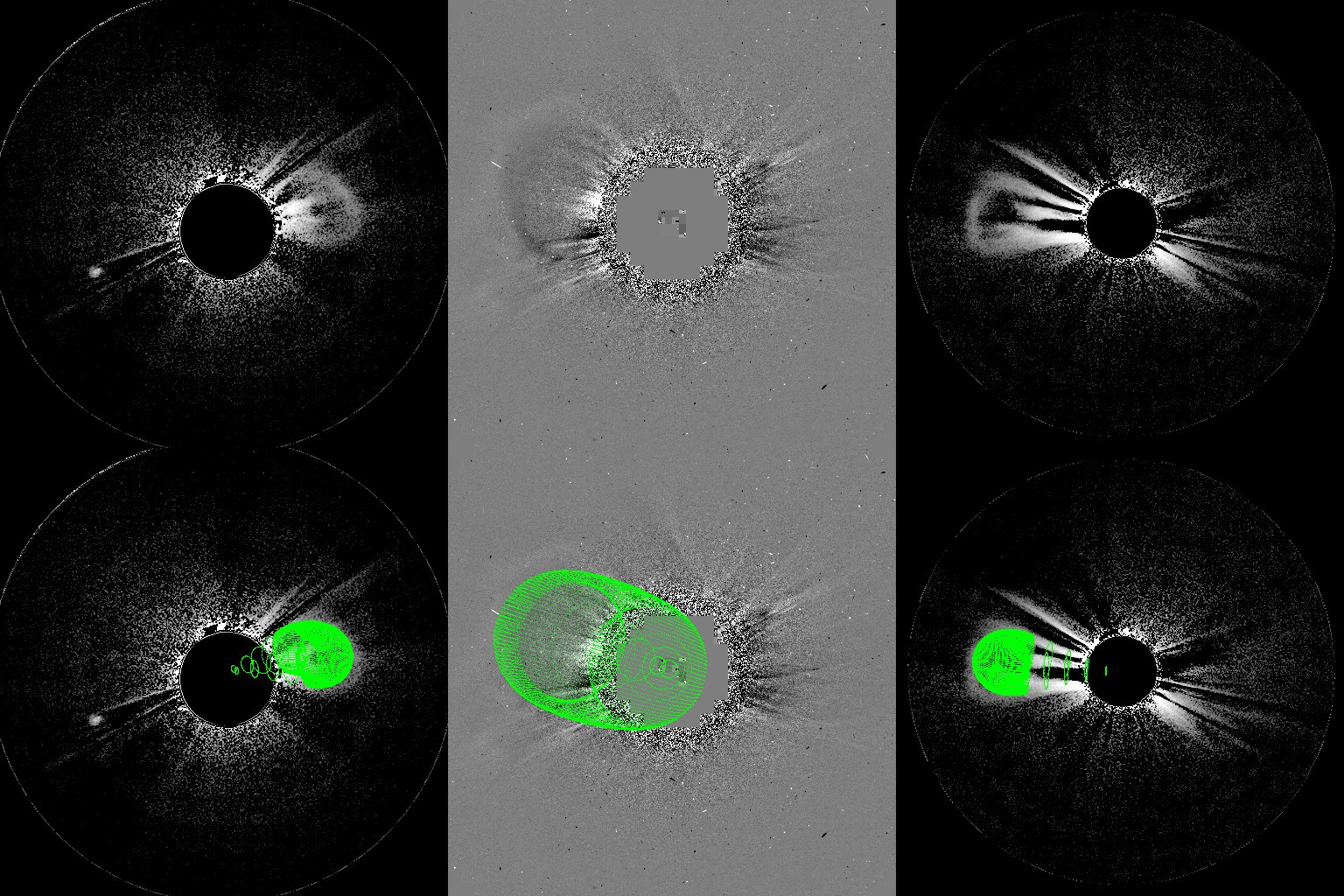}
\caption{CME 4 16-Jun-2010 19:24 UT height = 10}
\label{2010-06-16_1}
\end{center}
\end{figure} \end{landscape} 



\begin{landscape} \begin{figure}
\begin{center}
\includegraphics[width=20cm]{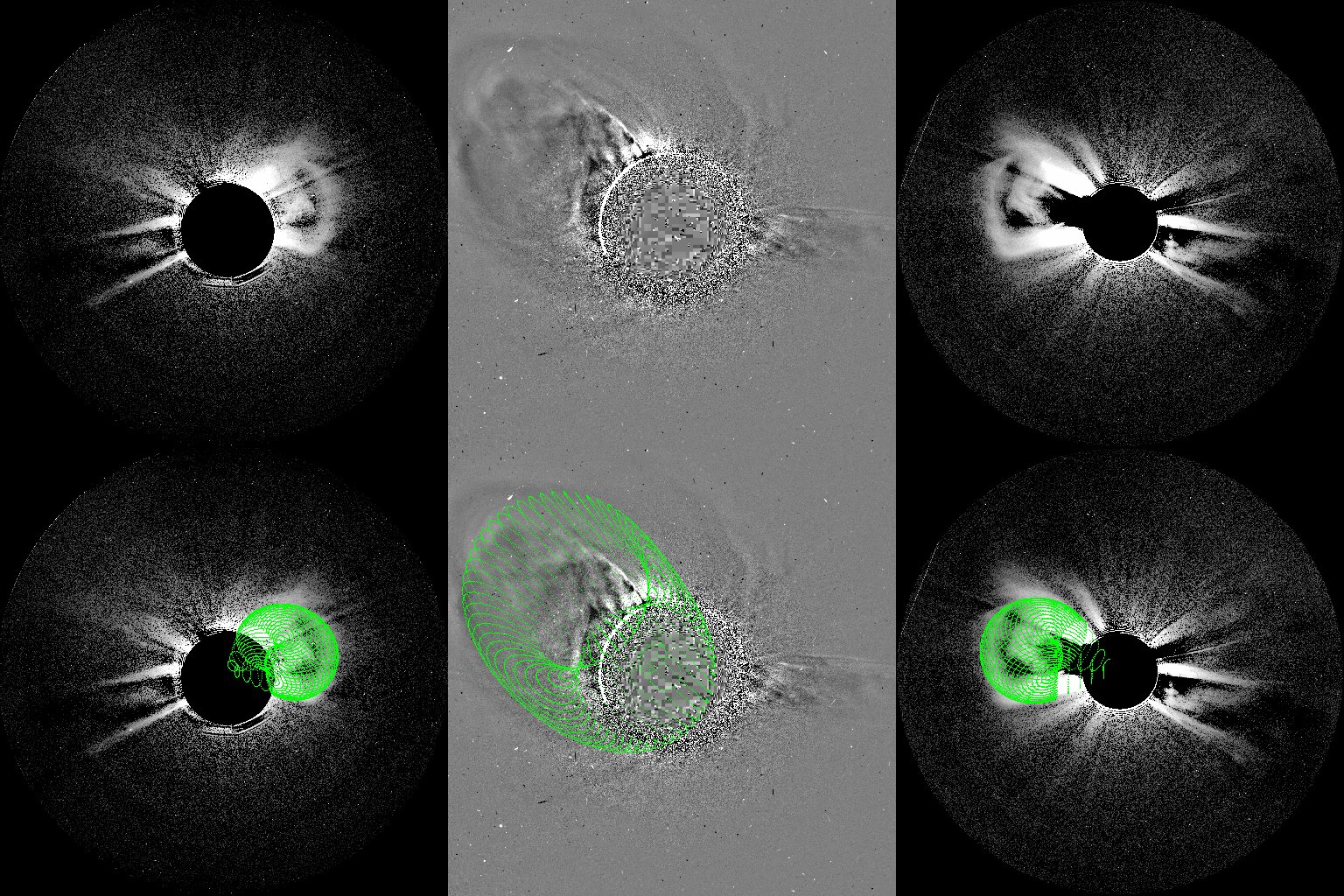}
\caption{CME 5 11-Sep-2010 05:24 UT height = 10}
\label{2010-09-11_1}
\end{center}
\end{figure} \end{landscape} 



\begin{landscape} \begin{figure}
\begin{center}
\includegraphics[width=20cm]{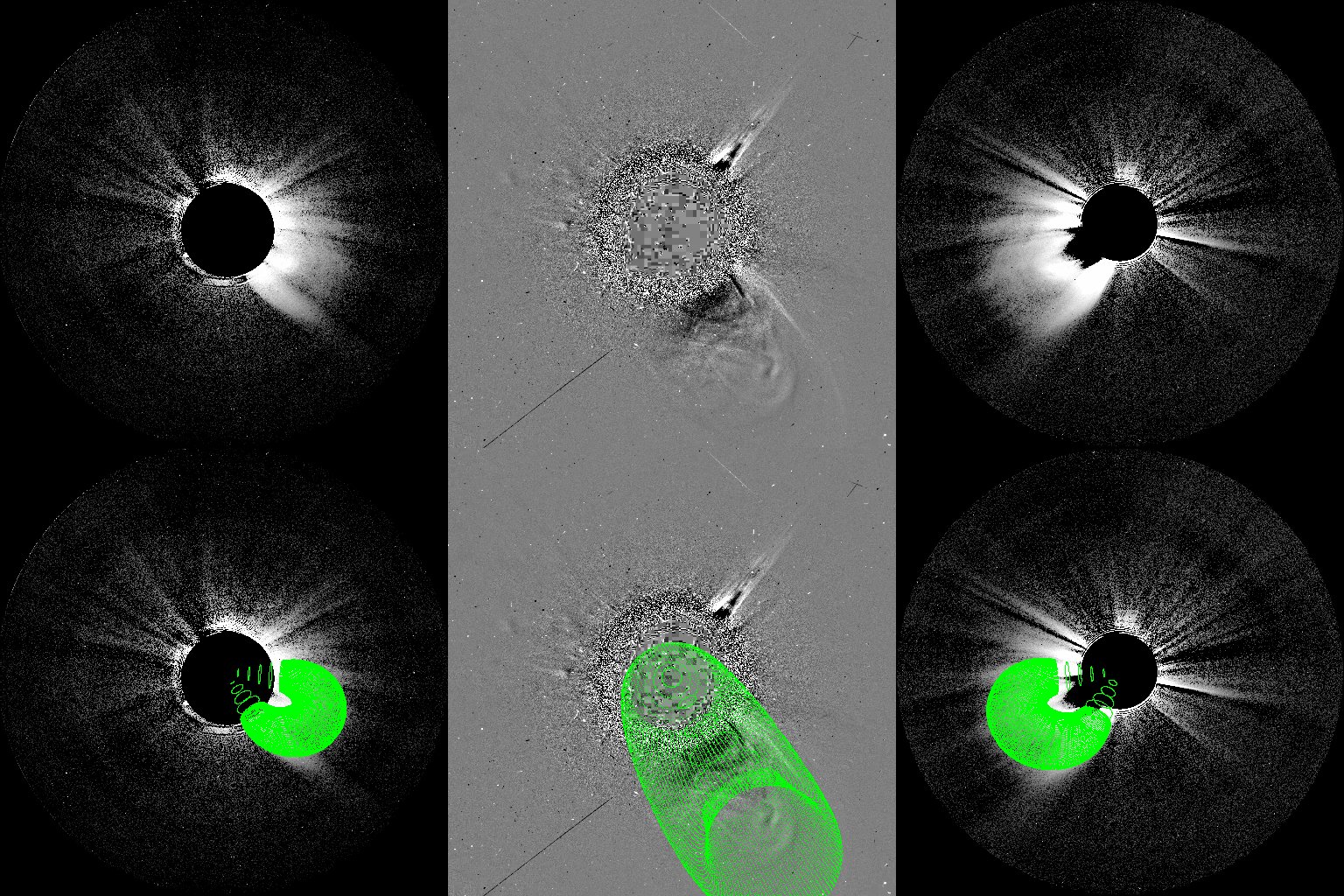}
\caption{CMR 6 26-Oct-2010 11:54 UT height = 10}
\label{2010-10-26_1}
\end{center}
\end{figure} \end{landscape} 



\begin{landscape} \begin{figure}
\begin{center}
\includegraphics[width=20cm]{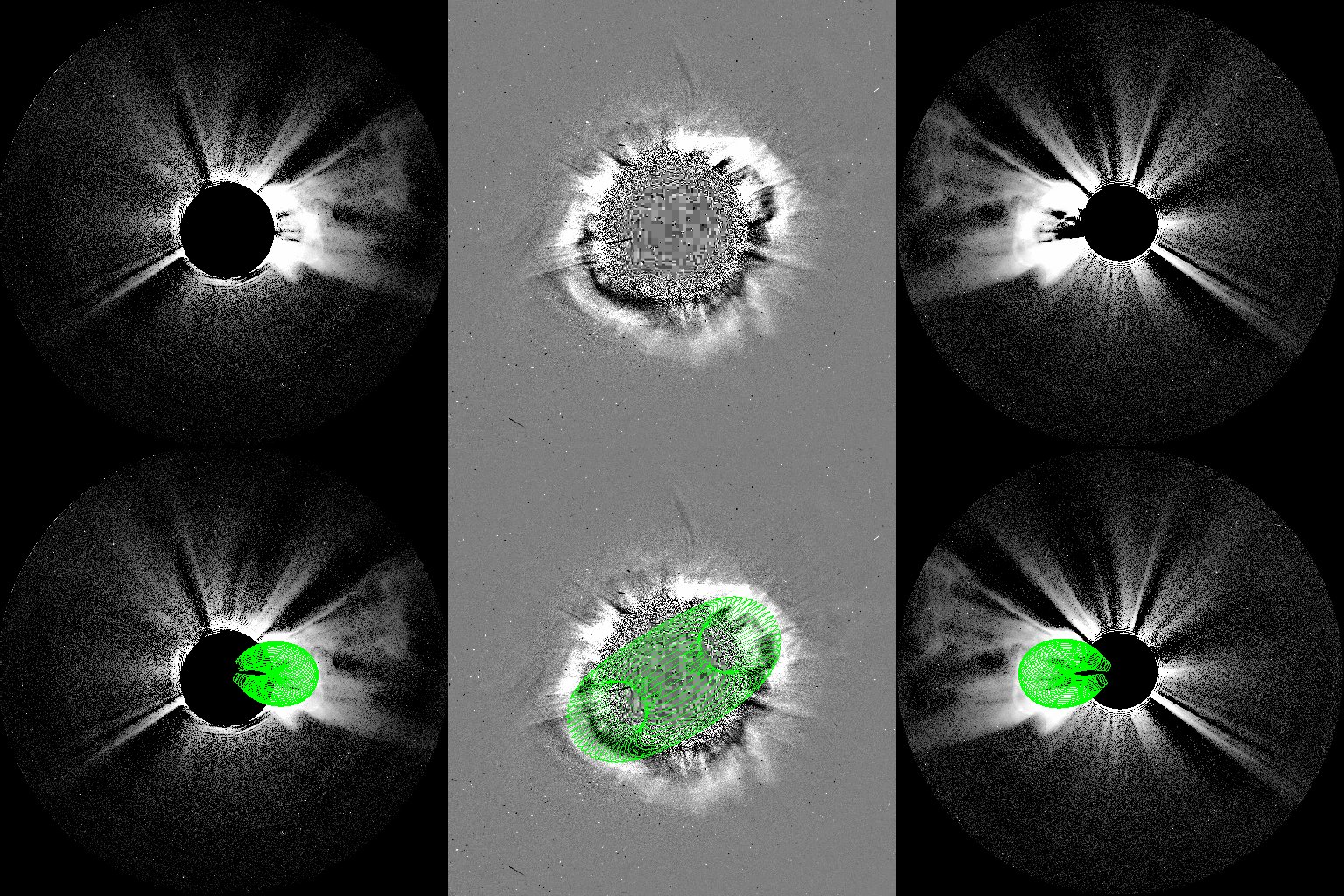}
\caption{CME 7 15-Feb-2011 02:54 UT height = 10}
\label{2011-02-15_1}
\end{center}
\end{figure} \end{landscape} 



\begin{landscape} \begin{figure}
\begin{center}
\includegraphics[width=20cm]{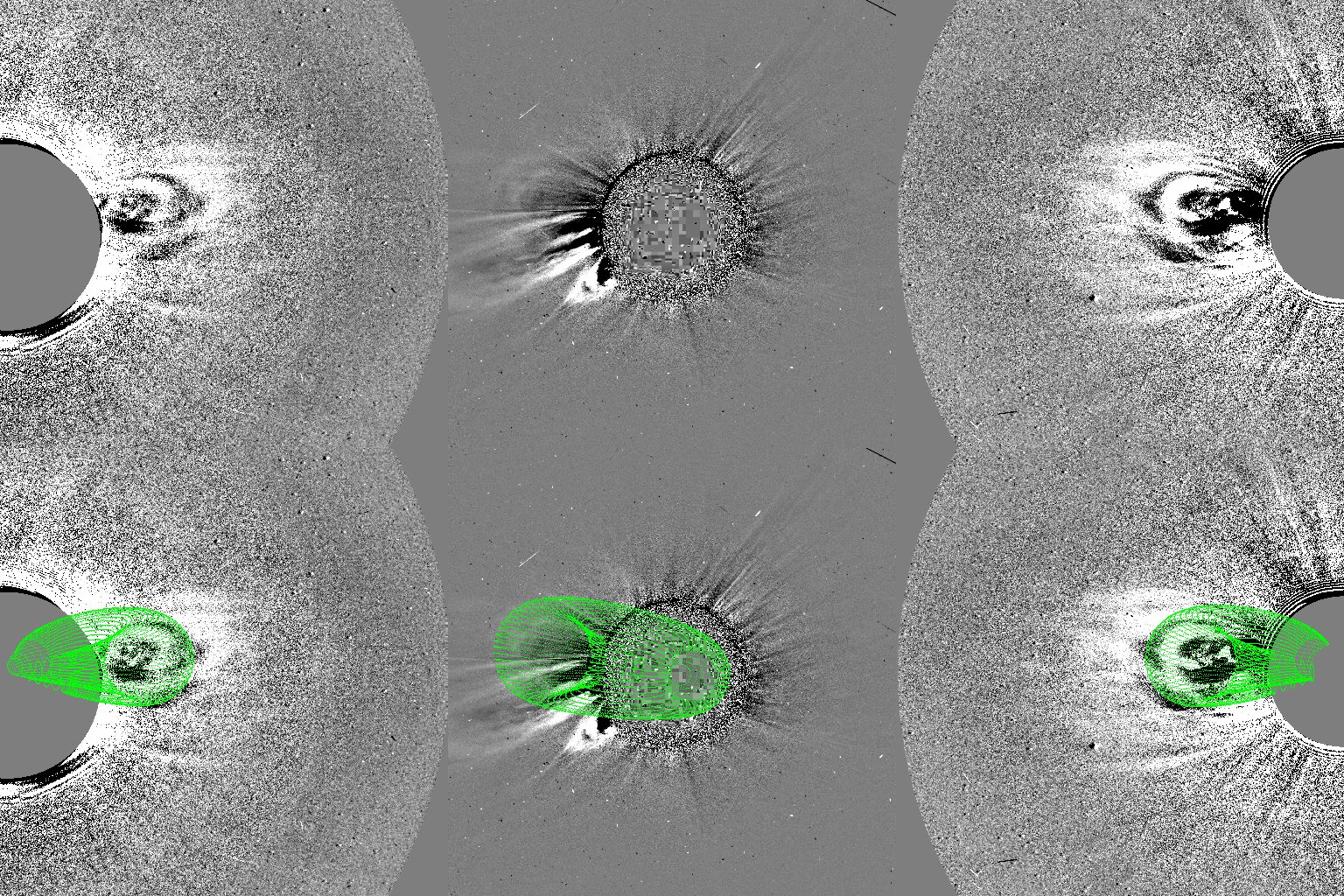}
\caption{CME 8 25-Mar-2011 11:54 UT height = 10}
\label{2011-03-25_1}
\end{center}
\end{figure} \end{landscape} 



\begin{landscape} \begin{figure}
\begin{center}
\includegraphics[width=20cm]{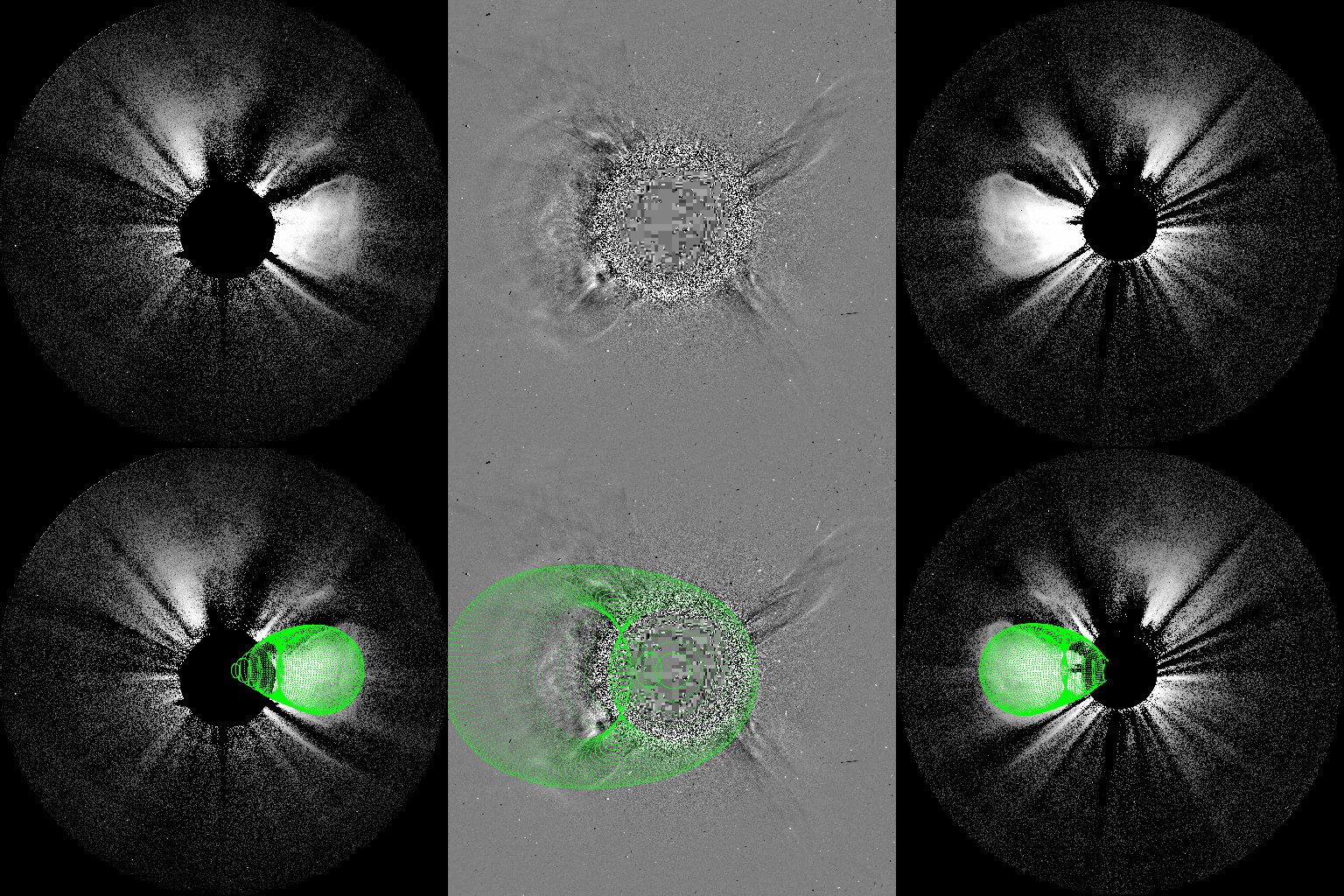}
\caption{CME 9 01-Jun-2011 20:39 UT height = 10}
\label{2011-06-01_1}
\end{center}
\end{figure} \end{landscape} 



 \appchapter[In Situ Data]{In Situ Data}\label{appinsitu}
In this Appendix, we give the in situ data used in this study. In the caption for each figure, we give the CME number and date of the in situ detection. All the data is from the WIND spacecraft and has been smooth to four points per hour. The light blue block demarcates the boundaries of the CME. The vertical dashed lines indicate the region of the data that was used for the EFR fit. 

\begin{figure}
\begin{center}
\includegraphics[width=16cm]{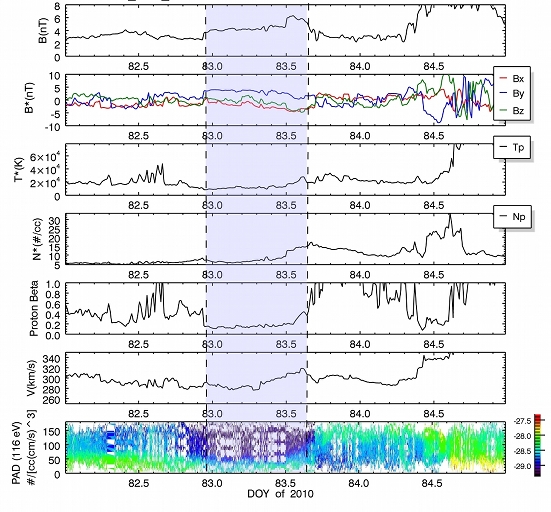}
\caption{CME 1 23-March-2010}
\label{2010-03-23}
\end{center}
\end{figure}  

\begin{figure}
\begin{center}
\includegraphics[width=16cm]{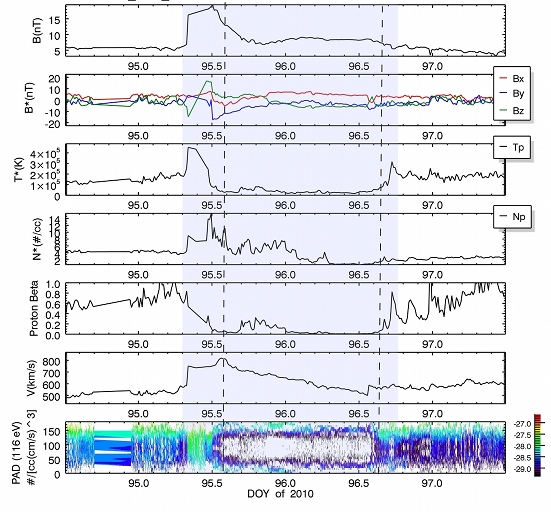}
\caption{CME 2 05-April-2010}
\label{2010-04-05}
\end{center}
\end{figure}  

\begin{figure}
\begin{center}
\includegraphics[width=16cm]{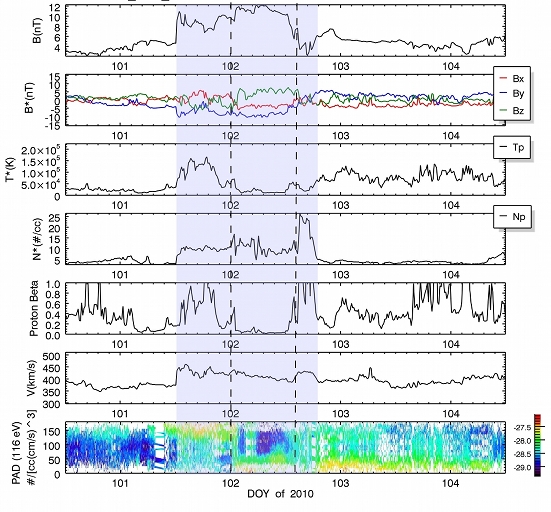}
\caption{CME 3 11-April-2010}
\label{2010-04-11}
\end{center}
\end{figure}  

\begin{figure}
\begin{center}
\includegraphics[width=16cm]{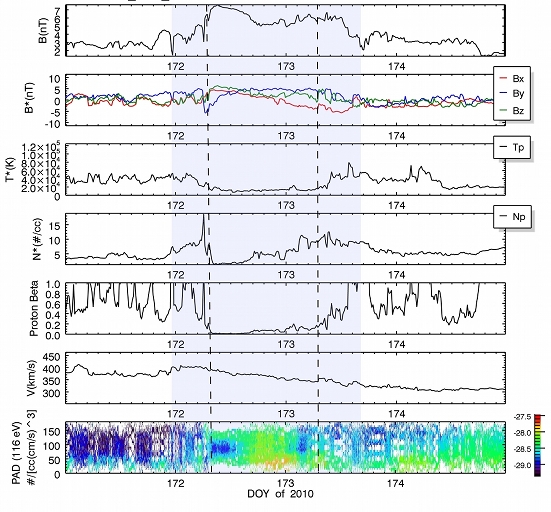}
\caption{CME 4 20-June-2010}
\label{2010-06-16}
\end{center}
\end{figure}  

\begin{figure}
\begin{center}
\includegraphics[width=16cm]{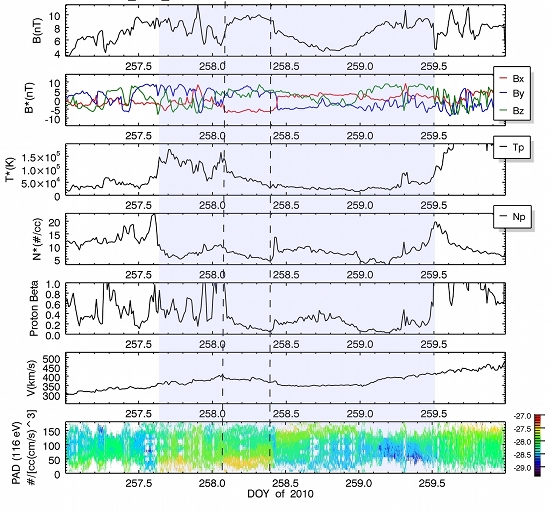}
\caption{CME 5 15-September-2010}
\label{2010-09-15}
\end{center}
\end{figure} 

\begin{figure}
\begin{center}
\includegraphics[width=16cm]{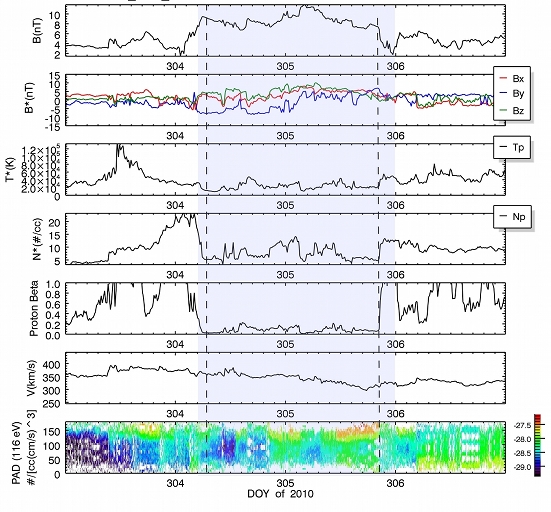}
\caption{CME 6 31-October-2010}
\label{2010-10-31}
\end{center}
\end{figure} 

\begin{figure}
\begin{center}
\includegraphics[width=16cm]{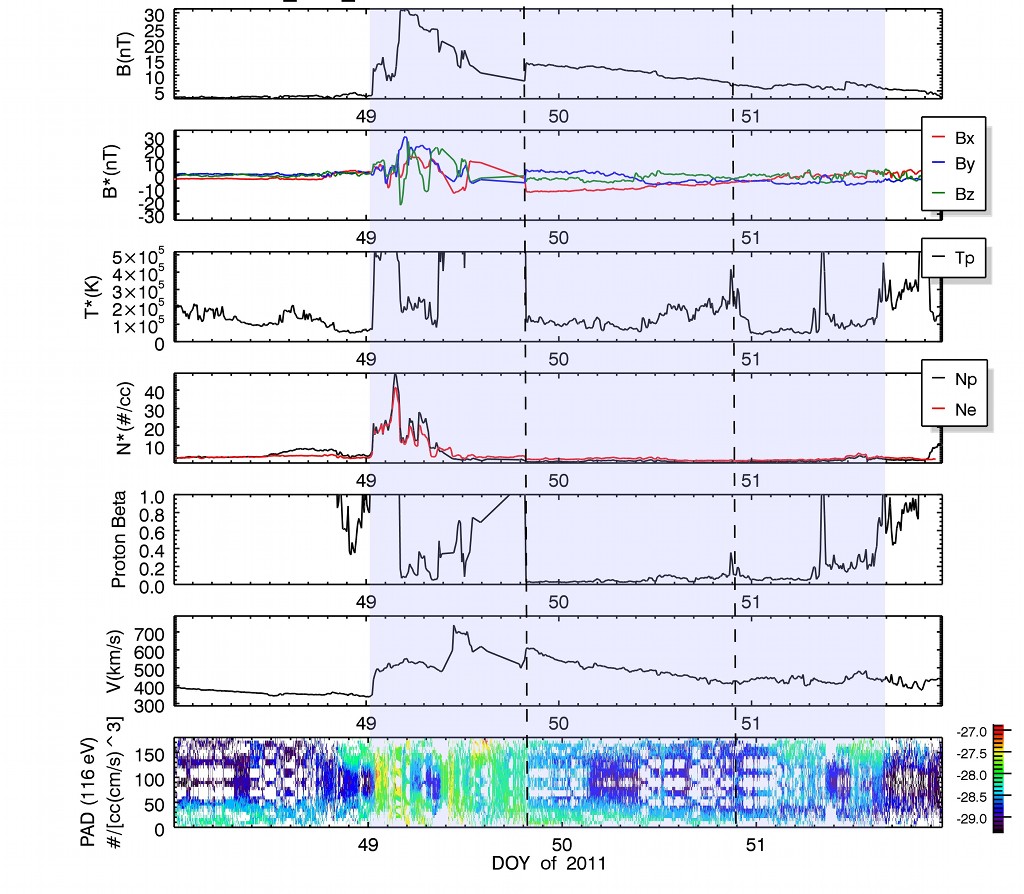}
\caption{CME 7 18-Feburary-2011}
\label{2011-02-18}
\end{center}
\end{figure} 

\begin{figure}
\begin{center}
\includegraphics[width=16cm]{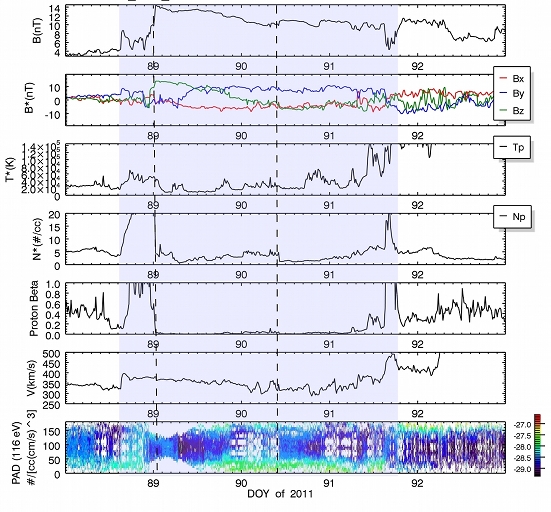}
\caption{CME 8 30-March-2011}
\label{2011-03-30}
\end{center}
\end{figure} 

\begin{figure}
\begin{center}
\includegraphics[width=16cm]{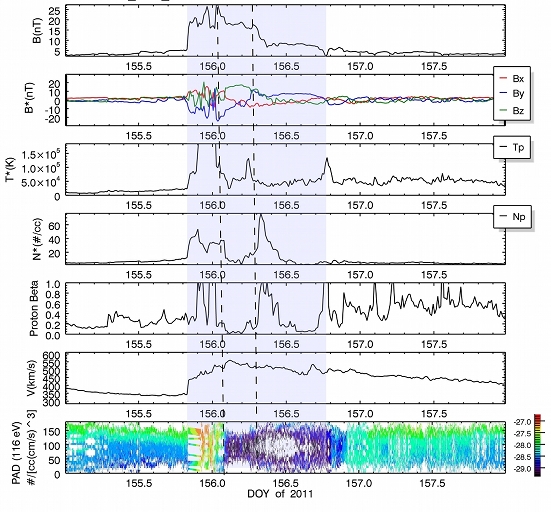}
\caption{CME 9 04-June-2011}
\label{2011-06-04}
\end{center}
\end{figure}



\thing
\bibliographystyle{kluwer}
\bibliography{biblio}

\cvpage

Robin Colaninno graduate from Guilford College, Greensboro, North Carolina in 2002 with a Bachelors of Science in Physics and a concentration in Mathematics for Physical Science Majors. As part of her degree, she submitted a thesis, \textit{UBVRI Photometry of the Galilean Satellites} to the Physics Department. She presented her undergraduate thesis at the 16th National Conference on Undergraduate Research in Whitewater, Wisconsin in 2002. While at Guildford College, she also helped setup the college's new observatory which she was the first to use for her thesis research. After graduating from Guilford College, she was employed by the Solar Physics Branch of the Naval Research Laboratory in 2003. At NRL, she has worked on a variety of aspects of solar physics. She has been the first author of two published papers and co-author on six others. Her paper on CME mass published in 2009 has 20 journal citations. She has also presented her work at numerous professional conferences. In 2008, she received Outstanding Student Paper Award for her poster presented at the America Geophysical Union Fall Meeting, San Francisco, California. In conjunction with her work at NRL, she has been persuing  a Physics PhD. at George Mason University since 2005. 

\end{document}